%% file: LLAGN.tex
\documentclass[fleqn,usenatbib]{mnras}

\usepackage{newtxtext,newtxmath}

\usepackage[T1]{fontenc}
\usepackage{ae,aecompl}

\usepackage{graphicx}	
\usepackage{amsmath}	
\usepackage{amssymb}	
\usepackage{caption}
\usepackage{graphicx}
\usepackage[utf8x]{inputenc}
\usepackage{lscape}
\usepackage{amsmath}
\usepackage[utf8x]{inputenc}
\usepackage{amssymb}
\usepackage{amsfonts}
\usepackage{tabularx}
\usepackage{natbib}
\usepackage{graphicx}
\usepackage{epsfig}
\usepackage{verbatim}
\usepackage[flushleft]{threeparttable}
\include{journaldefs}

\usepackage[caption=false]{subfig}
\usepackage{url}
\usepackage{textcomp}
\usepackage[squaren]{SIunits}
\usepackage{float}
\usepackage[T1]{fontenc} 
\usepackage{aecompl}
\usepackage{adjustbox}
\usepackage{hyperref}

\def\Tab{Table}

\def\deg{$^{o}\,$}
\def\arcm{$^{\prime}\,$}
\def\arcs{$^{\prime\prime}\,$}
\def\mujybm   {${\rm \mu}$Jy\,beam$^{-1}$}
\def\mjybm   {${\rm m}$Jy\,beam$^{-1}$}
\def\mujy   {${\rm \mu}$Jy}

\title[Low-luminosity Seyfert radio cores]{From radio-quiet to radio-silent: low luminosity Seyfert radio cores}

\author[E. Chiaraluce et al.]{
\href{http://orcid.org/0000-0002-4090-1327}{E. Chiaraluce}$^{1,2}$\thanks{elia.chiaraluce@inaf.it}, \href{https://orcid.org/0000-0002-5182-6289}{G. Bruni}$^1$, \href{http://orcid.org/0000-0003-0543-3617}{F. Panessa}$^1$, \href{https://orcid.org/0000-0002-8657-8852}{M. Giroletti}$^3$, \href{https://orcid.org/0000-0003-4470-7094}{M. Orienti}$^3$,
\newauthor 
\,\,\href{http://orcid.org/0000-0001-8744-7259}{H. Rampadarath}$^4$, \href{https://orcid.org/0000-0002-6689-9317}{F. Vagnetti}$^2$, \href{https://orcid.org/0000-0002-6562-8654}{F. Tombesi}$^{2,5,6,7}$
\smallskip
\\
$^{1}$ INAF - Istituto di Astrofisica e Planetologia Spaziali, via del Fosso del Caveliere 100, I-00133 Roma, Italy\\
$^{2}$ Dipartimento di Fisica, Univerisità di Roma Tor Vergata, via della Ricerca Scientifica 1, I-00133 Roma, Italy\\
$^3$ INAF-Istituto di Radioastronomia, via Gobetti 101, 40129 Bologna, Italy \\
$^4$ Jodrell Bank Centre for Astrophysics, School of Physics and Astronomy, University of Manchester, Turing Building, Oxford Road,\\ \,\,\,\,Manchester M13 9PL, UK \\
$^5$ Department of Astronomy, University of Maryland, College Park, MD, 20742, USA.\\
$^6$ X-ray Astrophysics Laboratory, NASA/Goddard Space Flight Center, Greenbelt, MD, 20771, USA.\\
$^7$ INAF – Osservatorio Astronomico di Roma, Via Frascati 33, 00044, Monte Porzio Catone (Roma), Italy. 
}

\date{Accepted 2019 February 25. Received 2019 February 25; in original form 2018 December 21}

\pubyear{2019}

\begin{document}
\label{firstpage}
\pagerange{\pageref{firstpage}--\pageref{lastpage}}
\maketitle

\begin{abstract}
A strong effort has been devoted to understand the physical origin of radio emission from low-luminosity AGN (LLAGN), but a comprehensive picture is still missing.
We used high-resolution ($\le1\,$arcsec), multi-frequency (1.5, 5.5, 9 and 14 GHz) NSF's Karl G. Jansky Very Large Array (VLA) observations to characterise the state of the nuclear region of ten Seyfert nuclei, which are the faintest members of a complete, distance-limited sample of 28 sources. With the sensitivity and resolution guaranteed by the VLA-A configuration, we measured radio emission for six sources (NGC3185, NGC3941, NGC4477, NGC4639, NGC4698 and NGC4725), while for the remaining four (NGC0676, NGC1058, NGC2685 and NGC3486) we put upper limits at tens \rm{\mujybm} level, below the previous $0.12\,$\mjybm  level of Ho $\&$ Ulvestad (2001), corresponding to luminosities down to $L\,\le\,10^{19}$ W\,Hz$^{-1}$ at 1.5 GHz for the highest RMS observation. Two sources, NGC4639 and NGC4698, exhibit spectral slopes compatible with inverted spectra ($\alpha\,\le\,$0, $S_{\nu}\,\propto\,{\nu}^{-\alpha}$), hint for radio emission from an optically-thick core, while NGC4477 exhibits a steep (+0.52$\pm$0.09) slope. The detected sources are mainly compact on scales $\le$ arcseconds, predominantly unresolved, except NGC3185 and NGC3941, in which the resolved radio emission could be associated to star-formation processes. A significant X-ray - radio luminosities correlation is extended down to very low luminosities, with slope consistent with inefficient accretion, expected at such low Eddington ratios. Such sources will be one of the dominant Square Kilometre Array (SKA) population, allowing a deeper understanding of the physics underlying such faint AGN.
\end{abstract}




\section{Introduction}
There has been increasing evidence in literature that radio-quiet Active Galactic Nuclei (AGN) are radio emitters at some level \citep[e.g.][]{HU01, Nagar2000}. Several works \citep[e.g.][]{HU01, Nagar2000, GirolettiPanessa2009,PG13,Baldi2018} have been dedicated to understand the physical origin of radio emission in Low-Luminosity AGN (LLAGN), a category comprising low-luminosity Seyferts, LINERs and transition nuclei \citep{Nagar2000}, and defined as AGN with $L_{H\alpha}\le$10$^{40}$\,ergs\,s$^{-1}$ by \citet{HoFilippenkoSargent1997a}. Although differences between the two classes can be found in the X-ray and radio bands, they share some properties, e.g. the absence of a Big Blue Bump in the Optical/UV, which clearly make them distinguishable from the classical Seyferts and Quasars \citep{Ho2008}.  
The faintness of their radio emission \citep[radio luminosities between 10$^{32}$ and 10$^{40}$\,ergs\,s$^{-1}$,][]{Baldi2018} requires high-sensitivity observations to carry out a robust study. LLAGN often show flat-spectrum radio cores, usually explained in terms of free-free emission/absorption or Advection-Dominated Accretion Flow \citep[ADAF,][]{NarayanYi1994}, synchrotron self-absorbed emission from a scaled-down version of a more powerful AGN jet \citep[e.g.][]{Falcke1999,Nagar2000}, possibly coupled with an ADAF \citep{FalckeMarkoff2000}, or a standard thin accretion disk \citep{Ghisellini2004}. A flat spectrum is also observed from Star-Forming (SF) regions, accompanied by diffuse and low surface brightness emission \citep{OrientiPrieto2010}. Synchrotron emission from shocked regions in low collimated outflows has been claimed as the origin of radio emission in quasar \citep{Zakamska2014} and, in a few cases, the outflow structure has been mapped at high resolution in the local Universe \citep{Kharb2006}. 

Observations with very-long baseline interferometry (VLBI) have established the occurrence of compact cores in a fraction of LLAGN \citep[e.g.][]{AndersonUlvestad2005,WrobelHo2006} down to milli-arcseconds (mas) scales, with high brightness temperatures ($T_B\ge$10$^{8}$\,K), and extended features resembling jet-like outflows \citep[e.g.][]{Falcke2000,Bontempi+2012}. High brightness temperature coupled with flat-spectrum radio cores at mas scales have been interpreted as signatures of non-thermal synchrotron self-absorbed (SSA) emission originating from the base of a jet. However, some high-resolution observations led to different conclusions, like in the case of NGC~1068 and NGC~4477, in which the nuclear radio emission has been attributed to thermal free-free emission from an X-ray heated corona \citep{Gallimore2004,Bontempi+2012}.

In the last two decades, a strong effort has been devoted to the understanding of the origin of the radio emission (that typically traces the ejected material) and its relation to the X-ray emission (tracing the disc corona accreting system) in black holes at different luminosity levels, leading to the formulation of the 'Fundamental Plane' of black-hole activity \citep{Merloni2003,FalckeKordingMarkoff2004}, suggesting a disk-jet coupling from low to high mass black holes. Moreover, \citet{Panessa2007} found a correlation between the nuclear $2-10$\,keV X-ray luminosity and radio luminosity (at 2, 6 and 20\,cm) for a number of local Seyfert galaxies, suggesting again a strong physical coupling. These relations are interesting in light of a possible AGN - X-ray binaries (XRB) analogy. According with this analogy LLAGN, similarly to 'hard state' XRB, would follow an inefficient accretion track \citep[e.g.][]{FalckeKordingMarkoff2004}. This analogy has been suggested by \citet{GuCao2009} for a sample of LINERs and local Seyferts and, more recently, by \citet{PG13}, who found a slope for the X-ray - radio luminosity relation comparable with the $\sim\,$0.7 found by \citet{Gallo2003} in low-hard state XRBs for a sample of local Seyfert galaxies.

In spite of the hints provided by the above mentioned efforts, however, a conclusive picture has not emerged, yet. Moreover, a number of LLAGN have remained undetected \citep[e.g.][]{HU01}, posing intriguing questions about the nature of their nuclei. 
In this work we present multi-frequency NSF's Karl G. Jansky Very Large Array (VLA) observations for ten Seyfert galaxies belonging to the complete, distance limited sample of \citet{Cappietal06}, refined by \citet{PG13}, for which radio nuclei were not detected in previous radio studies. For these ten sources, we take advantage of the improved sensitivity of the upgraded VLA, reaching lower flux limits thanks to a sensitivity level as low as $\sim$27 \mujybm (3$\,\sigma$), characterising their radio spectra and the morphology of the radio emission. Having completed the radio information for the whole sample of \citet{Cappietal06}, we provide detection rates and investigate the correlation between the X-ray and radio luminosity down to very low regimes.

The paper is organised as follows. In Section 2, we introduce the sample used for our analysis; in Section 3, we describe the observations and the data reduction procedure; in Section 4, we show the results obtained in the data analysis; in Section 5, we discuss the results and in Section 6 we summarise our findings and considerations.

Throughout the paper, we use a flat $\Lambda-$CDM cosmological model with $\Omega_{\rm{M}}$=0.3, $\Omega_\Lambda=$0.7 and a Hubble constant of 70 km s$^{-1}$ Mpc$^{-1}$ \citep{Bennett2003}.

\section{The Sample}


\begin{table*}
\caption{The targets used in this study. \textit{Columns:} (1) Target name; (2) $\&$ (3) J2000 optical positions from \citet{HU01}; (4) optical classification, S1 are Seyfert 1 sources, S2 are Seyfert 2 sources and T are transition objects (The ':' indicates an uncertain classification); (5) distance in Mpc. Columns 4 and 5 were obtained from \citet{Panessa2006}; the positions of NGC~4639 were obtained from \citet{Hakobyan+12}.}
\begin{tabular}{ccccc}
\hline
\hline
\multicolumn{1}{c}{Name} & RA  & Dec  & Seyfert Type & Distance \\ 
 & (J2000) & (J2000) &  & (Mpc) \\ 
 (1) & (2) & (3) & (4) & (5) \\ 
 \hline
NGC~676 & 01 48 57.38  $\pm$ 0.11 & +05 54 25.7 $\pm$ 1.7 & S2: & 19.5 \\ 
NGC~1058 & 02 43 30.24 $\pm$ 0.22 & +37 20 27.2 $\pm$ 2.6 & S2 & 9.1 \\ 
NGC~2685 & 08 55 34.79 $\pm$ 0.39 & +58 44 01.6 $\pm$ 3.1 & S2/T2: & 16.2 \\ 
NGC~3185 & 10 17 38.66 $\pm$ 0.12 & +21 41 17.2 $\pm$ 1.7 & S2: & 21.3 \\ 
NGC~3486 & 11 00 24.10 $\pm$ 0.20 & +28 58 31.6 $\pm$ 2.7 & S2 & 7.4 \\ 
NGC~3941 & 11 52 55.42 $\pm$ 0.21 & +36 59 10.5 $\pm$ 2.5 & S2: & 12.2 \\ 
NGC~4477 & 12 30 02.22 $\pm$ 0.18 & +13 38 11.3 $\pm$ 2.6 & S2 & 16.8 \\ 
NGC~4639 & 12 42 52.36 $\pm$ 0.03 & +13 15 26.5 $\pm$ 0.1 & S1 & 22.9 \\ 
NGC~4698 & 12 48 22.98 $\pm$ 0.16 & +08 29 14.8 $\pm$ 2.4 & S2 & 16.8 \\ 
NGC~4725 & 12 50 26.69 $\pm$ 0.16 & +25 30 02.3 $\pm$ 2.2 & S2: & 13.2 \\ 
\hline
\end{tabular}
\label{tab:targets}
\end{table*}


In this work we present results of VLA A-configuration observations for a group of ten Seyferts ('Reference sample', Table \ref{tab:targets}), nine Type~2 and one Type~1, the faintest members of the sample of 28 Seyfert galaxies in \citet{Cappietal06} ('Parent sample'). Indeed, the sources in the Reference Sample are characterised by an average Eddington ratio $\langle\log\,L_{2-10\,\rm{keV}}/L_{\rm{Edd}}\,\rangle\,\sim\,$-5.6, nearly an order of magnitude lower than the average Eddington ratio for the Parent sample ($\,\sim\,-4.7$). However, the original sample was initially constituted by only 27 sources: NGC~3982 was excluded because of lack of XMM-\textit{Newton} observations. \citet{PG13} updated the sample to 28 galaxies including NGC~3982. This last sample, which we will call 'Parent sample', is complete and distance limited (D$\le$23$\,$Mpc) and sources have been chosen as the nearest ones in the optically selected sample of 52 Seyfert galaxies given in \citet{HoFilippenkoSargent1997a}, that was extracted from the Palomar optical spectroscopic survey of nearby galaxies \citep{FilippenkoSargent1985,HoFilippenkoSargent1995}, comprising 486 northern ($\delta$>$0\,^\circ$) galaxies provided with homogeneous spectral classification \citep{HoFilippenkoSargent1997a}. This sample is complete to $B_T=$12.0\,mag and 80 per cent complete to $B_T=$12.5\,mag \citep{Sandage1979}, and it has also other several advantages, concerning selection biases and the wide range of radio luminosities, see \citet{HU01} for a deeper discussion.

Recently, \citet{Baldi2018}, in their LeMMINGs 1.5-GHz parsec-scale survey, have revised the optical classification of \citet{HoFilippenkoSargent1997} using the spectroscopic diagnostic diagrams based on criteria by \citet{Kewley2006} and \citet{Buttiglione2010}. The basic difference in this classification is in the transition from Seyferts to LINERs, as low $O[III]/H\beta$ Seyferts are classified as LINERs, and the use of a more stable index, which is the average of low-ionisation line ratios. Moreover, they removed the 'Transition Galaxies' class \citep[for the details see][]{Baldi2018}. We reconsidered the Parent and Reference sample optical classification in the light of the revised classification of \citet{Baldi2018} finding that approximately 40 per cent of sources in the Parent sample (11/28) can be classified as LINERs, and the same percentage holds if we select only the Reference sample (4/10), i.e., NGC~3941, NGC~1058, NGC~4477 and NGC~4639. Note however, that these objects populate a region very close to the Seyferts/LINERS boundaries. Considering the related uncertainties, we prefer to keep the Seyfert classification, in order to be consistent with previous works of our group and literature \citep[e.g.][]{PG13}.

The Palomar sample has been the subject of extensive observational campaigns, through VLA 
\citep[e.g.][]{Nagar2000,Nagar2005,HU01}, VLBI \citep[e.g.][]{Falcke2000,Nagar2005,PG13} and, more recently, of the LeMMINGs 1.5-GHz parsec-scale survey \citep[][D. R. A. Williams' thesis]{Baldi2018} and the 15-GHz survey performed by \citet{Saikia2018}.

In particular, \citet{HU01} conducted a radio continuum survey at 6 and 20\,cm with the VLA (in configuration B and A, respectively) at $\sim$1\,arcsec resolution for the sample of 52 Seyfert galaxies in \citet{HoFilippenkoSargent1997a}. They detected radio emission in the 85 per cent of cases at 6\,cm and in the 71 per cent at 20\,cm for the whole sample, but when considering the subsample of 28 sources in the Parent sample, the VLA
detection rates are 82 and 64 per cent at 6 and 20\,cm, respectively \citep{PG13}. The morphology is predominantly compact on arcseconds scales, either unresolved or slightly resolved, and the nuclear spectral indices range from steep to flat/inverted. Among the sources of the reference sample listed in Table \ref{tab:targets}, five, namely NGC~0676, NGC~1058, NGC~2685, NGC~3486 and NGC~4725 have not been detected by \citet{HU01} at the 3$\sigma$ sensitivity threshold of $\sim$0.12$-$0.14\,\mjybm at either 20 or 6\,cm. The remaining sources have not been detected at 20\,cm, at 6\,cm they are considered marginal detections, as they have peak intensities of $\sim$3.5$-$6$\,\sigma$, and their morphology is classified as ambiguous.

At very-long baseline interferometry (VLBI) milli-arcseconds scales, \citet{GirolettiPanessa2009} analysed five sources, NGC~4051, NGC~4388, NGC~4501, NGC~5033 and NGC~5273, among the sources in the Parent sample, detecting radio emission at $\sim$100\,\mujybm level, except for NGC~5273, with physical parameters consistent with different underlying physical processes. \citet{Bontempi+2012} performed a dual-frequency, 1.7 and 5\,GHz, analysis with the European VLBI Network (EVN) of the faintest Seyferts nuclei in the Parent sample. They did not detect radio emission at the 3$\sigma$ sensitivity level of $\sim$20-120$\,$\mujybm for the four sources NGC~3185, NGC~3941, NGC~4639 and NGC~4698 which have been identified by \citet{HU01} as marginal detections. Only NGC~4477 has been detected at 5 GHz, although with a low significance, exhibiting flux density compatible with that quoted by \citet{HU01}, and physical parameters which would rule out synchrotron self-absorption (SSA) as emission mechanism in favour of thermal free-free emission.

Moreover, \citet{PG13} conducted a census of physical properties of these 28 galaxies using published VLA data (at 20 and 6\,cm, mostly from \citealt{HU01}) and new and published VLBI data (at 20 and 6\,cm) for 23/28 galaxies. They found that the brightness temperature as derived from VLBI observations is higher, on average, in type 1 than in type 2, and this evidence has been interpreted as hint for a free-free domination in type 2. They also found that there is a significant radio-X-rays luminosity relation when considering VLA scales, but at VLBI scales they did not find any correlation.

In the present work we present new VLA - configuration A observations for the ten faintest members of the \citet{Cappietal06} sample ('Reference sample') and, coupled with previous data available for the remaining sources in the \citet{Cappietal06} sample, we will establish for the first time, radio detection rates for a complete, distance limited sample of nearby Seyfert galaxies with homogeneous observations.
 

\section{Observations and Data Reduction}
\label{sec:procedure}

\begin{center}
\centering
\begin{table}\footnotesize
\centering
\caption{List of calibrators (flux and phase) per observation group. \textit{Columns:} (1) Target name; (2) Observation date; (3) Absolute flux density scale calibrator; (4) Phase Calibrator}

\begin{tabular}{cccc}
\hline
 & & \multicolumn{2}{c}{Calibrators} \\
\cline{3-4} \\
Target & Obs Date & Flux & Phase \\ 
  &  (dd/mm/yy) &  &  \\ 
 (1) & (2) & (3) & (4)  \\ 
 \hline
 \hline
NGC~0676 & 09/11/12 & 3C48 & J0149+0555 \\ 
NGC~1058 & 09/11/12 & 3C48 & J0230+4032 \\
\hline 
NGC~2685 & 05/01/13 & 3C286 & J0854+5757 \\ 
NGC~3185 & 05/01/13 & 3C286 & J1014+2301 \\
\hline 
NGC~3486 & 30/12/12 & 3C286 & J1102+2757 \\ 
NGC~3941 & 30/12/12 & 3C286 & J1146+3958 \\
\hline 
NGC~4477 & 17/12/12 & 3C286 & J1254+1141 \\ 
NGC~4639 & 17/12/12 & 3C286 & J1254+1141 \\
\hline 
NGC~4698 & 17/12/12 & 3C286 & J1239+0730 \\ 
NGC~4725 & 17/12/12 & 3C286 & J1221+2813 \\ 
\hline
\end{tabular}

\label{tab:radObs}
\end{table}
\end{center}

\begin{table*}
\caption{Total observing time on each target per frequency band L (1.5 GHz), C (5.5 GHz), X (9 GHz) and Ku (14 GHz), and the expected (theoretical) sensitivity.} 
\centering
\begin{tabular}{ccccccccc}
\hline
  & \multicolumn{2}{c}{L-band}  & \multicolumn{2}{c}{C-band}  & \multicolumn{2}{c}{X-band}  & \multicolumn{2}{c}{Ku-band}
 \\ 
Target & Time & $\sigma_{\mathrm{th}}$ & Time & $\sigma_{\mathrm{th}}$& Time & $\sigma_{\mathrm{th}}$& Time & $\sigma_{\mathrm{th}}$ \\ 
 & (min) & (\mujybm) & (min) & (\mujybm) & (min) & (\mujybm) & (min) & (\mujybm)\\ 
\hline
\hline
 NGC~0676 & 6 & 25 & 7 & 9 & 8 & 9 & 8 & 11\\ 
 NGC~1058 & 6 & 25 & 7 & 9 & 8 & 9 & 8 & 11\\ 
 NGC~2685 & 6 & 25 & 6 & 10 & 8 & 9 & 8 & 11\\ 
 NGC~3185 & 6 & 25 & 6 & 10 & 8 & 9 & 8 & 11\\ 
 NGC~3486 & 6 & 25 & 7 & 9 & 8 & 9 & 8 & 11\\ 
 NGC~3941 & 8 & 20 & 8 & 9 & 8 & 9 & 8 & 11\\ 
 NGC~4477 & 8 & 20 & 8 & 9 & 8 & 9 & 10 & 10\\ 
 NGC~4639 & 8 & 20 & 8 & 9 & 8 & 9 & 10 & 10\\  
 NGC~4698 & 6 & 25 & 7 & 9 & 8 & 9 & 8 & 11\\ 
 NGC~4725 & 6 & 25 & 7 & 9 & 8 & 9 & 8 & 11\\ 
\hline
\hline
\multicolumn{9}{l}{\textbf{}}\\
\multicolumn{9}{l}{{}}\\
\end{tabular}
\label{table:rmsTable}
\end{table*}

A total of 10 hours of observations, divided over 4 days in November, December 2012 and
January 2013, were obtained with the VLA in A array configuration. The ten targets of the Reference sample were observed in five groups, with 2 hours dedicated to each group. \Tab~\ref{tab:radObs} lists the observations, sub-divided by group (based upon observing date). 

The observations were conducted in L (1.5 GHz), C (5.5 GHz), X (9 GHz) and Ku (14 GHz) bands, and each source has been observed for a total of 32 minutes, while the flux density calibrator has been observed for a total of 18 minutes per group. The scheduling blocks have been organised as follows. For a single observing block, the observations at different frequencies have been switched from one source to the other, in order to optimise the \textit{(u,v)} plane coverage. Therefore, the exposure time of a source at a certain frequency have been split into several scans, each one bracketed by the observation of the phase calibrators, which are thus observed every 2-3 min depending on the scheduling block. 

Table \ref{table:rmsTable} lists, for each source and for each frequency, the Time-On-Source (TOS) together with the expected theoretical sensitivity. The total observing bandwidths were 1024\,MHz for L band, while 2048\,MHz were used for the C, X and Ku bands. All bands were subdivided into 16 spectral windows of 64 channels. 

The data calibration and reduction were performed using the Common Astronomy Software Applications (\textsc{casa} 5.3.0 version, \citealt{McMullin+07}).
The full, unaveraged datasets (five in total) were downloaded from the NRAO science data archive\footnote{\url{https://archive.nrao.edu/archive/archiveproject.jsp}} as SDM-BDF datasets with flags generated during the observations applied. Each dataset contained the calibrators (flux, phase and bandpass) and two targets for the four frequency bands. The calibration was performed using the \textsc{casa} calibration pipeline 5.1.1-5. After calibration, the resulting plots of the calibrators were inspected for RFI and the different bands were split into separate MS files.

Since the observations covered wide bandwidths, the calibrators were imaged using the multi-frequency synthesis (mfs) algorithm \citep{CCW90, RC11}, with nterms=2. The integrated flux densities of the flux calibrators (3C48 dataset 1, and 3C286 all other datasets) were measured using the \textsc{casa} task \textsc{imfit}, and compared to the modelled flux densities given by \citet{PerleyButler13}. All measured flux densities were found to be within the 5 per cent of the tabulated flux densities, so we adopted an average 5 per cent flux calibration error. The accuracy of phase calibrators is typically of order of $\sim$1 mas, so our estimates of radio positions in the four bands are dominated by uncertainties associated to the estimation of the peak of elliptical gaussian fit performed by the \textsc{casa} task \textsc{imfit}.

We adopted different imaging strategies for the L-band and for the C, X and Ku bands, performed using the \textsc{casa} task, \textsc{tclean}. 

For the L-band data, we used the deconvolution procedure of \citet{Hogbom+1974} in \textsc{tclean} with image size of 8192 pixels (0.26 arcsec per pixel); we adopted wide-imaging technique \textsc{w-projection} in order to take into account the non-coplanarity of baselines and we applied a primary-beam correction in a non-interactive mode. The L-band maps have then been visually inspected and, in cases in which artifacts were still present due to outlier sources, we rerun the clean algorithm with outliers field in an interactive mode. The position of the outliers have been checked via the FIRST Survey\footnote{Faint Images of the Radio Sky at Twenty cm \citep{Becker1995}}.

Considering the C, X and Ku bands, each target field has been imaged in \textsc{tclean} using a multi-term, multi-frequency synthesis algorithm (with nterms=2) and image size of 2048 pixels (0.07, 0.04 and 0.03 arcsec per pixel for C, X and Ku bands, respectively). As before, in cases in which artifacts in final maps were still present, we rerun the clean algorithm with outliers field.

In all the four frequency bands, initial imaging of each target field has been performed using the full array angular resolution (i.e. no tapering), with Briggs \citep{Briggs1995} weighting with robust parameter equal to 0.5.

The positions, peak intensities, integrated flux densities, deconvolved sizes and position angle (PA) of the sources were estimated by fitting a two-dimensional Gaussian in the image plane via the \textsc{casa} task \textsc{imfit}. We determined the rms noise of each map from a source-free annular region around the source. 
The resulting average RMS in the four bands is approximately 40, 12, 10 and 9 \mujybm in the L, C, X and Ku bands, respectively.
The uncertainty in the final flux density measurements are affected by fitting errors from \textsc{imfit}, and flux calibration error of 5 per cent, which are added in quadrature and adopted as the error measurements. The positional accuracy of the detected radio nuclei is limited by the positional accuracy of the phase calibrators, typically few mas, and by the accuracy of the gaussian fit to the source brightness distribution as performed by \textsc{imfit}. The uncertainty on radio position expressed in columns 8 and 9 in Table \ref{table:fluxTable} is therefore the sum in quadrature of the two contributions.

\section{Results}

\begin{table*}
\caption{Imaging results. \textit{Columns:} (1) Target name; (2) Frequency band; (3) Image noise rms [\mujybm]; (4) Integrated flux density (\mujy); (5) Peak intensity (\mujybm); (6) Deconvolved FWHM dimensions (major $\times$ minor axis) for the fitted source, determined from an elliptical Gaussian fit source size (arcsec); (7) Source position angle (deg); (8) $\&$ (9) Detected source position in epoch J2000 (hh:mm:ss and \deg:\arcm:\arcs).} 
\centering
\begin{adjustbox}{width=1\textwidth,center=\textwidth}
\begin{tabular}{cccccccccc}
\hline
\hline
Target & Band & $\sigma_{\rm image}$ & F$_{\mathrm{total}}$ & F$_{\mathrm{peak}}$ & $\theta_{\mathrm{M}} \times \theta_{\mathrm{m}}$ & P.A. & $\alpha_{\rm J2000}$ & $\delta_{\rm J2000}$\\ 
(1) & (2) & (3) & (4) & (5) & (6) & (7) & (8) & (8)\\ 
 \hline
NGC0676 
 & L  & 89 & $\cdots$ & $<$267 & $\cdots$  & $\cdots$ & $\cdots$ & $\cdots$\\ 
 & C  & 18 & $\cdots$ & $<$54 & $\cdots$ & $\cdots$ & $\cdots$ & $\cdots$\\ 
 & X  & 10.5 & $\cdots$ & $<$31.5 & $\cdots$ & $\cdots$ & $\cdots$ & $\cdots$\\ 
 & Ku & 9 & $\cdots$ & $<$27 & $\cdots$ & $\cdots$ & $\cdots$ & $\cdots$\\
 \hline 
NGC1058 
 & L &  32  & $\cdots$ & $<$96 & $\cdots$  & $\cdots$ & $\cdots$&$\cdots$\\ 
 & C & 11   & $\cdots$ & $<$33 &  $\cdots$  & $\cdots$ & $\cdots$&$\cdots$\\ 
 & X & 9    & $\cdots$ & $<$27 & $\cdots$  & $\cdots$ &$\cdots$ &$\cdots$\\ 
 & Ku & 9   & $\cdots$ & $<$27 &  $\cdots$ & $\cdots$ &$\cdots$ &$\cdots$\\ 
 \hline
NGC2685 
 & L   & 36  & $\cdots$ & $<$108 & $\cdots$ & $\cdots$ & $\cdots$& $\cdots$\\ 
 & C   & 11 & $\cdots$ & $<$33 & $\cdots$  & $\cdots$ & $\cdots$& $\cdots$\\ 
 & X    & 9 & $\cdots$ & $<$27 & $\cdots$   & $\cdots$  & $\cdots$ & $\cdots$\\ 
 & Ku   & 9.5 & $\cdots$ & $<$28.5 & $\cdots$ & $\cdots$ & $\cdots$ & $\cdots$ \\ 
\hline
NGC3185 
 & L  & $27$ & 9250$\pm$470 & 1130$\pm$96 & 5.05$\times$3.8  & $\cdots$ & $\cdots$ & $\cdots$\\ 
 & C  &  16 & 3240$\pm$160 & 583$\pm$48 & 3.7$\times$3.5 & $\cdots$ & $\cdots$ & $\cdots$\\ 
 & X  & 14  & 668$\pm$47   & 116$\pm$34 & 3.7$\times$1.9  & $\cdots$  & $\cdots$ & $\cdots$ \\ 
 & Ku & 10 & $\cdots$  &$<$30  & $\cdots$ & $\cdots$ & $\cdots$ & $\cdots$\\ 
\hline
NGC3486 & L  & 34 & $\cdots$ & $<$102 & $\cdots$ & $\cdots$  & $\cdots$&$\cdots$\\ 
 & C & 10 & $\cdots$ & $<$30 & $\cdots$ & $\cdots$  & $\cdots$&$\cdots$\\ 
 & X & 8 & $\cdots$ &  $<$24 & $\cdots$  & $\cdots$ & $\cdots$&$\cdots$\\ 
 & Ku  & 9 & $\cdots$ & $<$27 & $\cdots$ & $\cdots$  & $\cdots$&$\cdots$\\ 
\hline
NGC3941 & 
 L & 30 & 320$\pm$70 & 221$\pm$30 & 1.50$\pm$0.59$\times$0.12$\pm$0.56 & 120$\pm$28 & 11:52:55.397$\pm$0.149 & +36.59.10.854$\pm$0.073 \\ 
 & C & 11 & 46$\pm$16 & 54$\pm$9 & <0.8$\times$0.55 & $\cdots$ & 11:52:55.364$\pm$0.161 & +36.59.10.861$\pm$0.05 \\ 
 & X & 12  &$\cdots$  &$<$36  &$\cdots$  & $\cdots$ & $\cdots$& $\cdots$ \\ 
 & Ku & 14 & $\cdots$ & <42 & $\cdots$ & $\cdots$  & $\cdots$ & $\cdots$\\ 
\hline

NGC4477 & L  & 44 & 210$\pm$76 & 209$\pm$42 & $<$0.60 $\times$0.5 & $\cdots$ & 12:30:02.23$\pm$0.141 & +13.38.11.686$\pm$0.085 \\ 
& C & 10 & 91$\pm$15 & 72$\pm$7 & $<$0.17 $\times$ 0.15 & $\cdots$ & 12:30:02.197$\pm$0.017 & +13.38.11.546$\pm$0.018\\ 
& X & 9 & 67$\pm$9 & 84$\pm$5 & $<$0.10 $\times$ 0.09 & $\cdots$ & 12:30:02.195$\pm$0.004 & +13.38.11.570$\pm$0.004 \\ 
& Ku & 8 & 61$\pm$9 & 93$\pm$7 & $<$0.07 $\times$ 0.06 & $\cdots$ & 12:30:02.196$\pm$0.003 & +13.38.11.547$\pm$0.003 \\ 
\hline

NGC4639 
& L  & 27 & 353$\pm$55 & 303$\pm$28 & $<$0.60 $\times$0.56 & $\cdots$ & 12:42:52.380$\pm$0.057  & +13.15.26.735$\pm$0.044\\
& C  & 10 & 404$\pm$17 & 393$\pm$10 & $<$0.17$\times$0.15  & $\cdots$ & 12:42:52.378$\pm$0.003 & +13.15.26.730$\pm$0.004\\ 
& X & 9 & 488$\pm$15 & 477$\pm$9 & $<$0.10 $\times$ 0.09  & $\cdots$ & 12:42:52.378$\pm$0.002
& +13.15.26.735$\pm$0.002\\ 
& Ku & 8 & 586$\pm$13 & 610$\pm$8 & $<$0.07 $\times$ 0.06 & $\cdots$ & 12:42:52.379$\pm$0.001 & +13.15.26.734$\pm$0.001\\ 
\hline

NGC4698 
 & L & 38 & 150$\pm$53  & 119$\pm$33 & $<$0.78 $\times$ 0.58 & $\cdots$ & 12:48:22.902$\pm$0.167 & +08.29.14.711$\pm$0.047\\ 
 & C & 10 & 238$\pm$17 & 241$\pm$10 & $<$0.21 $\times$ 0.15 & $\cdots$ & 12:48:22.910$\pm$0.008 & +08.29.14.667$\pm$0.006\\ 
 & X & 9 & 259$\pm$15 & 275$\pm$9 & $<$0.13 $\times$ 0.10 & $\cdots$ & 12:48:22.909$\pm$0.003 & +08.29.14.673$\pm$0.003\\ 
 & Ku & 9 & 258$\pm$16 & 267$\pm$9 & $<$0.08 $\times$ 0.06 & $\cdots$ & 12:48:22.909$\pm$0.003 & +08.29.14.664$\pm$0.002 \\ 
\hline

NGC4725 
& L  & 39 & $\cdots$ & $<$117 & $\cdots$ & $\cdots$ &$\cdots$ &$\cdots$\\ 
& C & 11 & 66$\pm$15 & 84$\pm$10 & $<$0.18 $\times$ 0.15 & $\cdots$ & 12:50:26.570$\pm$0.016 & +25.30.02.691$\pm$0.014\\ 
& X & 9 & 118$\pm$14 & 133$\pm$9 & $<$0.10 $\times$ 0.10 & $\cdots$ & 12:50:26.569$\pm$0.007 & +25.30.02.749$\pm$0.004\\ 
& Ku  & 8 & 81$\pm$11 & 105$\pm$8 & $<$0.07 $\times$ 0.06 & $\cdots$ & 12:50:26.569$\pm$0.005 & +25.30.02.748$\pm$0.003\\ 
\hline
\hline

\end{tabular}
\end{adjustbox}
\label{table:fluxTable}
\end{table*}

We detected radio emission in six out of ten sources: NGC~3185, NGC~3941, NGC~4477, NGC~4639, NGC~4698 and NGC~4725. Images of the detected sources (contours and coloured maps) are shown in Figs \ref{fig:contourMaps3185} - \ref{fig:contourMaps4725}. A source is considered detected if the peak intensity is $\ge3\,\sigma$. However, a source exhibiting a peak intensity 3$\,\le\,S_{\rm{P}}\,<\,6\,\sigma$ is considered as marginal detection, following the same criterion as in \citet{HU01}. Following this criterion, the detection rates are 5/10 in the L-band (with NGC~4477 and NGC~4698 as marginal detections), 6/10 in the C-band (with NGC~3941 as marginal detection), 5/10 in the X band (with NGC~3185 as marginal detection) and 4/10 in Ku band. 

As can be seen from images shown in Figs \ref{fig:contourMaps3185} - \ref{fig:contourMaps4725}, four out of six sources (NGC~4477, NGC~4639, NGC~4698 and NGC~4725) are are not resolved in our VLA observations, and for them we could only set an upper limit to their sizes which correspond to half of the beam size, in analogy with \citet{HU01}.

Four sources, NGC~0676, NGC~1058, NGC2685 and NGC~3486, have not been detected at any wavelength at 3$\,\sigma$ sensitivity thresholds ranging from 270 \mujybm in L-band down to 27 \mujybm in Ku band, and in these cases we provide upper limits on peak intensities defined as three times the rms noise.

Initially, we detected NGC~3185 and NGC~3941 only in the L-Band (at 1.5 GHz), with un-tapered maps exhibiting diffuse radio emission (see Figs. \ref{fig:contourMaps3185} and \ref{fig:contourMaps3941}). In order to further investigate this issue, the C, X and Ku datasets of these sources have been re-imaged with a uv-taper equal to the L-band clean beam size (via the parameter \textsc{uvtaper} with sub-parameter \textsc{outertaper}=['1.21arcsec','1.15arcsec','-74.63deg'] in \textsc{tclean}), applied to match their respective L-band observation. This has been done in order to be more sensitive to the extended radio emission. Then, the final uv-tapered images have been smoothed to the final resolution of the L-band beam (via the \textsc{restoringbeam} parameter of the \textsc{casa} task \textsc{tclean}). In the case of NGC~3941, which is fainter with respect to NGC~3185, we used a \textsc{briggs} weighting with robust parameter equal to 2: this value ensures that the weighting is close to a natural weighting. Natural weighting guarantees a better surface brightness sensitivity but degraded resolution, and for this reason it is more appropriate when considering diffuse emission. We have thus recovered radio emission in C band for NGC~3185 and NGC~3941 at a level above 5$\sigma$, and radio emission in NGC~3185 in X band at a level $\sim$ 3.5$\sigma$ (images are shown in Figs. \ref{fig:contourMaps3185} and \ref{fig:contourMaps3941}).

In the case of NGC~3185, in order to estimate source parameters we considered interactively defined boundaries via the \textsc{casa viewer}. The uncertainties associated to peak and integrated flux densities are calculated as ${\sigma_S}=\sqrt{N\times{(rms)^2}+(0.05{\times}S)^2}$, where N is the number of beam areas covered by a source of flux density S, and it is taken into account an uncertainty of 5 per cent in the absolute flux density scale.

In Table \ref{table:fluxTable} we summarise our findings.

\subsection{Detection rates}

In Table \ref{table:Tab} we list the sources belonging to the Parent sample, and in boldface we indicate the sources of the Reference sample, for which we provided new estimates of the parameters based on the new VLA data. 

\begin{table*}\footnotesize
\caption{The Parent sample of 28 sources. Objects in boldface are sources in the Reference sample, for which we report new VLA observations.} 
\centering
\begin{adjustbox}{width=0.95\textwidth,center=\textwidth}
\begin{tabular}{ccccccccccccc}
\hline
Target & D & Seyfert Type & Hubble Type & $L_{5}$ & $L_{1.4}$ & $L_{X}$ & $\log{R_X}$ & $\alpha$ & $M_{BH}$ & $L_X/L_{Edd}$ \\ 
 & (Mpc) & & & VLA & VLA &  &  & VLA & ($M_{\odot}$) &  \\ 
 (1) & (2) & (3)  & (4) & (5) & (6) & (7) & (8) & (9) & (10) & (11) \\ 
\hline
\hline
  \textbf{NGC~676} & 19.5 & S2: & S0/a:spin & \textbf{<35.1} & \textbf{<35.2} & 40.8 & \textbf{<-5.7} & $-$ & $-$ & $-$  \\
  \textbf{NGC~1058} & 9.1 & S2 & Sc & \textbf{<34.2} & \textbf{<34.1} & <37.5 & \textbf{$\lessgtr$-3.35} & $-$ & 4.9 & -5.4\\
  NGC~1068 & 14.4 & S1.9 & Sb & 38.8 & 38.6 & 42.8 & -4.0 & 0.7 & 7.2 & -2.5\\
  \textbf{NGC~2685} & 16.2 & S2/T2: & SB0+ pec & \textbf{<34.7} & \textbf{<34.7} & 39.9 & \textbf{-5.2} & $-$  & 7.1 & -5.3\\
  NGC~3031 & 3.5 & S1.5 & Sab & 36.8 & 36.15 & 40.2 & -3.5 & -0.16 & 7.8 & -5.6\\
  NGC~3079 & 17.3 & S2 & SBc spin & 38.2 & 37.6 & 42.6 & -4.4 & -0.18 & 7.7 & -3.13\\
  \textbf{NGC~3185} & 21.3 & S2: & Sbc & $-$ & $-$ & 40.8 & $-$ & <-0.12 & 6.1 & -3.4\\
  NGC~3227 & 20.6 & S1.5 & SABa pec & 37.7 & 37.6 & 41.7 & -4.0 & 0.82 & 7.6 & -3.9\\
  \textbf{NGC~3486} & 7.4 & S2 & SABc & \textbf{<34.0} & \textbf{<34.0} & 38.9 & \textbf{<-4.9} & $-$  & 6.1 & -5.4\\
  \textbf{NGC~3941} & 12.2 & S2: & SB0 & \textbf{34.7} & \textbf{34.7} & 38.9 & \textbf{-4.2} & \textbf{+1.52} & 8.2 & -7.4\\
  NGC~3982 & 20.5 & S1.9 & SABb: & 36.7 & 36.3 & 41.2 & -4.5 & 0.39 & 6.1 & -3.0\\
  NGC~4051 & 17.0 & S1.2 & SABbc & 36.6 & 36.3 & 41.3 & -4.7 & 0.55 & 6.1 & -2.9\\
  NGC~4138 & 13.8 & S1.9 & S0+ & 35.6 & 35.2 & 41.3 & -5.3 & -0.32 & 7.6 & -4.6\\
  NGC~4151 & 20.3 & S1.5 & SABab: & 38.3 & 38.1 & 42.5 & -4.2 & 0.6 & 7.2 & -2.8\\
  NGC~4258 & 7.2 & S1.9 & SABbc & 35.8 & 35.2 & 40.9 & -5.1 & -0.01 & 7.6 & -4.8\\
  NGC~4388 & 16.7 & S1.9 & Sb: spin & 37.0 & 36.8 & 41.7 & -4.7 & 0.69 & 6.8 & -3.2\\
  NGC~4395 & 2.6 & S1 & Sm: & 34.4 & 34.2 & 39.8 & -5.4 & 0.66 & 5.0 & -3.3\\
  NGC~4472 & 16.7 & S2:: & E2 & 37.5 & 37.2 & <39.3 & >-1.8 & 0.42 & 8.8 & -7.6\\
  \textbf{NGC~4477} & 16.8 & S2 & SB0? & \textbf{35.1} & \textbf{35.0} & 39.6 & \textbf{-4.5} & \textbf{+0.52} & 7.9 & -6.4\\
  NGC~4501 & 16.8 & S1.9 & Sb & 36.2 & 35.9 & 39.6 & -3.4 & 0.44 & 7.9 & -6.4\\
  NGC~4565 & 9.7 & S1.9 & Sb? spin & 36.2 & 35.5 & 39.4 & -3.2 & -0.19 & 7.7 & -6.4\\
  NGC~4579 & 16.8 & S1 & SABb & 37.8 & 36.9 & 41.0 & -3.2 & 0.56 & 7.8 & -4.8\\
  \textbf{NGC~4639} & 22.9 & S1.0 & SABbc & \textbf{36.1} & \textbf{35.4} & 40.2 & \textbf{-4.1} & \textbf{-0.40} & 6.9 & -4.7\\
  \textbf{NGC~4698} & 16.8 & S2 & Sab & 35.6 & 34.7 & 39.2 & \textbf{-3.6} & \textbf{-0.13} & 7.3 & -6.2\\
  \textbf{NGC~4725} & 13.2 & S2: & SABab pec & \textbf{34.9} & \textbf{34.5} & 38.9 & \textbf{-4.0} & <-0.5 & 7.5 & -6.7\\
  NGC~5033 & 18.7 & S1.5 & Sc & 36.8 & 36.5 & 41.1 & -4.3 & 0.51 & 7.3 & -4.3\\
  NGC~5194 & 8.4 & S2 & Sbc pec & 35.6 & 35.4 & 40.9 & -5.3 & 0.6 & 7.0 & -4.14\\
  NGC~5273 & 16.5 & S1.5 & S0 & 36.2 & 35.9 & 41.4 & -5.2 & 0.49 & 6.5 & -3.2\\
\hline
\hline
\multicolumn{11}{l}{Column (1), name of the source; Column (2), distance of the source; Columns (3) and (4); Seyfert and Hubble type; Columns (5) and (6),} \\
\multicolumn{11}{l}{log VLA 5 and 1.4 GHz peak luminosities in erg/s, respectively; Column (7), log 2-10 keV X-ray luminosity; Column (8), radio-loudness} \\
\multicolumn{11}{l}{parameter as defined by \citet{TerashimaWilson2003}; Column (9), spectral index defined as $S_{\nu}\,\propto\,\nu^{-\alpha}$; Column (10), BH} \\
\multicolumn{11}{l}{mass; Column (11), Eddington ratio.} \\
\multicolumn{11}{l}{Columns (2), (3), (4), (7) and (10) are taken from \citet{Panessa2006}.  } \\
\multicolumn{11}{l}{For NGC~4725 and NGC~3185 we kept the upper limit on spectral index provided by \citet{PG13}.} \\
\end{tabular}
\end{adjustbox}
\label{table:Tab}
\end{table*}

We provide the detection rates at 5 and 1.4 GHz for the Parent sample, together with other average values for relevant quantities, indicated in Table \ref{tab:full_sample}. The Parent sample is constituted by 19 type-2 Seyferts and 9 type-1 Seyferts. We found a detection rate at 1.5 GHz of 15/19 (79 per cent) and 9/9 (100 per cent), respectively, while at 5 GHz we found detection rates of 16/19 (84 per cent) for type 2 objects and 9/9 (100 per cent) for type 1 objects. Similar detection rates are obtained when considering detections at both frequencies. The detection rate for the Parent sample at 5 and 1.4 GHz is 24/28 (86 per cent). Indeed, as already shown in Section 4, four sources, namely NGC~0676, NGC~1058, NGC~2685 and NGC~3486 remain undetected at our tens \mujybm sensitivity level. Clearly, given the new detections, such rates are higher with respect to \citet{PG13}.

All sources in the Parent sample, except NGC~1068, NGC~2685, NGC~3185, NGC~3486, NGC~4477, NGC~4501, NGC~4639, NGC~4698 and NGC~4725, are part of the 15-GHz survey performed by \citet{Saikia2018} at sub-arcsec resolution with the VLA-A configuration. Only five sources, namely NGC~3982, NGC~4051, NGC~4395, NGC~5194 and NGC~5273, have been detected by them, with peak intensities ranging from 0.5\,\mjybm{} (NGC~3982) to 0.17\,\mjybm{} (NGC~4395). If we consider only the Reference sample, the three sources in common, NGC~0676, NGC~1058 and NGC~3941, have not been detected by \citet{Saikia2018} and the upper limits on the peak intensities are in agreement with our limits listed in Table \ref{table:fluxTable}.

\begin{table*}\footnotesize
\caption{Summary of statistical results for the Parent sample.}
\centering
\begin{adjustbox}{width=1\textwidth,center=\textwidth}

\begin{tabular}{cccccccc}
\hline
Target & Det rate 1.4 GHz & Det rate 5 GHz & Det. rate at either freq. & $\langle\,L_5^{VLA}\,\rangle$ & $\langle\,\alpha\,\rangle$ & $\alpha$ flat:steep & $\langle\,\log{R_X}\,\rangle$ \\ 
 \hline
 \hline
Full & 24/28 (86\%)  & 24/28 (86\%)  & 24/28 (86\%) & 36.1$\pm$0.3$^{*}$ & 0.30$\pm$0.10 & 12:11  & -4.40$\pm$0.15$^{*}$ \\
Type 1 & 9/9 (100\%) & 9/9 (100\%) & 9/9 (100\%) & 36.7$\pm$0.4 & 0.40$\pm$0.13 & 1:2 & -4.3$\pm$0.2 \\
Type 2 & 15/19 (79\%) & 16/19 (84\%) & 15/19 (79\%) & 35.8$\pm$0.3$^{*}$ & 0.28$\pm$0.13 & 9:5 & -4.5$\pm$0.2$^{*}$ \\ 
\hline
\hline
\multicolumn{7}{l}{Notes: $^{*}$, cases in which the mean may be ill-defined, see Section 4.2} \\
\multicolumn{7}{l}{In the calculation of mean values we neglected NGC~3185; in the calculation of mean radio-loudness parameter we} \\
\multicolumn{7}{l}{excluded NGC~1058 and NGC~4472, for which we have upper limits on X-ray luminosities.} \\
\multicolumn{7}{l}{The flat-to-steep threshold for the spectral index is 0.5.} \\
\end{tabular}
\end{adjustbox}
\label{tab:full_sample}
\end{table*}

\subsection{Radio Spectra}

\begin{figure*}\scriptsize
\centering
\subfloat{\includegraphics[scale=0.35]{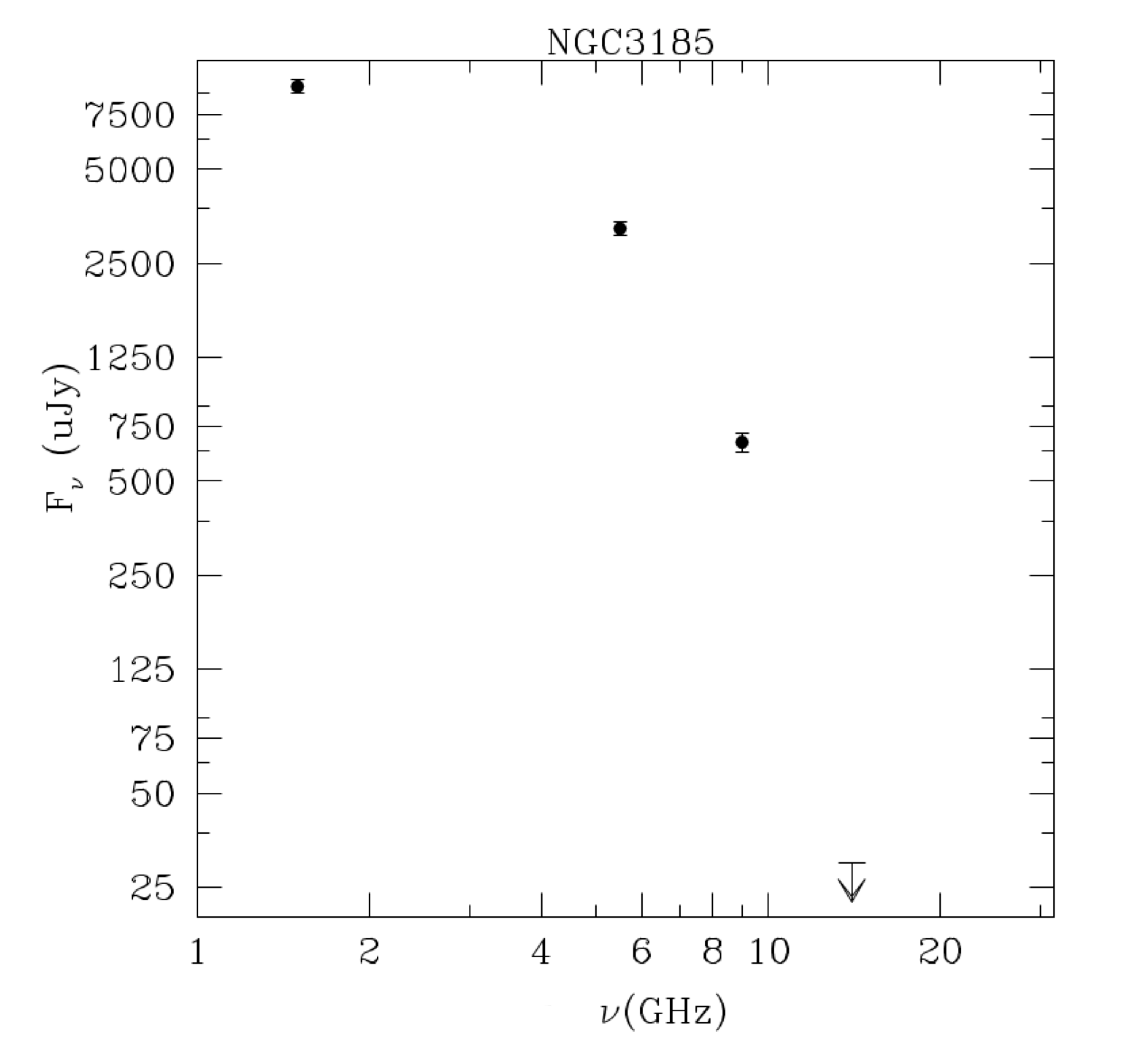}}
\subfloat{\includegraphics[scale=0.35]{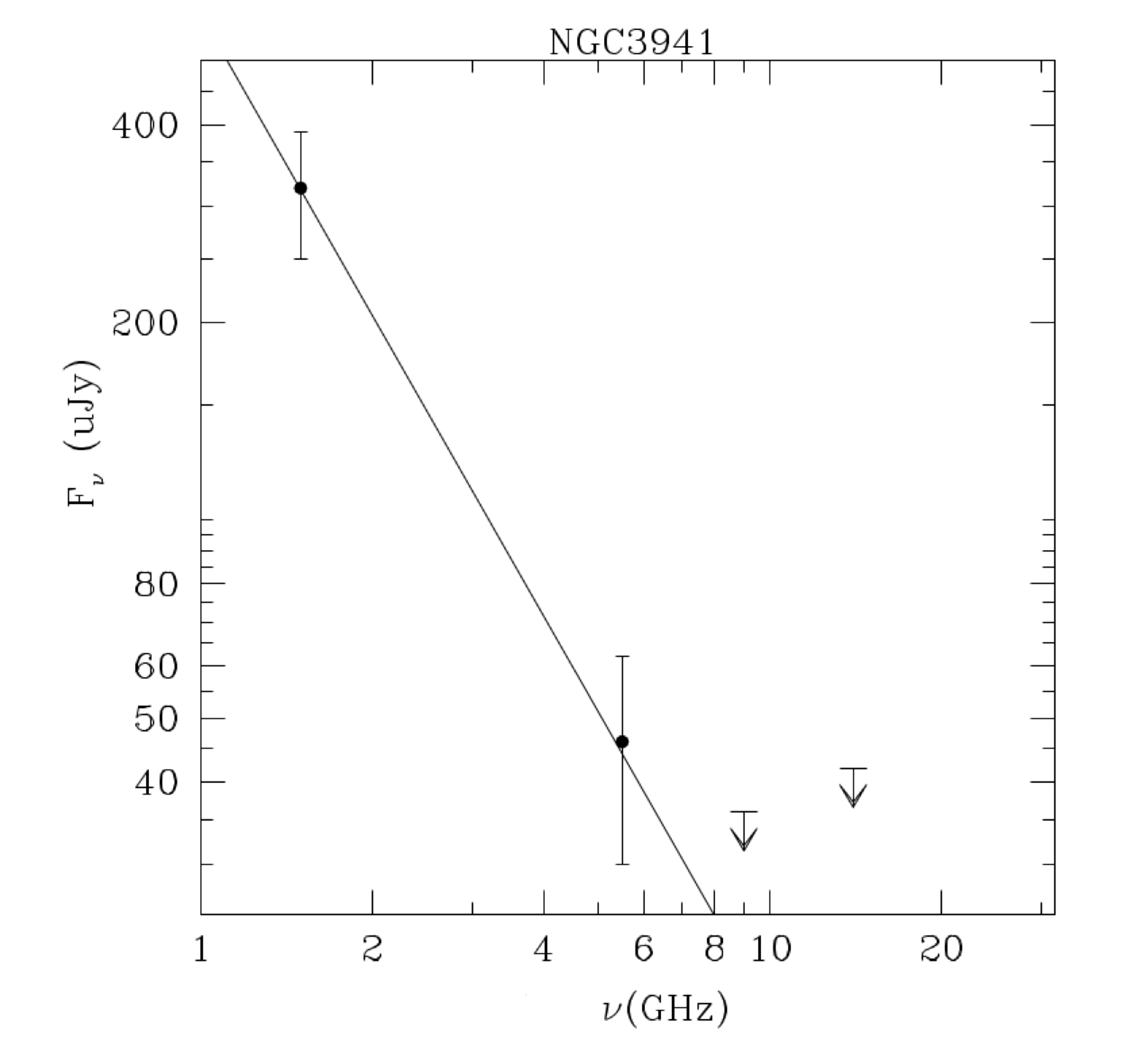}}\\
\subfloat{\includegraphics[scale=0.35]{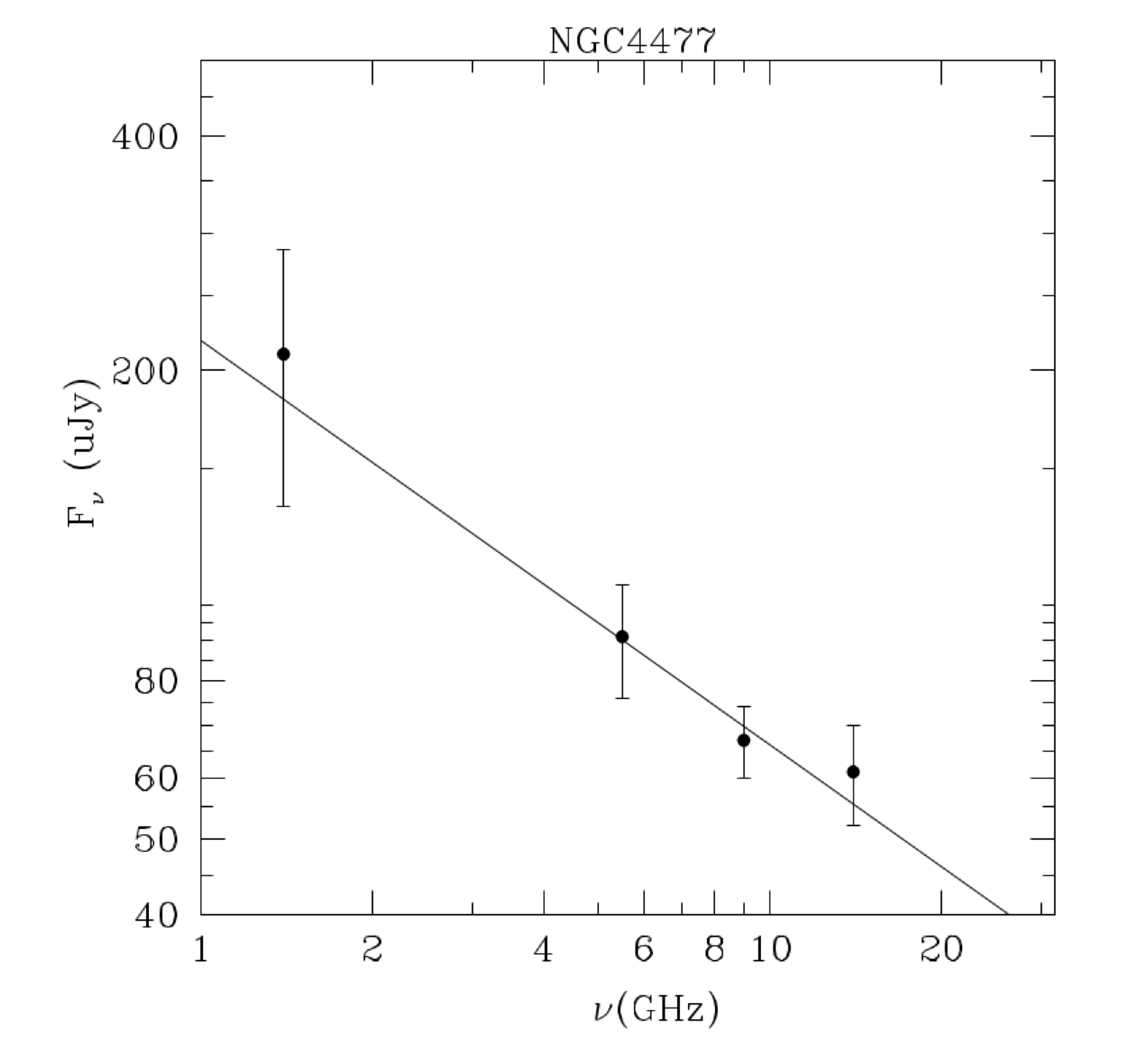}}
\subfloat{\includegraphics[scale=0.35]{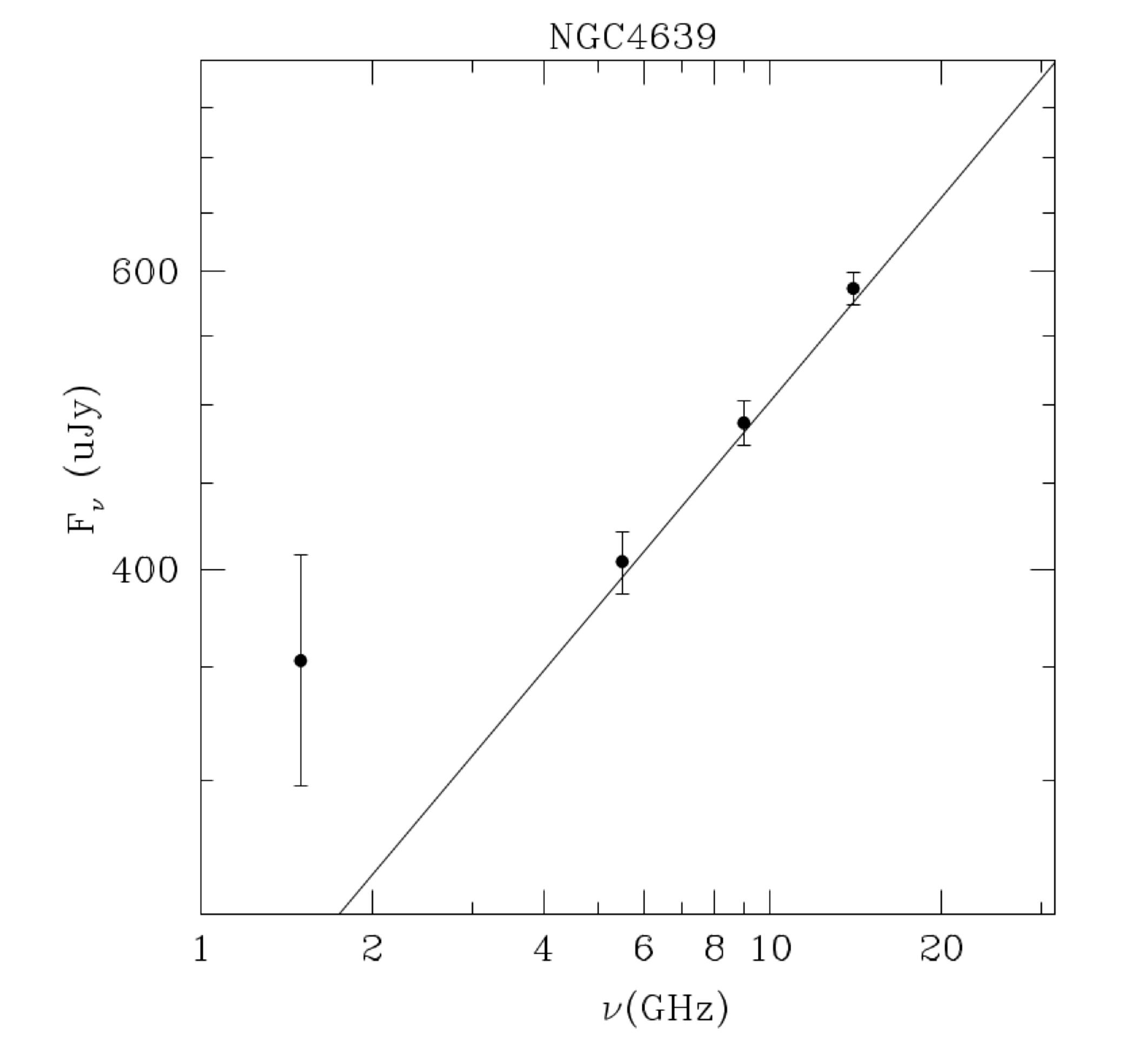}}\\
\subfloat{\includegraphics[scale=0.35]{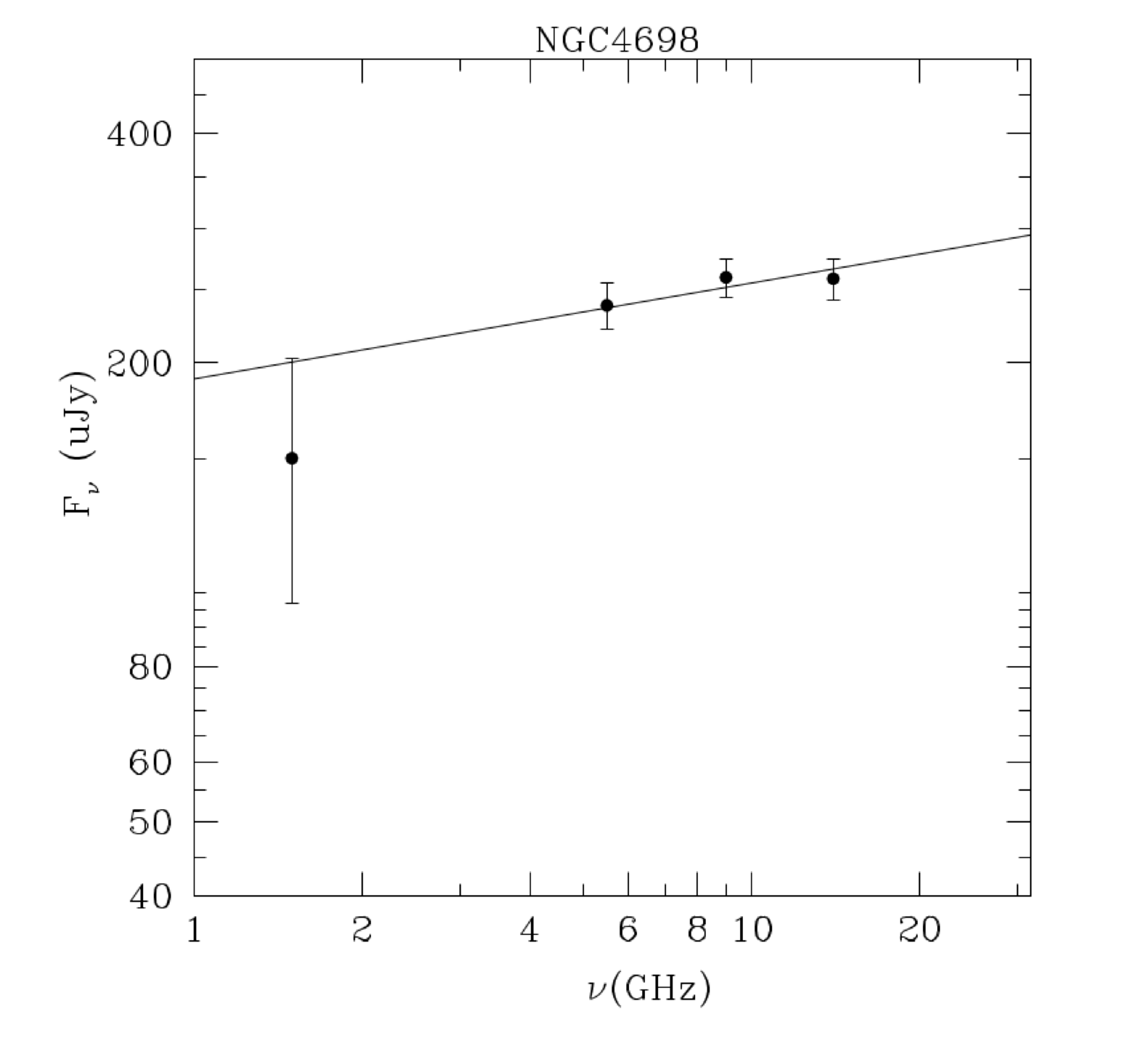}}
\subfloat{\includegraphics[scale=0.35]{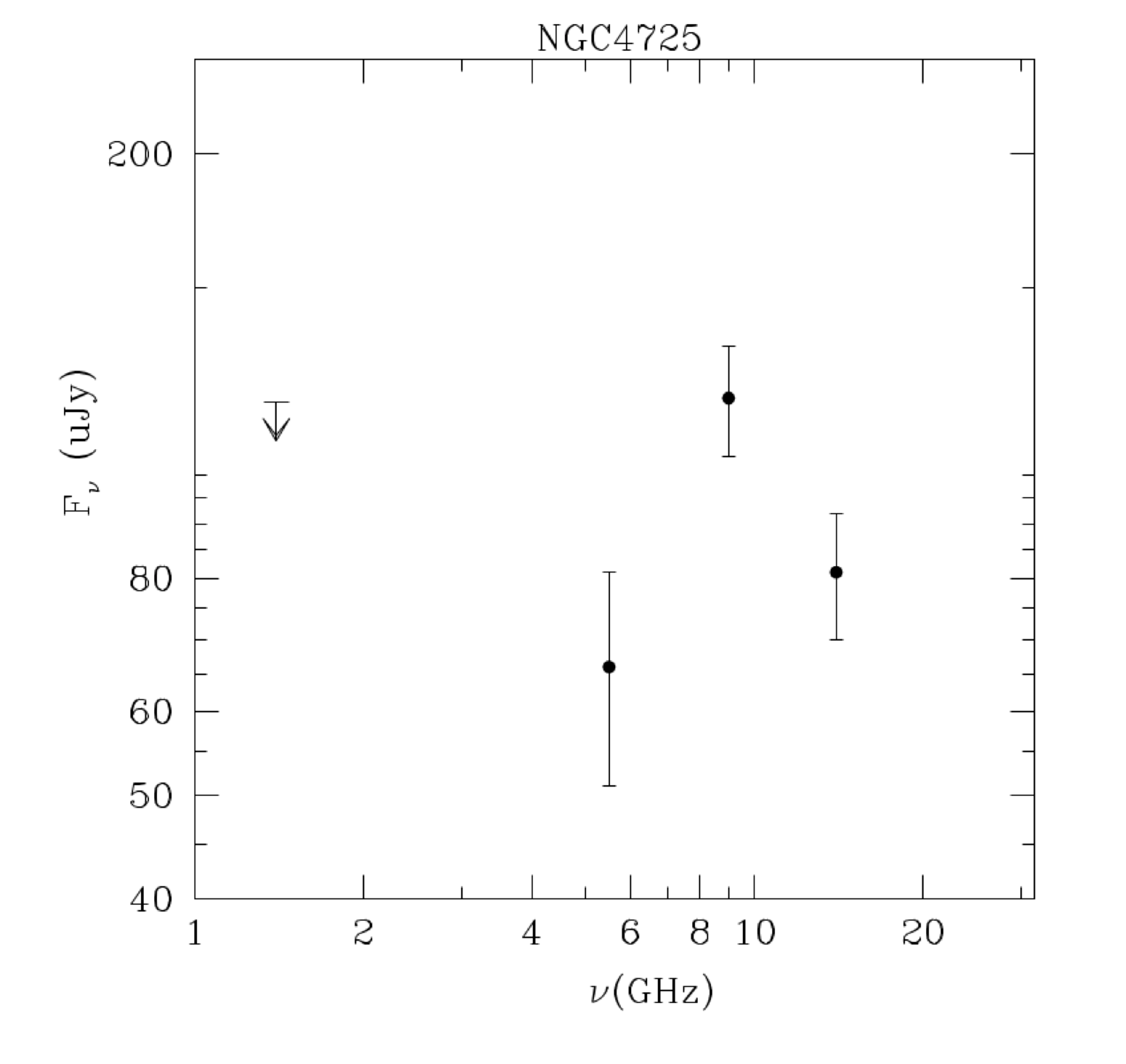}}

\caption{Radio spectra of LLAGNs in the Reference sample that were detected in at least two bands. The flux densities and their corresponding errors are listed in \Tab~\ref{table:fluxTable}, upper limits are represented by arrows and, when possible, a power-law ${S_\nu}\,{\propto}\,{{\nu}^{-\alpha}}$ fit is performed. The black solid lines are power-law least squares fits to the data (for NGC~4639 the 1.5 GHz point has not been considered).}
\label{fig:radiospectra}
\end{figure*}

We have built radio spectra for the sources which have been detected in two or more frequencies: NGC~3185, NGC~3941, NGC~4477, NGC~4639, NGC~4698 and NGC~4725 (Fig. \ref{fig:radiospectra} - upper limits are represented as arrows), considering all the available data points. We performed a weighted linear least squares fit using the \textsc{Sci-Py} \citep{scipy} routine \textsc{curve{\textunderscore}fit}, weighting for the uncertainties in the flux densities. 

In three cases,  NGC~4477, NGC~4639 and NGC~4698, the targets were found to fit a typical power-law ${S_{\nu}^{\rm{I}}}\,{\propto}\,{\nu^{-\alpha}}$ to the spectrum\footnote{We define a steep spectrum as one having $\alpha\,\ge\,$0.5, and a flat one as $\alpha\,\le\,$0.5, following \citet{PG13}, with $S_{\nu}\,\propto\,{\nu}^{-\alpha}$. We also define an inverted spectrum as one having $\alpha\,\le\,$0}. 
For NGC~4639 we considered only the 5.5, 9 and 14 GHz data points, as the 1.5 GHz data point seems to deviate with respect to the power-law behaviour, being probably associated with a different emitting component which dominates at lower frequencies. We found an inverted spectral index for NGC~4639 and NGC~4698 of $\alpha\,=$-0.40$\pm$0.01 and $\alpha\,=$-0.13$\pm$0.07, respectively, and a steep spectral index of $\alpha\,=$+0.52$\pm$0.09 for NGC~4477. 



We computed the spectral index also for NGC~3941 with the only two frequency points at 1.4 and 5 GHz, and we found a spectral slope of +1.52$\pm$0.33, in which the uncertainty in the slope has been estimated as $\sqrt{{{(\sigma_{f_1}/S_{f_1})}^2}+{{(\sigma_{f_2}/S_{f_2})^2}}}/ln({f_2/f_1})$, where $\sigma_{f_{1,2}}$ and $S_{f_{1,2}}$ are the uncertainties on the flux density and the flux density at the two frequencies \citep{HU01}. We remark that initially this source was detected only at 1.4 GHz and that we recovered emission at 5 GHz by applying a uv-tapering equal to the L-band clean beam size. 

The spectral index for NGC~3185 has not been computed as the radio emission from this source has a complex morphology and it may be due to other components, like star-formation.

In Table \ref{tab:full_sample}, we indicate the average values of the slope for the objects in the Parent sample in order to test if there are significant differences between the general population and the type-1 and type-2 sub-populations. We used the routine \textsc{KMESTM} in the Astronomy Survival Analysis software package \citep[ASURV,][]{Isobe1990,Lavalley1992}, which gives the Kaplan-Meier estimator for the distribution function of a randomly censored sample \footnote{We caution that such averaged values may suffer from statistical biases, in the case in which the lowest value in the sample is an upper limit, for details see the ASURV documentation.}. Using the aforementioned Kaplan-Meier estimator, we computed an average spectral index and the flat-to-steep ratio for the full sample and for the type-1-only and type-2-only sub-populations. We find that the Parent sample exhibits a nearly equal number of flat and steep spectral indices (12:11 flat-to-steep ratio), with an average spectral slope compatible with being flat (0.30$\pm$0.10). However, if we consider type 1 and type 2 objects separately, the two mean values in our case are consistent one with each other within errors: 0.40$\pm$0.13 and 0.28$\pm$0.13, respectively.

\subsection{The nuclear Radio and X-rays luminosity correlation} 

In order to study the relation between nuclear radio luminosity and the X-rays one, we correlated the 5 GHz VLA peak luminosity with the 2-10 keV X-rays luminosity (corrected for Compton-thin and Compton-thick absorption) for the Parent sample \citep{PG13} including our new data.

\begin{figure*}\scriptsize
\centering
\includegraphics[scale=0.42]{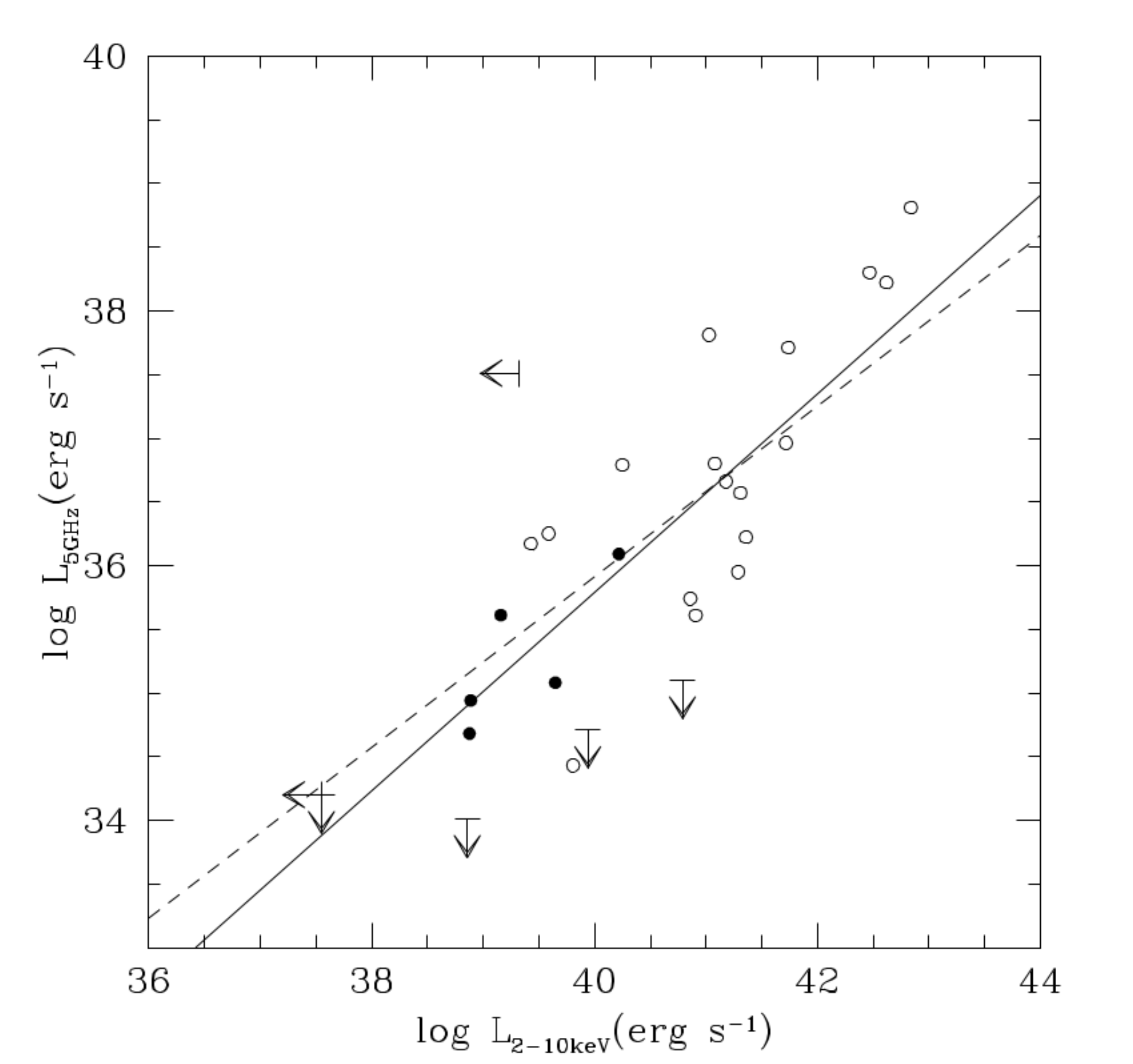}
\caption{The 5 GHz VLA luminosity as a function of the 2-10$\,$ keV X-rays luminosity for the sources in the Parent sample (NGC~3185 is excluded). The straight black line is the Schmitt's regression line to our sample, the black dashed line is the slope from \citet{PG13}. Filled circles represent detected sources in this work; sources which have not been detected and for which we provided upper/lower limits are indicated as arrows; empty circles represent objects not in the Reference sample but present in the Parent sample.}
\label{fig:radio_x}
\end{figure*}

Given the presence of dual censored data, we performed a Schmitt's linear regression algorithm to compute the correlation slope and a generalised Kendall's tau test for the significance of the correlation using the ASURV package. In Fig. \ref{fig:radio_x} we show a plot of 5 GHz VLA luminosity as a function of 2-10 keV luminosity. NGC~3185 has been excluded from this fit, for its complex morphology not ascribable to a core emission (see the L-band map in Fig. \ref{fig:contourMaps3185}). 

The slope of the Schmitt's regression is $Log\,{L_{5\,\rm{GHz}}[\rm{erg\,s^{-1}}]}=(0.8\pm0.1)Log\,{L_{2-10\,\rm{keV}}[\rm{erg\,s^{-1}}]}$. The generalised Kendall's tau test for correlation results in a probability P that a correlation is not present of 1.8$\times$10$^{-6}$ (z-value$\,\sim\,$4.8, $\ge\,$4.5$\,\sigma$). Even though our sources are all nearby (d$\,\le\,$23$\,$Mpc), in order to eliminate distance effects, we performed correlation test for the associated flux-flux relation, finding a P$\,\sim\,$ 2.6$\times$10$^{-6}$ for the null hypothesis (z-value$\,\sim\,4.7$, $\ge\,$4.5$\,\sigma$), so the correlation is significant. For three out of four undetected sources, the derived upper limits are nearly an order of magnitude smaller than the fluxes derived via the luminosity-luminosity relation.

\begin{figure*}\scriptsize
\centering
\subfloat{\includegraphics[scale=0.42]{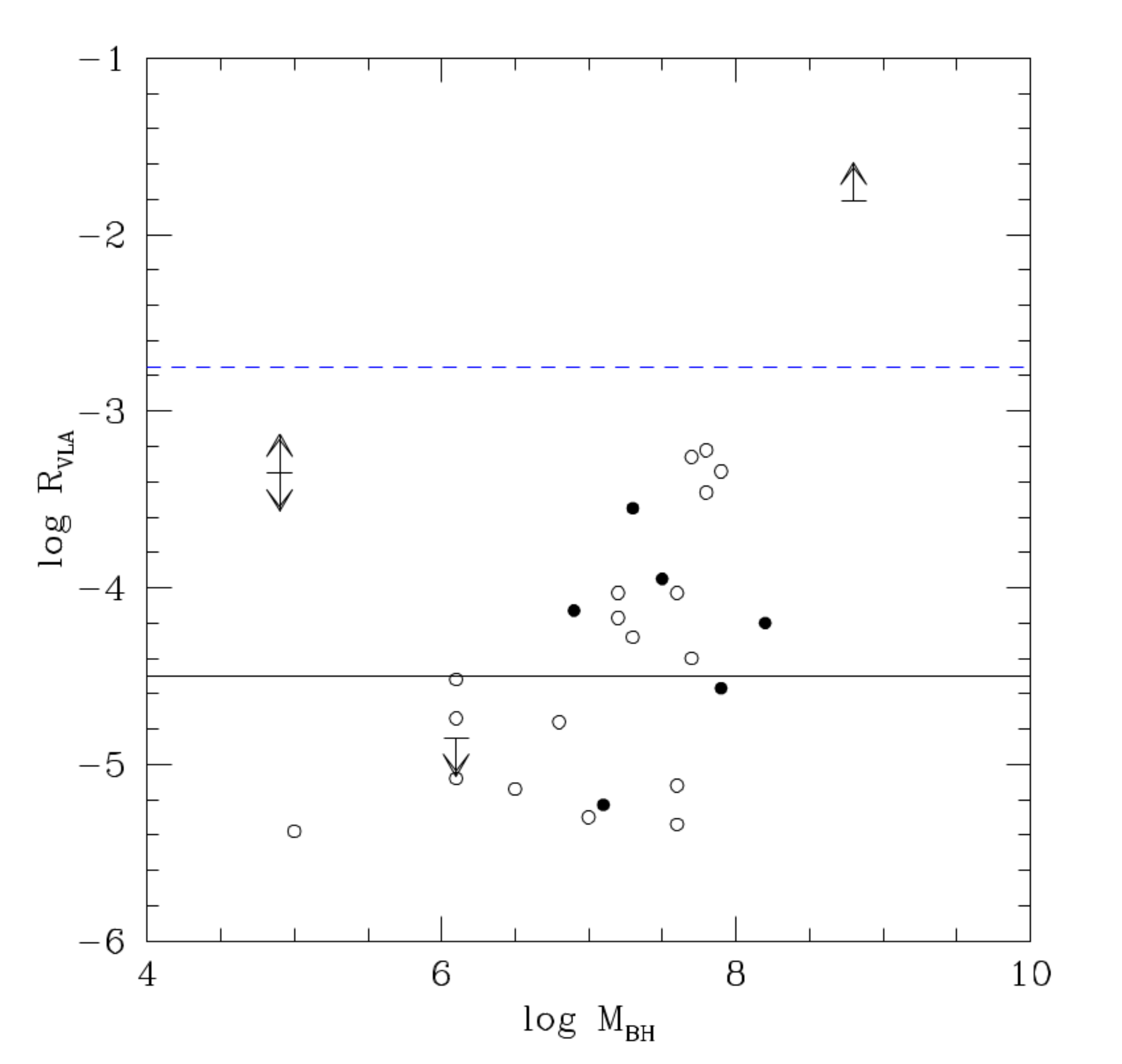}}
\subfloat{\includegraphics[scale=0.42]{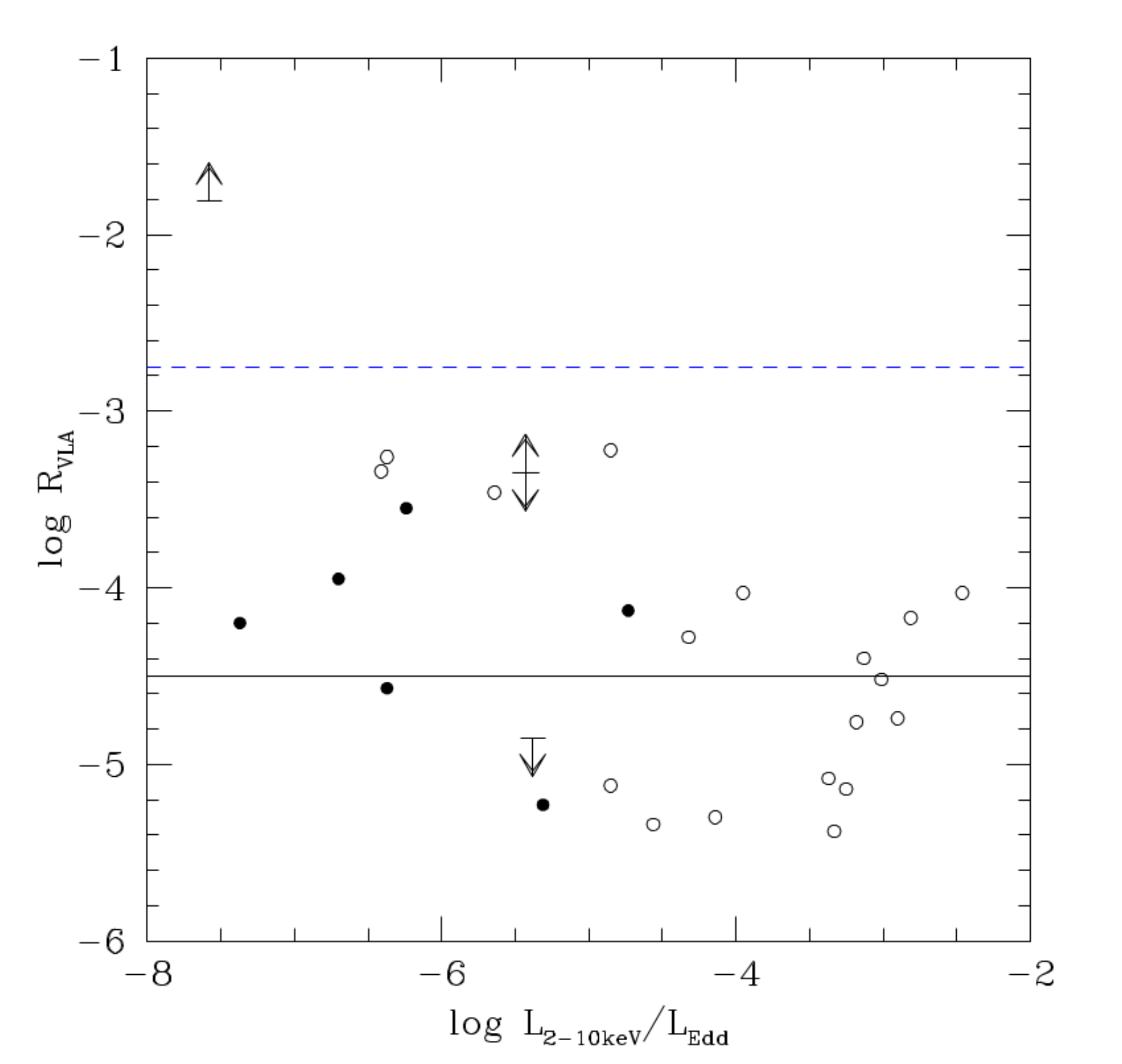}}\\
\caption{Left panel: the VLA radio-loudness parameter $R_X$, as defined by \citet{TerashimaWilson2003}, versus the BH mass. The black line is the \citet{TerashimaWilson2003} limit of $R_X=-4.5$, while the blue dashed line is the $R_X=-2.755$ limit for LLAGN derived by \citet{Panessa2006}.
Right panel: the VLA radio-loudness parameter $R_X$, as defined by \citet{TerashimaWilson2003}, versus the Eddington ratio (NGC~3185 and NGC~0676 are excluded). The black line is the \citet{TerashimaWilson2003} limit of $R_X=-4.5$, while the blue dashed line is the $R_X=-2.755$ limit for LLAGN derived by \citet{Panessa2006}. Filled circles represent detected sources in this work; sources which have not been detected and for which we provided upper/lower limits are indicated as arrows; empty circles represent objects not studied in the Reference sample but present in the Parent sample.}
\label{fig:radio_x2}
\end{figure*}

\subsection{Radio loudness and Accretion efficiency}

 The distribution of VLA X-ray radio-loudness parameter \footnote{$\log{R_X}=\log{L(5\,\rm{GHz})/L(2-10\,\rm{keV})}$, as defined by \citet{TerashimaWilson2003}} for the detected sources in the Parent sample is such that there is a prevalence of radio-loud sources, based on the radio-loudness limit $\log{R_X}\,=\,$-4.5 by \citet{TerashimaWilson2003}, but if we consider the -2.75 limit of \citet{Panessa2007}, then all our sources (except NGC~4472) are radio-quiet.  

We note that the BH masses, as in \citet{Panessa2006}, are taken from literature, and therefore they have been derived with different methods: maser kinematics, gas kinematics, stellar kinematics, reverberation mapping and mass-velocity dispersion, with reverberation mapping and stellar kinematics measurements which have been identified by \citet{WooUrry2002} as the most reliable ones. As noted in \citet{Panessa2006}, the BH mass estimates should be affected by a typical 0.3 - 0.5 dex uncertainty, mainly due to the mass-velocity relation scatter.

In Fig. \ref{fig:radio_x2}, we show $R_X$ as function of BH mass and Eddington ratio ($\log{L_{2-10\,\rm{keV}}/L_{\rm{Edd}}}$), left and right panel, respectively. We performed a Kendall's tau test for the significance of the correlation between the radio-loudness parameter and the black-hole mass (Fig. \ref{fig:radio_x2}, left panel), finding that there is a weak trend (z-value$\sim$3.17 and a probability P$\sim$0.0015), while no significant correlation is found with the Eddington ratio (z-value$\sim$1.28 and a  P$\sim$0.2). We have also investigated the correlation between the spectral index and black-hole mass and Eddington ratio as recently reported in \citet{LaorBaldiBehar2019}, however we do not find significant correlations with either the black-hole mass or the Eddington ratio. In order to deepen this issue, a larger sample of sources would allow a more statistically robust study. 

The population of sources in between the two radio-loudness thresholds (see Fig. \ref{fig:radio_x2}) could be referred to as Radio-Intermediate (RI) AGN, see for instance \citet{FalckeSherwoodPatnaik1996}. For the PG quasar sample, \citet{FalckeSherwoodPatnaik1996} find that, when considering intermediate values of the radio loudness parameter R (in the Kellerman definition), an approximately equal fraction of flat-spectrum sources emerges. \citet{Kellerman1994} showed that these flat-spectrum radio-intermediate sources are compact on VLA scales.
Interestingly, in our Parent sample, 12 sources out of the 23 having $R_X$ and spectral index information can be considered as Radio-Intermediate, with 7 of them exhibiting a flat-spectrum with a compact VLA component (with the exception of NGC 3079). If we consider the $\log{R_X}<-4.5$ sources (9 out of 23), then only three are both compact and flat-spectrum.

In Table \ref{tab:full_sample} we indicate the average values for the 5 GHz luminosity and for the X-ray radio-loudness parameter $\log{R_X}=\log{L(5\,\rm{GHz})/L(2-10\,\rm{keV})}$ (as defined by \citet{TerashimaWilson2003}). If we consider the average 5 GHz VLA luminosity and the average radio-loudness parameter for type 1 and type 2 Seyferts, no significant difference is found.

\section{Discussion}

We detected radio emission in six out of ten sources, with detection rates of 5/10 in L-band, 6/10 in C-band, 5/10 in X-band and 4/10 in Ku-band. The new VLA A-configuration observations allowed us to reach sensitivity levels down to $\sim\,$270 \mujybm in L-band (highest RMS for NGC~0676) and $\sim\,$27 \mujybm in Ku-band, which translates into radio powers of $L\,\sim\,10^{19}$ and $L\,\sim\,10^{18}$ W\,Hz$^{-1}$ at 1.5 and 14 GHz respectively. This allows us to increase the detection rates for the 28 sources of the Parent sample compared to \citet{PG13}: from 64 to 86 per cent in L-band and from 82 to 86 per cent in C-band (see table \ref{tab:full_sample}), thus sampling among the lowest radio luminosities for AGN. In the C-band, we find a detection rate of 84 per cent for the type-2 Seyferts and 100 per cent for the type-1s.

Clues on the origin of the radio emission can be obtained by analyzing observational information, such as the spectral slope and the morphology. In two cases, namely NGC~4639 and NGC~4698, we have found spectral slopes of -0.40$\pm$0.01 and -0.13$\pm$0.07, respectively. Inverted or slightly-inverted radio spectra from very compact sources (like the above cases) are usually associated with optically thick synchrotron emission from the base of a jet \citep[e.g.][]{Blandford1979,Reynolds1982}, in which multiple self-absorbed components, peaking at different frequencies, overlap and flatten the overall spectrum with respect to the 2.5 SSA slope \citep{FalckeBiermann1995}. Alternatively, thermal processes like free-free emission may also produce flat/inverted spectral slopes \citep{Bontempi+2012}. The other detected sources exhibit different spectra: NGC~4477 has been found to fit an intermediate spectral index between flat and steep (0.52$\pm$0.09), NGC~3941 has a very steep ($\alpha\,$=$+1.52\,\pm\,0.33$) spectrum, both compatible with optically-thin synchrotron emission. NGC~4725 exhibits a peculiar spectrum and further observations are needed to derive meaningful information for this source.

The above consideration could be strengthened via brightness temperature arguments. Indeed, following \citet{Falcke1999}, the natural limit between thermal and non-thermal processes is roughly $T_B\,\sim\,10^{6}$\,K, in which a brightness temperature much greater than this value suggests a non-thermal origin. However, we cannot put strong limits on the brightness temperature from VLA observation, given the loose upper limit on the deconvolved size. The only VLBI detection of the Reference Sample is for NGC~4477 at 5 GHz \citep{Bontempi+2012}. A very inverted spectrum and an intermediate brightness temperature between thermal and non-thermal processes ($\log{T_B}[K]\sim$6.5) was found. The derived physical parameters, under the assumption that the spectrum is due to synchrotron-self absorption, would agree with a thermal emission originating in a compact, hot corona surrounding the accretion disc, as proposed by coronal models by \citet{LaorBehar2008}.

\begin{figure*}\scriptsize
\centering
\subfloat{\includegraphics[scale=0.3]{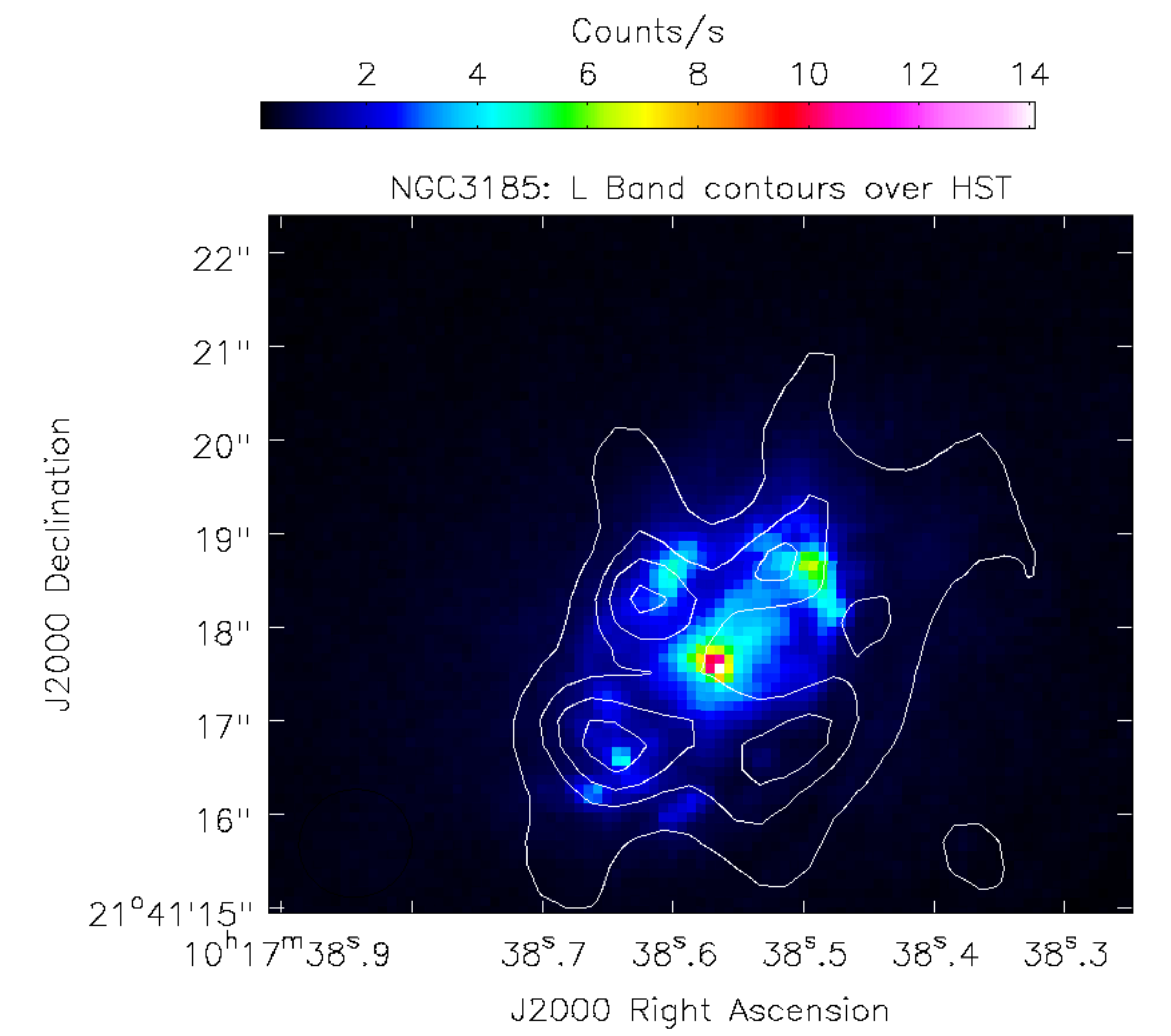}}
\subfloat{\includegraphics[scale=0.295]{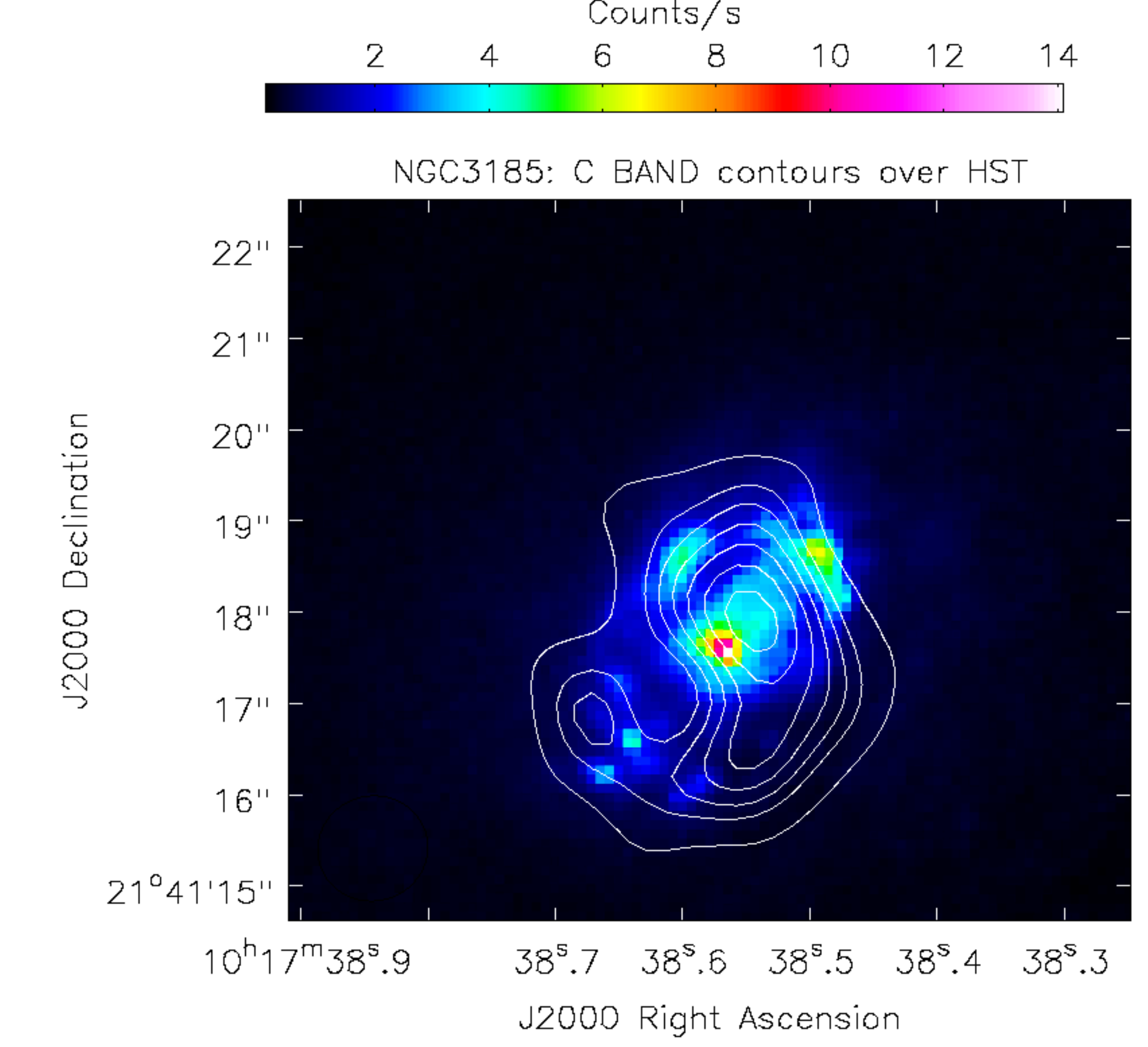}}\\

\caption{Left panel: L-band radio contours superimposed to HST image for NGC~3185; Right panel: C-band radio contours superimposed to HST image for NGC~3185. The archivial HST image has been taken via the WFPC2 detector and F450W filter.}
\label{fig:3185_HST}
\end{figure*}

We detected radio emission for six sources, but while the morphology of radio emission in four over six cases is compact on $\le\,$ arcseconds scales, predominantly unresolved, for NGC~3185 and NGC~3941 we find a more complex morphology. 
The estimated radio luminosities at this frequency are L$\,\sim\,5\,\times\,10^{20}$ and $\sim\,6\,\times\,10^{18}$\,W\,Hz$^{-1}$ for NGC~3185 and NGC~3941, respectively, well below the Radio-Loud/Radio-Quiet threshold of L$\,\sim\,10^{23}$\,W\,Hz$^{-1}$ defined by \citet{Condon1992}.
Considering NGC~3185, the structure observed in L-Band has a size of $\sim$5$\,\times\,$3.8 arcsec, which translates into a linear scale of $\sim\,$0.5$\,\times\,$0.4 kpc\footnote{If instead we use z-independent distance provided in the Nasa/Ipac Extra-galactic Database (NED) of 23.2 Mpc, then the linear scale would be $\sim\,$0.56$\,\times\,$0.42 kpc, so our linear scales estimates are affected by a $\le\,10$ per cent error, which do not affect our considerations.}. Analogous considerations can be made for the C and X Bands, leading to linear scales of $\sim\,$0.38$\,\times\,$0.36 kpc. There is evidence for radio emission spread over the host galaxy scale for a number of Seyferts when observed with adequate angular resolution and sensitivity \citep{Orienti2015}. In particular, circumnuclear starburst rings with knots of star formation are observed. \citet{OrientiPrieto2010} observed diffuse radio emission with knots of star formation for a number of Seyferts, on scales smaller than 1 kpc (NGC~5506, NGC~7469 and NGC~7582). Analogously, the morphology of the radio emission in the case of NGC~3185 as observed in our radio maps may be consistent with emission from circumnuclear rings of star formation. In order to check this hypothesis, we overlapped L-band and C-band radio contours to an archivial HST image\footnote{Taken from the Hubble Legacy Archive (HLA), \url{https://hla.stsci.edu/}} for NGC~3185, as shown in Fig. \ref{fig:3185_HST}. The radio contours overlap with optical emission, from which a nearly circular, ring-like structure emerges. However, further work is needed to understand the link between radio emission and star formation in the form of circumnuclear rings in this source. In particular, intermediate angular resolution scales radio observations (such as those of e-MERLIN) combined with H$\alpha$ maps would allow us to confirm the star-formation hypothesis. 
If we consider the Parent sample with our new data, then we find that the average spectral slope is compatible with flat (0.30$\pm$0.10), with a nearly equal number of flat and steep slopes. If we consider the two sub-populations of type-1 and type-2 Seyferts, then no significant differences are found with respect to the average spectral index and the average radio-loudness parameter, the only difference being that type-1 sources are slightly more luminous. Considering the X-ray radio loudness parameter, the black-hole mass and the Eddington ratio, we do not find significant differences between the type-1 and type-2 sub-classes. We note that, even though the sources in the Reference sample have an Eddington ratio nearly an order of magnitude smaller than the average for the Parent sample, the X-ray radio-loudness parameter does not exhibit anomalous values.

We also investigated the relation between the 5 GHz luminosity and the nuclear 2-10 keV X-ray luminosity. The latter is usually considered a tracer of accretion luminosity, while the former may have a core and a jet contribution. Nevertheless, we performed this analysis in order to understand if there is an interplay between the emitting components. We have found a significant correlation between the two quantities $\log L_\mathrm{5\ GHz} \propto (0.8\pm0.1) \log L_\mathrm{2-10\ keV}$ at scales traced by the VLA, in agreement with considerations of \citet{Panessa2007}. The increased sensitivity of our observations allows us to put stringent limits on the derived flux densities at very low luminosities, resulting in a steepening of the correlation slope with respect to the previous estimate of \citet{PG13}, i.e.,  $\log L_\mathrm{5\ GHz} \propto (0.67\pm0.01) \log L_\mathrm{2-10\ keV}$. However, our value is still in agreement with the general $\sim$ 0.7 slope of \citet{Gallo2003} found in the case of 'low-hard' state XRBs. During their outburst activity, XRBs experience transitions between different accretion and ejection states, likely in dependence of the efficiency of the accretion flow \citep{Fender2004}. In their low-hard state, XRBs show a radio spectrum which is flat/slightly inverted, and can be interpreted as synchrotron self-absorbed emission from an optically thick core, probably the base of a jet as may be occurring in AGN \citep{Blandford1979}.
The fact that LLAGNs do exhibit a slope of the radio - X-rays correlation comparable to that of low-hard state XRB has been interpreted as symptom of a common underlying physics \citep[e.g.][]{FalckeKordingMarkoff2004}. The two classes of low-accreting sources would therefore be associated with a radiatively inneficient accretion (RIAF) coupled with a jet \citep{Ho2008}, although there have been successful attempts in modelling in terms of jet-only and RIAF-only models \citep[e.g.][]{Kording2006}. Indeed, the observed low Eddington ratios favour an inefficient regime for our sources. 
However, we have to note that the jet origin of radio emission, associated to the low/hard state of XRBs, for the sources in our sample can not be confirmed. Indeed, a possible non-negligible contribution to radio emission might come from central star-formation regions (like in the case of NGC~3185). Moreover, the interpretation of spectral indices is not unique: different processes, such as low-power jets, outflows or star formation could produce steep or flat spectra, dependently on the conditions \citep[for a review see][]{Panessa2019}.

\begin{table}\footnotesize
\centering
\caption{Spectral index for the four sources NGC~3941, NGC~4477, NGC~4639 and NGC~4698. \textit{Columns}: (1) target name; (2) spectral index.}

\begin{tabular}{cc}
\hline
Target & Spectral index \\ 
  &  $\alpha$ \\  
(1) & (2) \\
 \hline
 \hline
NGC~3941 & +1.52$\pm$0.33 \\
NGC~4477 &  +0.52$\pm$0.09 \\ 
NGC~4639 & -0.40$\pm$0.01 \\
NGC~4698 & -0.13$\pm$0.07 \\ 
\hline
\end{tabular}
\label{tab:spec_ind}
\end{table}

\section{Conclusions}

We present new multi-frequency (L, C, X and Ku bands), high resolution (A configuration) VLA observations for ten Seyferts (Reference sample), which are the faintest members of the complete, distance limited sample of \citet{Cappietal06} (Parent sample). Below, we summarize our results: \\
- We detected radio emission in six out of ten sources, with four of them, namely NGC~0676, NGC~1058, NGC~2685 and NGC~3486 which remain radio-silent at 3$\,\sigma$ sensitivity levels ranging from $\sim\,$270 $\mu$Jy/beam (L-band) down to $\sim\,$27 $\mu$Jy/beam (Ku-Band), this translates into upper limits on luminosities of   L$\,\sim\,10^{19}$ and L$\,\sim\,10^{18}$\,W\,Hz$^{-1}$ at 1.5 and 14 GHz, respectively.\\
- The increased sensitivity of the new observations translates into a higher detection rate in the Parent sample, from 64 to 86 per cent in the L-band and from 82 to 86 per cent in the C-band, with respect to previous works \citep{PG13}, suggesting that all nuclei should be radio emitters when observed with the adequate sensitivity.  \\
- For the detected sources we computed radio spectral slopes, and we found: $\alpha\,=$+0.52 for NGC~4477, compatible with both a steep and flat spectrum; $\alpha\,=$-0.40 for NGC~4639 and $\alpha\,=$-0.13 for NGC~4698, slightly inverted spectra compatible with both compact optically-thick emission and thermal (e.g. free-free) emission; $\alpha\,=$+1.52 for NGC~3941, with the spectral index defined as $S_\nu^{\rm{I}}\,\propto\,\nu^{-\alpha}$. NGC~4725 exhibits a peculiar spectrum, that deserves further investigation. \\
- We studied the morphology of the radio emission for the ten faintest Seyferts, and we found it to be predominantly compact on scales $\le$ arcsec, which corresponds to linear scales smaller than $\sim$\,100\,pc, except in two cases, NGC~3185 and NGC~3941, which show complex morphology at sub-kpc scales, probably due to star formation processes. \\
- We did not find particular trends with neither the black-hole mass nor the Eddington ratio considering the Parent sample. Even though the sources in the Reference sample have an Eddington ratio an order of magnitude lower than the average for the Parent sample, they do not exhibits any specific trend for the X-ray radio-loudness parameter \citep[as defined by][]{TerashimaWilson2003} \\
- Considering the type-1 and type-2 sub-populations for the Parent sample, we did not find a clear dichotomy when considering the average radio-loudness parameter $\log{R_X}=\log{L(5\,\rm{GHz})/L(2-10\,\rm{keV})}$ and the average spectral index, the only difference being that type-1 Seyferts have a prevalence of steep spectra with respect to type-2. \\
- We correlated the radio 5 GHz luminosity with the 2-10 keV luminosity for the Parent sample, taking into account censored data. The slope we find of 0.8$\pm$0.1 is consistent with the slope of $0.7$ found by \citet{Gallo2003} for the low-hard state XRBs, which would suggest a possible common inefficient accretion physics.\\

These faint nuclei will constitute one of the dominant source population in the Square Kilometer Array (SKA) sky,
allowing to investigate several different physical processes that are possibly producing radio emission in these nuclei.

\section*{Acknowledgements}
We thank the referee for the comments which significantly contributed to improve the quality of the paper. 
EC thanks D.R.A. Williams for comments that improved the manuscript.
EC acknowledges  the  National  Institute  of  Astrophysics (INAF) and the University of Rome - Tor Vergata for the PhD scholarship in the XXXIII PhD cycle. GB acknowledges financial support under the INTEGRAL ASI-INAF agreement 2013-025-R.1. FP and MG acknowledge support from a grant PRIN-INAF SKA-CTA 2016.
FT acknowledges support by the Programma per Giovani Ricercatori – anno 2014 “Rita Levi Montalcini”.

\bibliographystyle{mnras}
\bibliography{ref} 




\appendix

\section{Radio Maps}

\begin{table*}\scriptsize
\centering
\begin{tabular}{cc}
{\large \textbf{NGC~3185}} & {\large \textbf{NGC~3185}} \\

\includegraphics[scale=0.4]{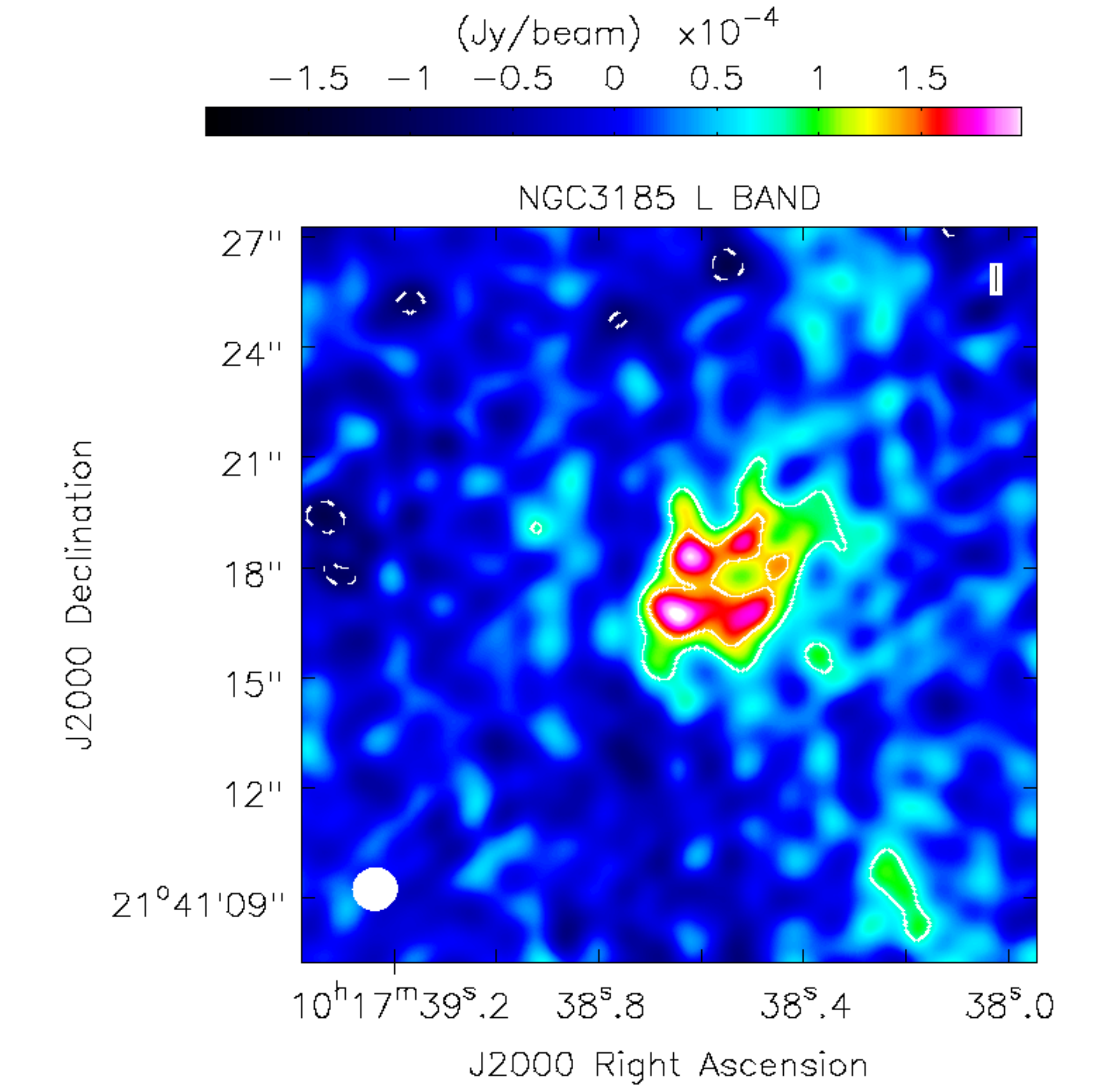} & 
\includegraphics[scale=0.4]{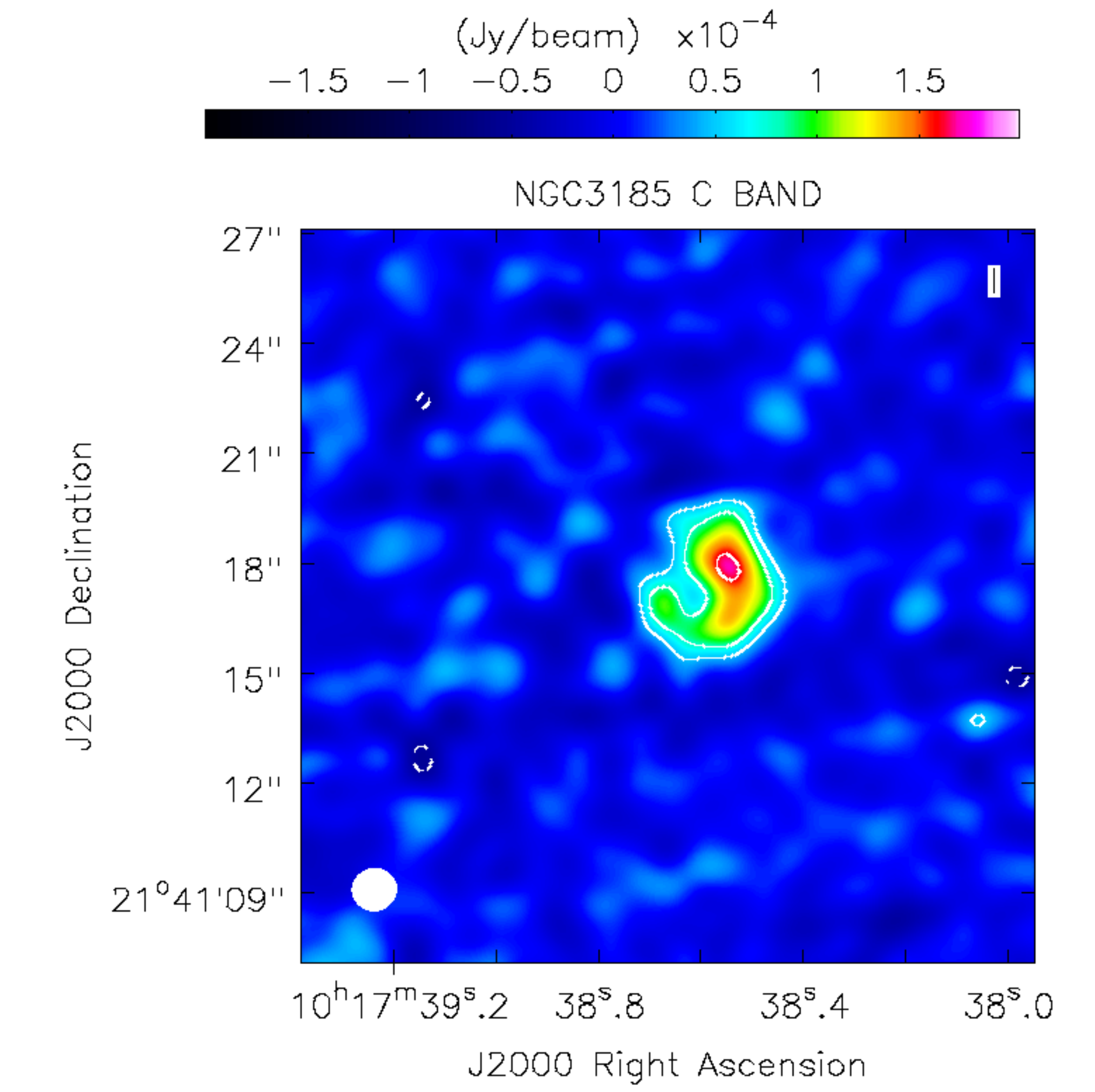} \\

\includegraphics[scale=0.4]{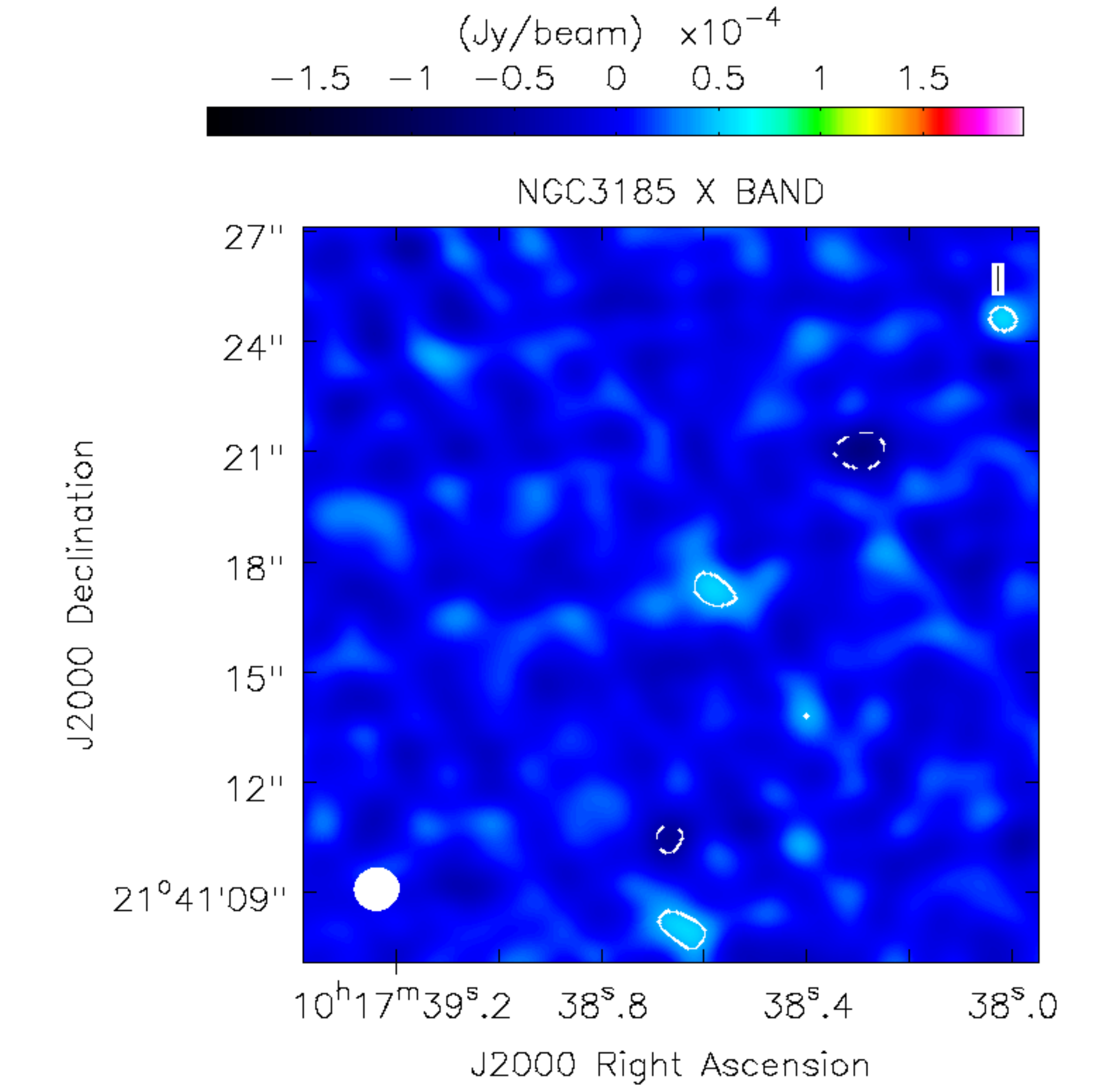} &
\includegraphics[scale=0.4]{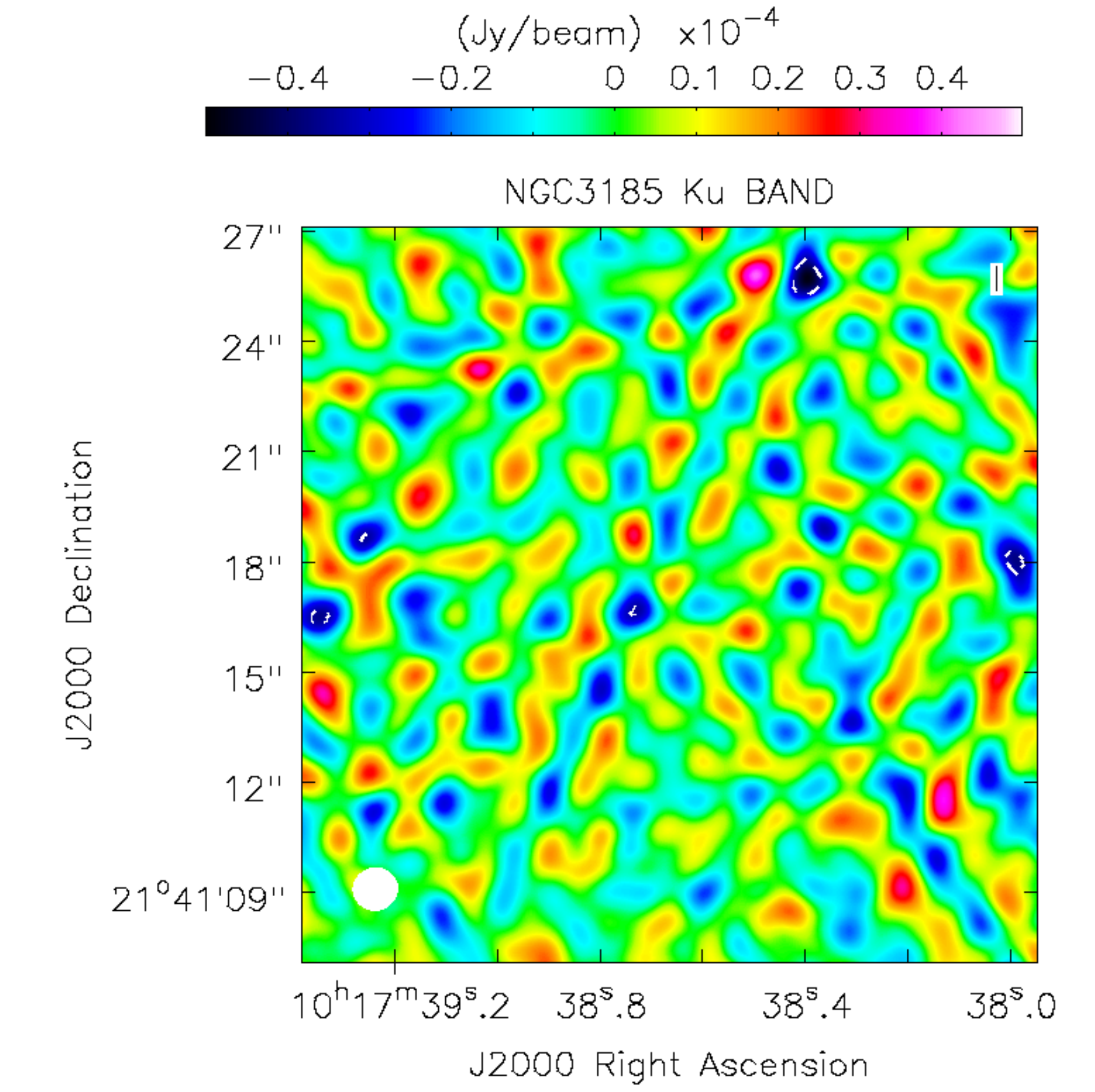}\\

\end{tabular}
 \captionof{figure}{UV-tapered contour and coloured maps of NGC~3185 in the four frequency bands: L-band (top-left), C-band (top-right), X-band (bottom-left) and Ku-band (bottom-right). The contours are displayed at [-3,3,5,10]$\times\,\sigma_{\rm image}$, where $\sigma_{\rm image}$ is the respective image noise rms given in \Tab~\ref{table:fluxTable}. The restoring beam is depicted as a white-filled ellipse in the left-hand corner of the map.}
\label{fig:contourMaps3185}
\end{table*}

\begin{table*}\scriptsize
\centering
\begin{tabular}{cc}
{\large \textbf{NGC~3941}} & {\large \textbf{NGC~3941}} \\

\includegraphics[scale=0.4]{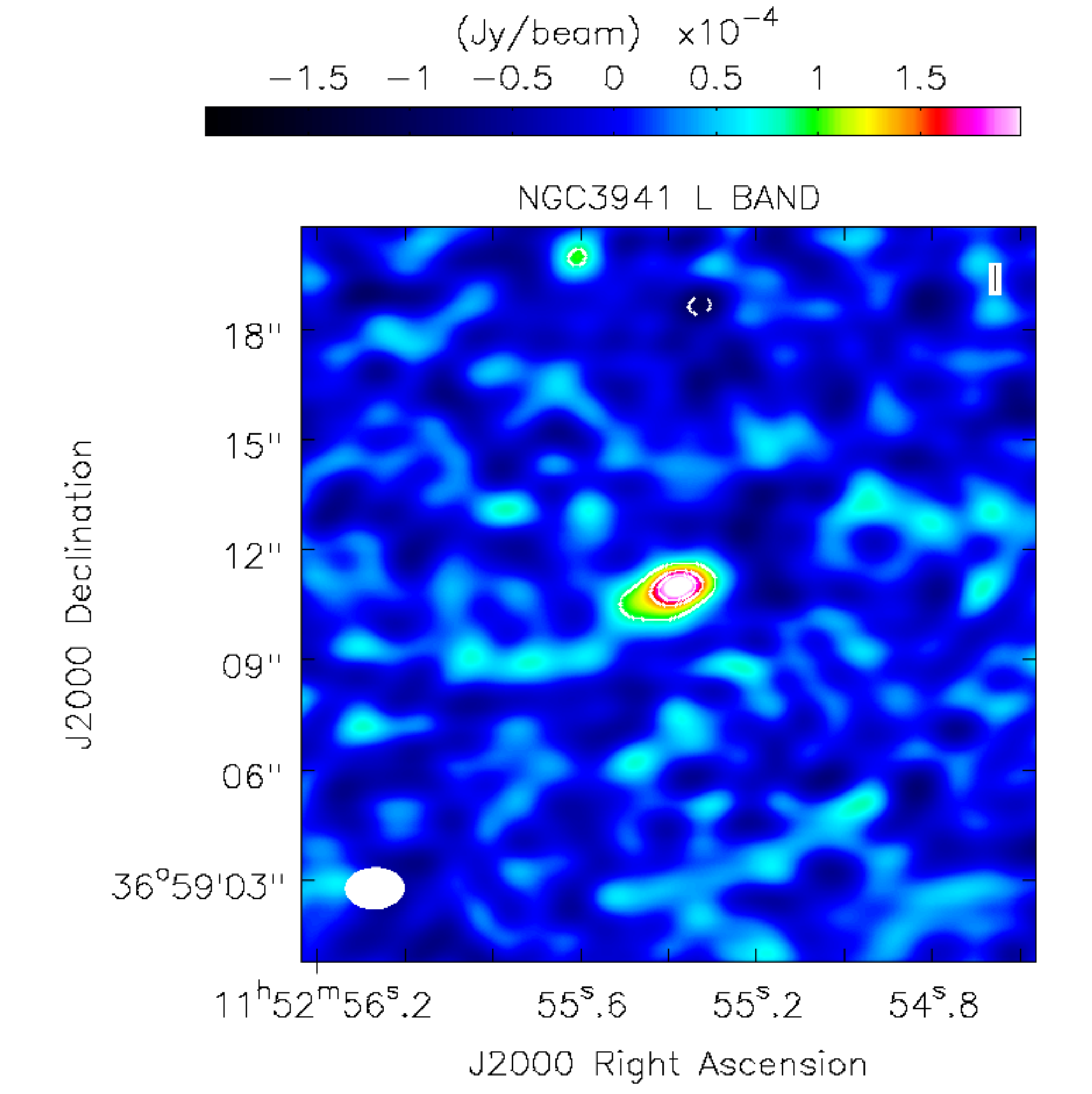} & 
\includegraphics[scale=0.4]{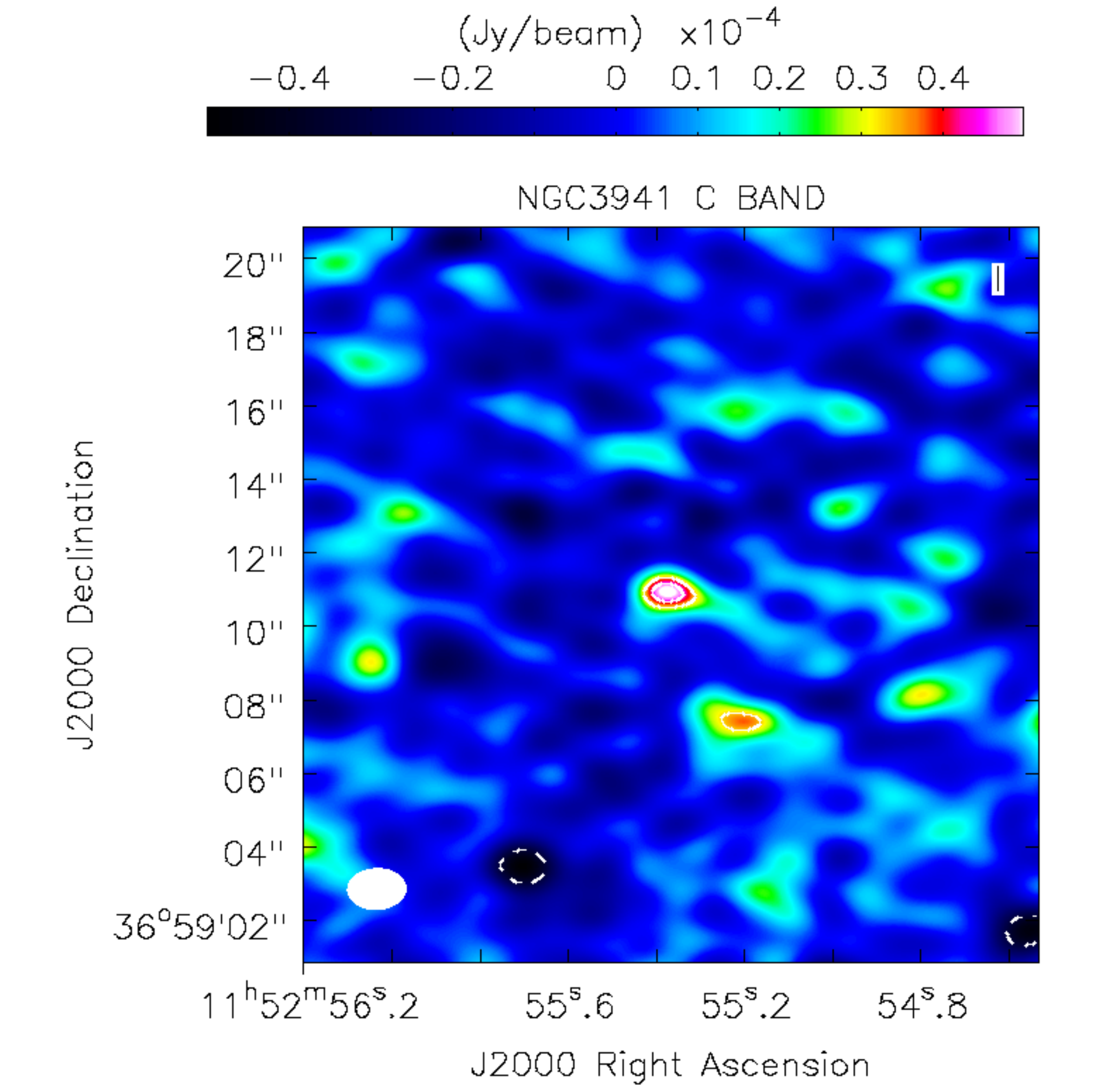} \\

\includegraphics[scale=0.4]{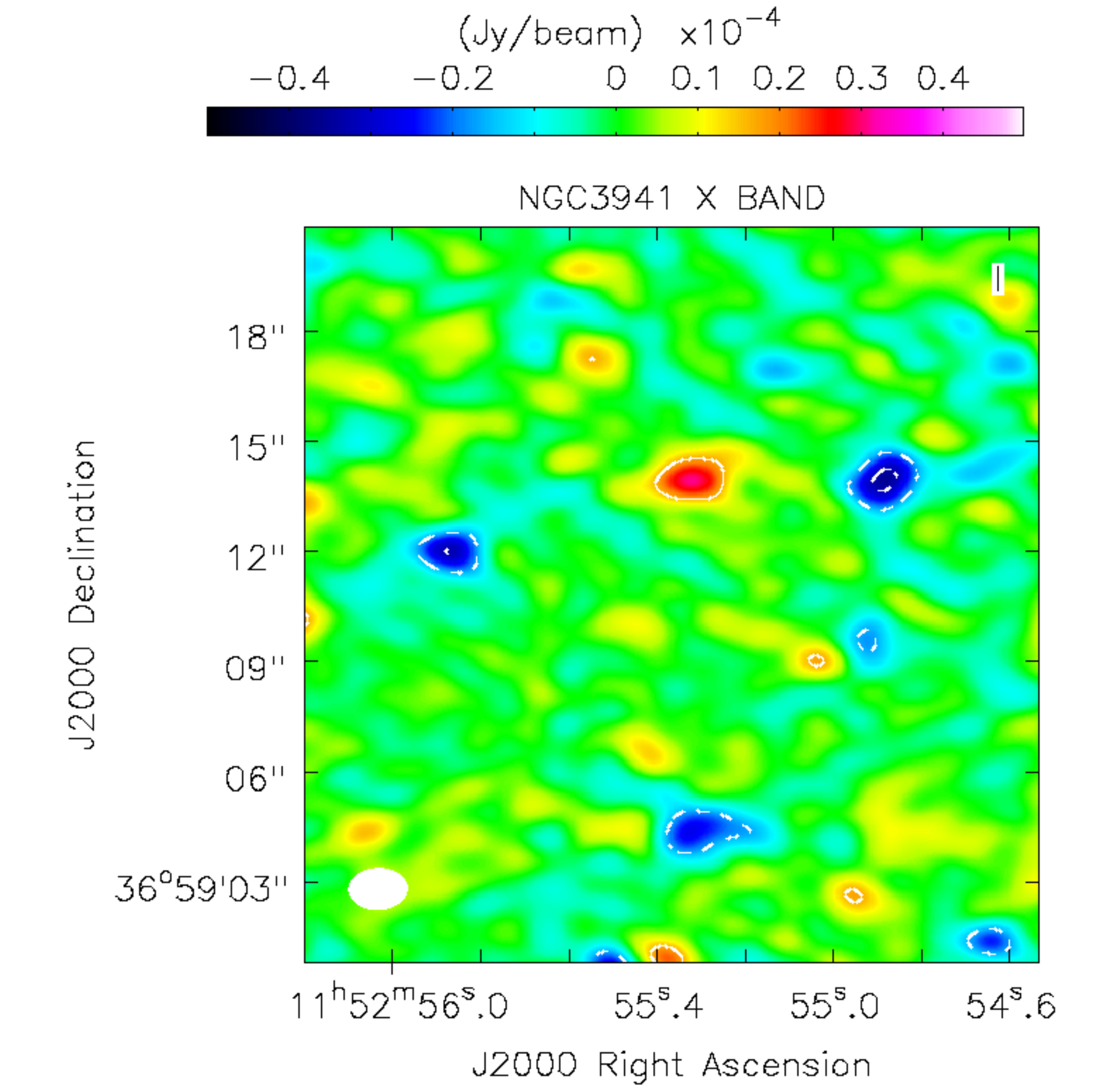} &
\includegraphics[scale=0.4]{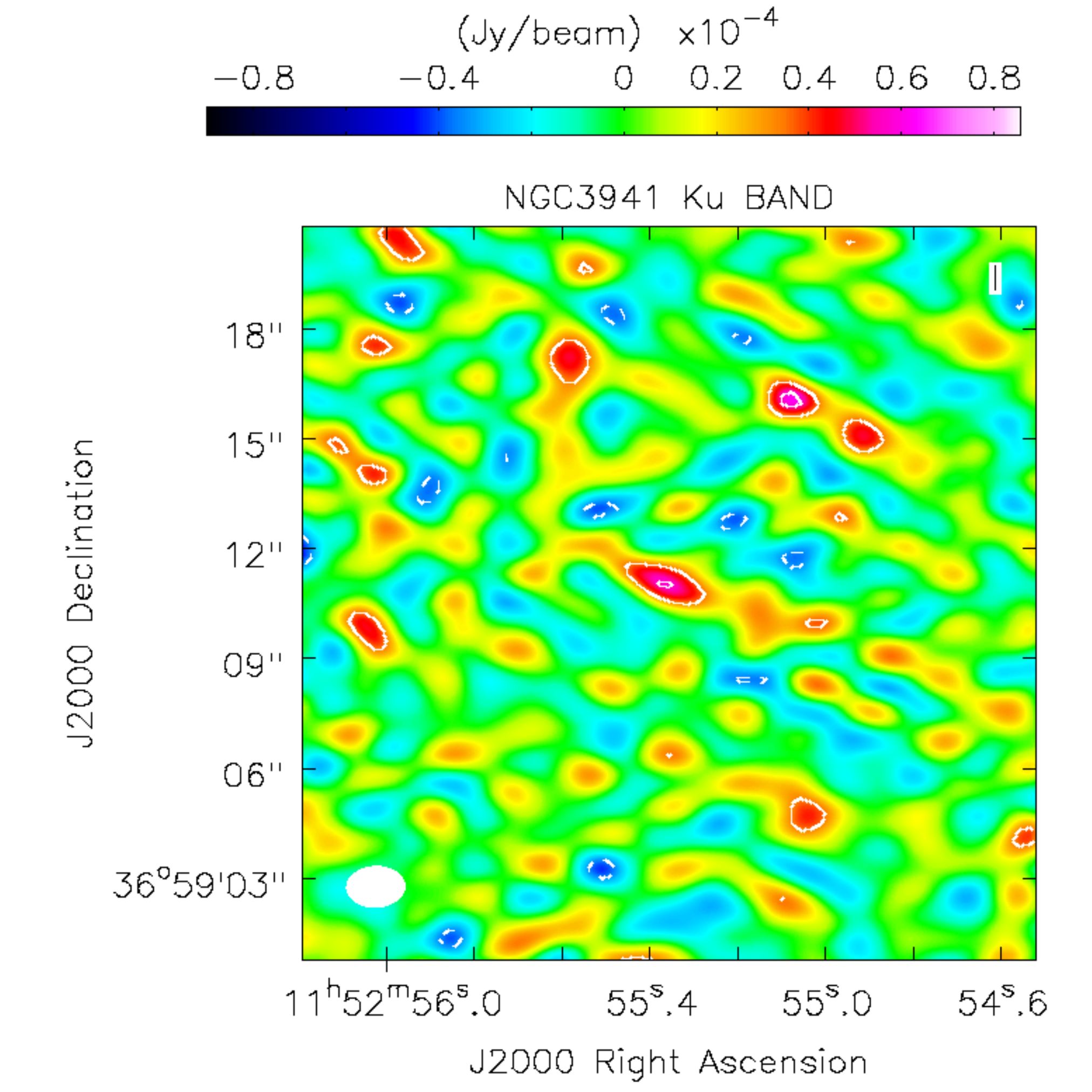}\\

\end{tabular}
 \captionof{figure}{UV-tapered contour and coloured maps of NGC~3941 in the four frequency bands: L-band (top-left), C-band (top-right), X-band (bottom-left) and Ku-band (bottom-right). The contours are displayed at [-3,3,5,10]$\times\,\sigma_{\rm image}$, where $\sigma_{\rm image}$ is the respective image noise rms given in \Tab~\ref{table:fluxTable}. The restoring beam is depicted as a white-filled ellipse in the left-hand corner of the map.}
\label{fig:contourMaps3941}
\end{table*}

\begin{table*}\scriptsize
\centering
\begin{tabular}{cc}
{\large \textbf{NGC~4477}} & {\large \textbf{NGC~4477}} \\

\includegraphics[scale=0.4]{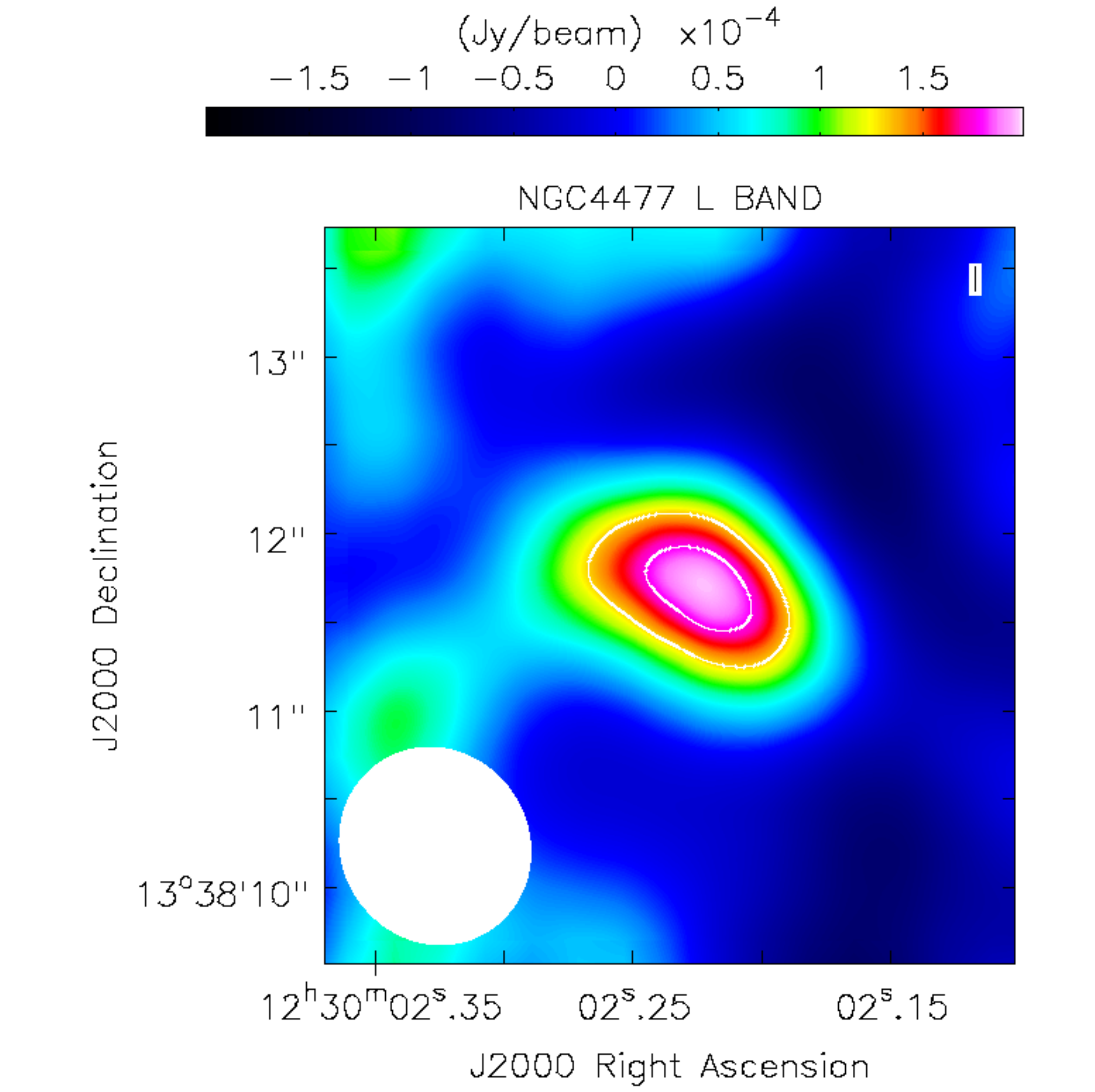} & 
\includegraphics[scale=0.4]{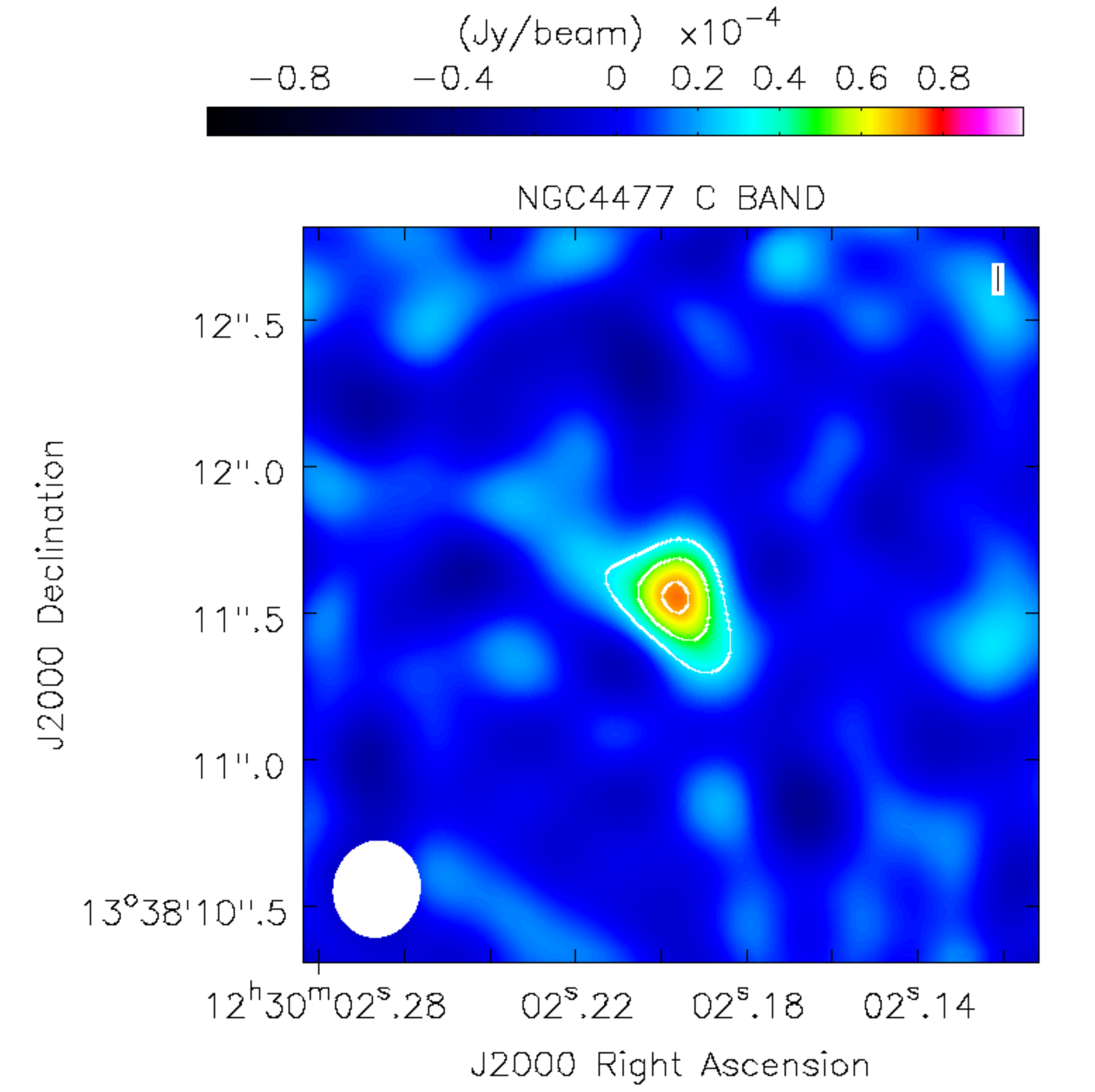} \\

\includegraphics[scale=0.4]{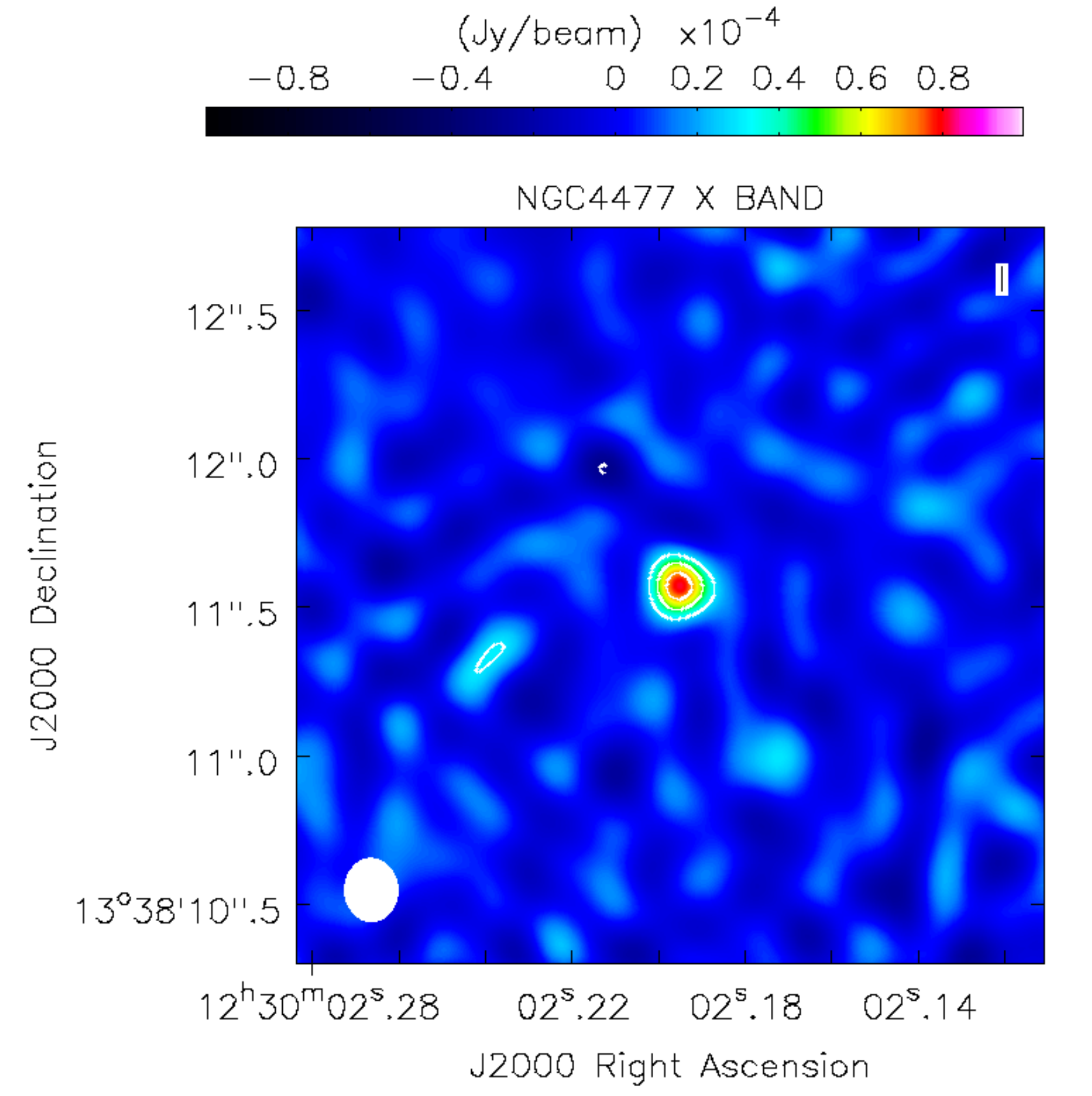} &
\includegraphics[scale=0.4]{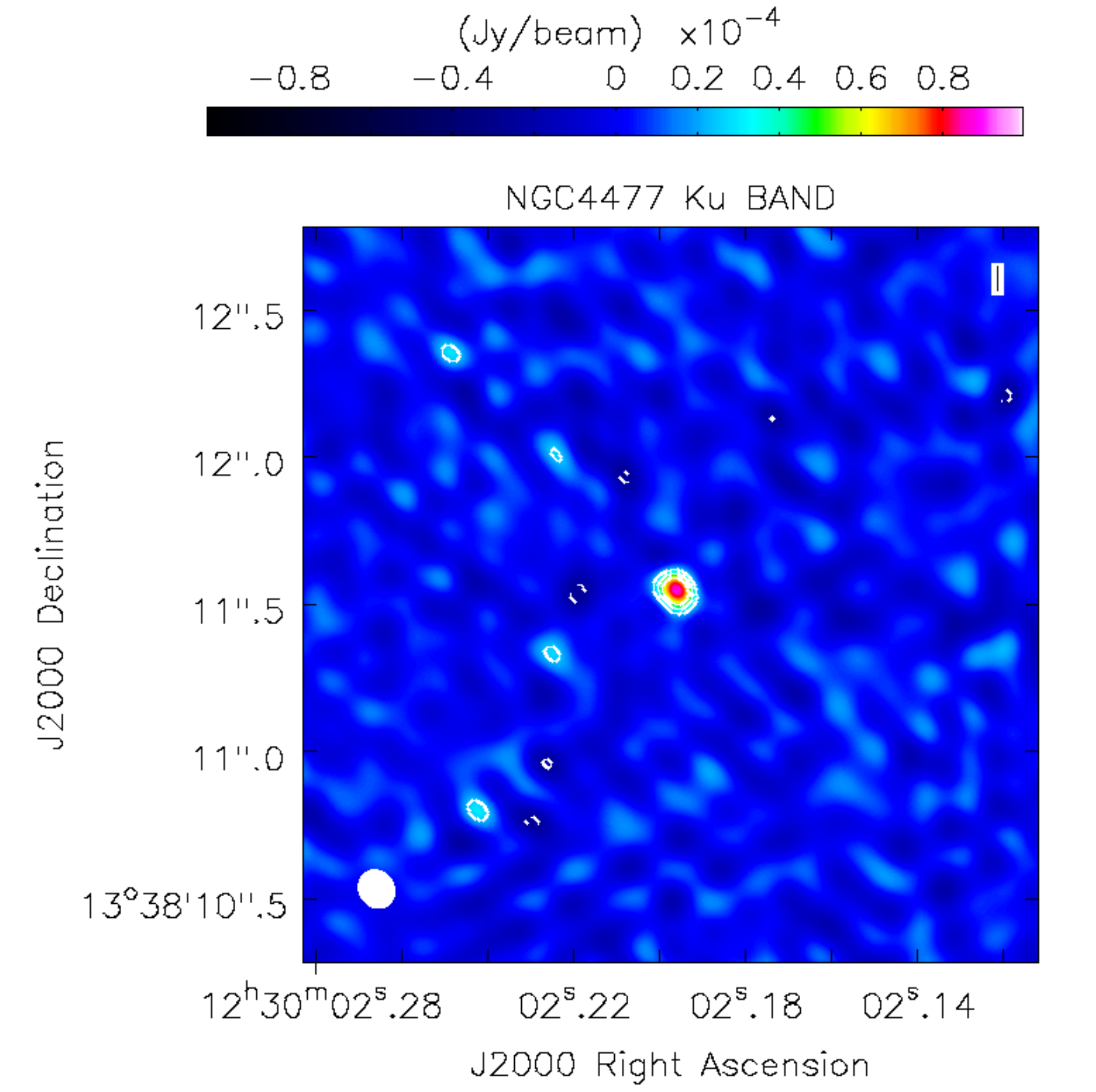}\\ 

\end{tabular}
 \captionof{figure}{Full resolution contour and coloured maps of NGC~4477 in the four frequency bands: L-band (top-left), C-band (top-right), X-band (bottom-left) and Ku-band (bottom-right). The contours are displayed at [-3,3,4,5]$\times\,\sigma_{\rm image}$ for the L band and at [-3,3,5,7]$\times\,\sigma_{\rm image}$, where $\sigma_{\rm image}$ is the respective image noise rms given in \Tab~\ref{table:fluxTable}. The restoring beam is depicted as a white-filled ellipse in the left-hand corner of the map.}
\label{fig:contourMaps4477}
\end{table*}

\begin{table*}\scriptsize
\centering
\begin{tabular}{cc}
{\large \textbf{NGC~4639}} & {\large \textbf{NGC~4639}} \\

\includegraphics[scale=0.4]{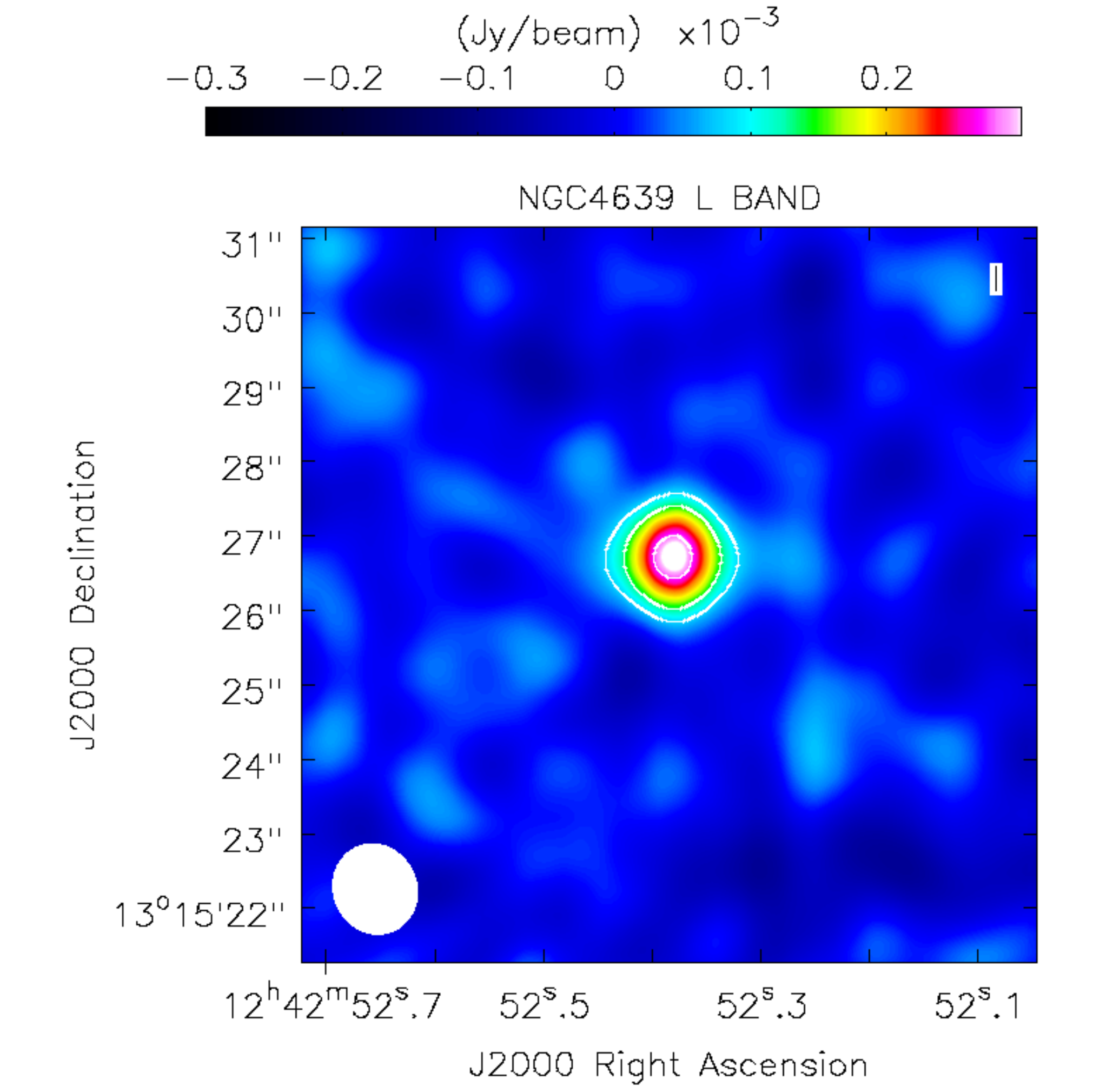} & 
\includegraphics[scale=0.4]{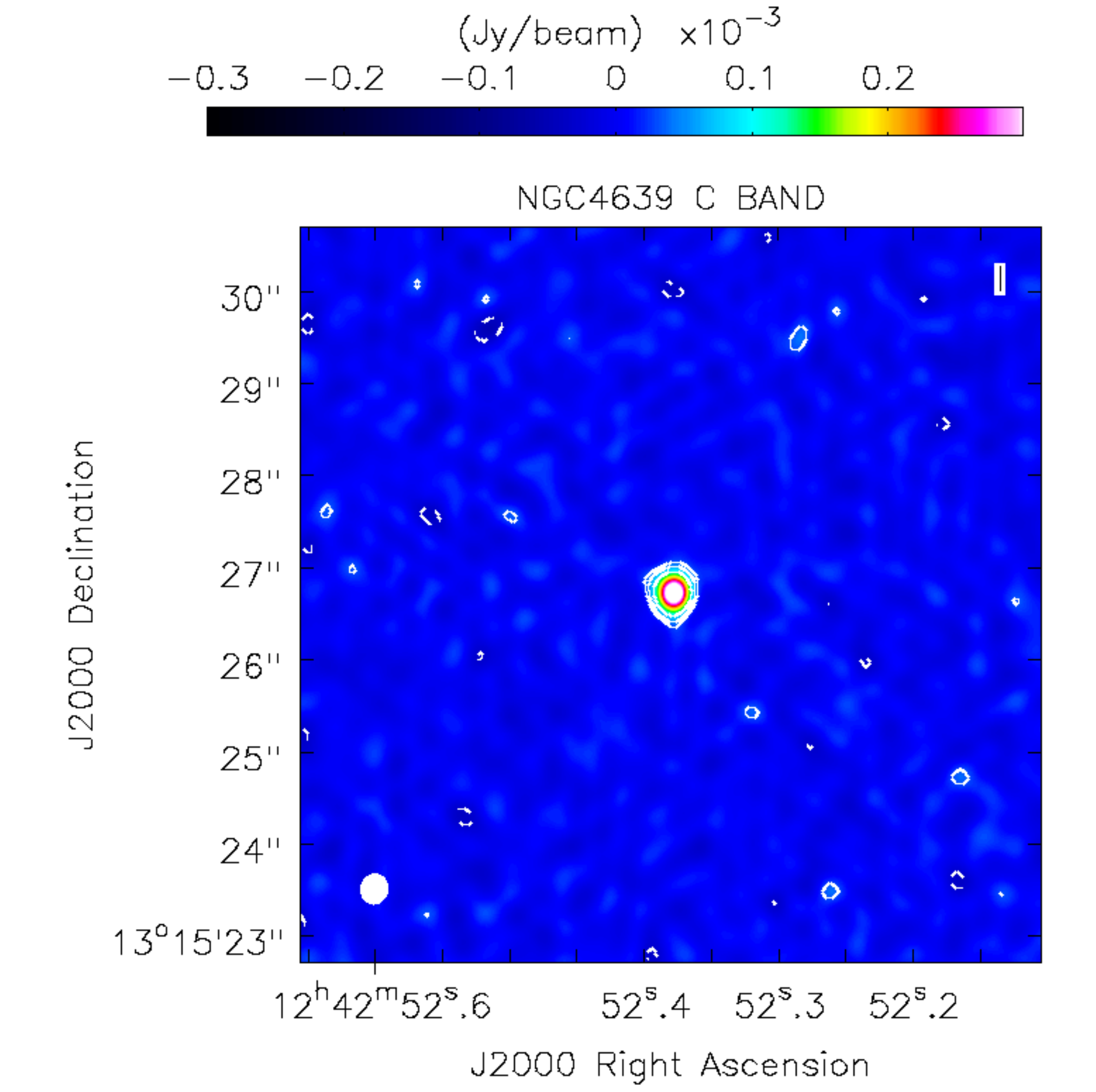} \\

\includegraphics[scale=0.4]{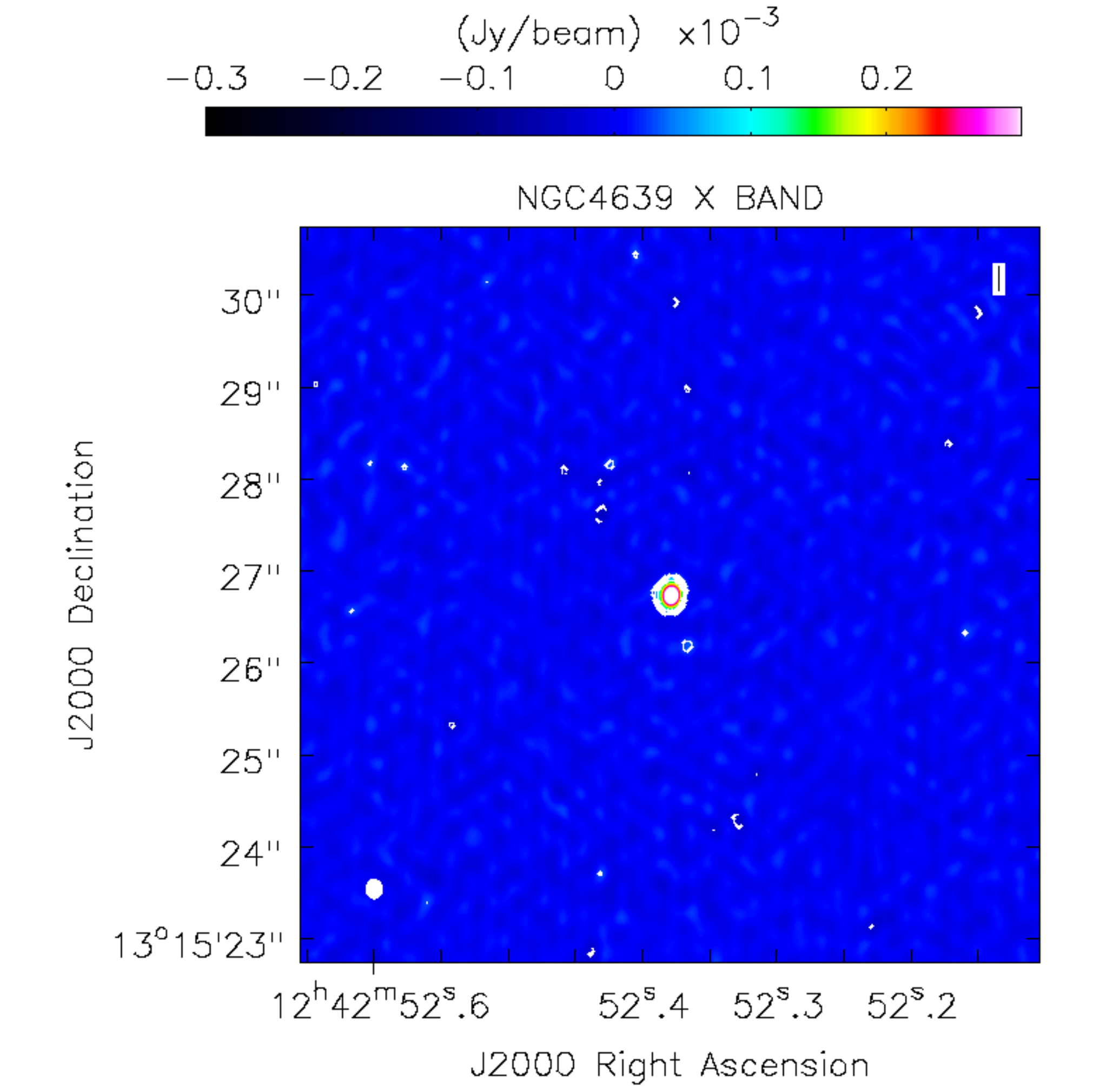} &
\includegraphics[scale=0.4]{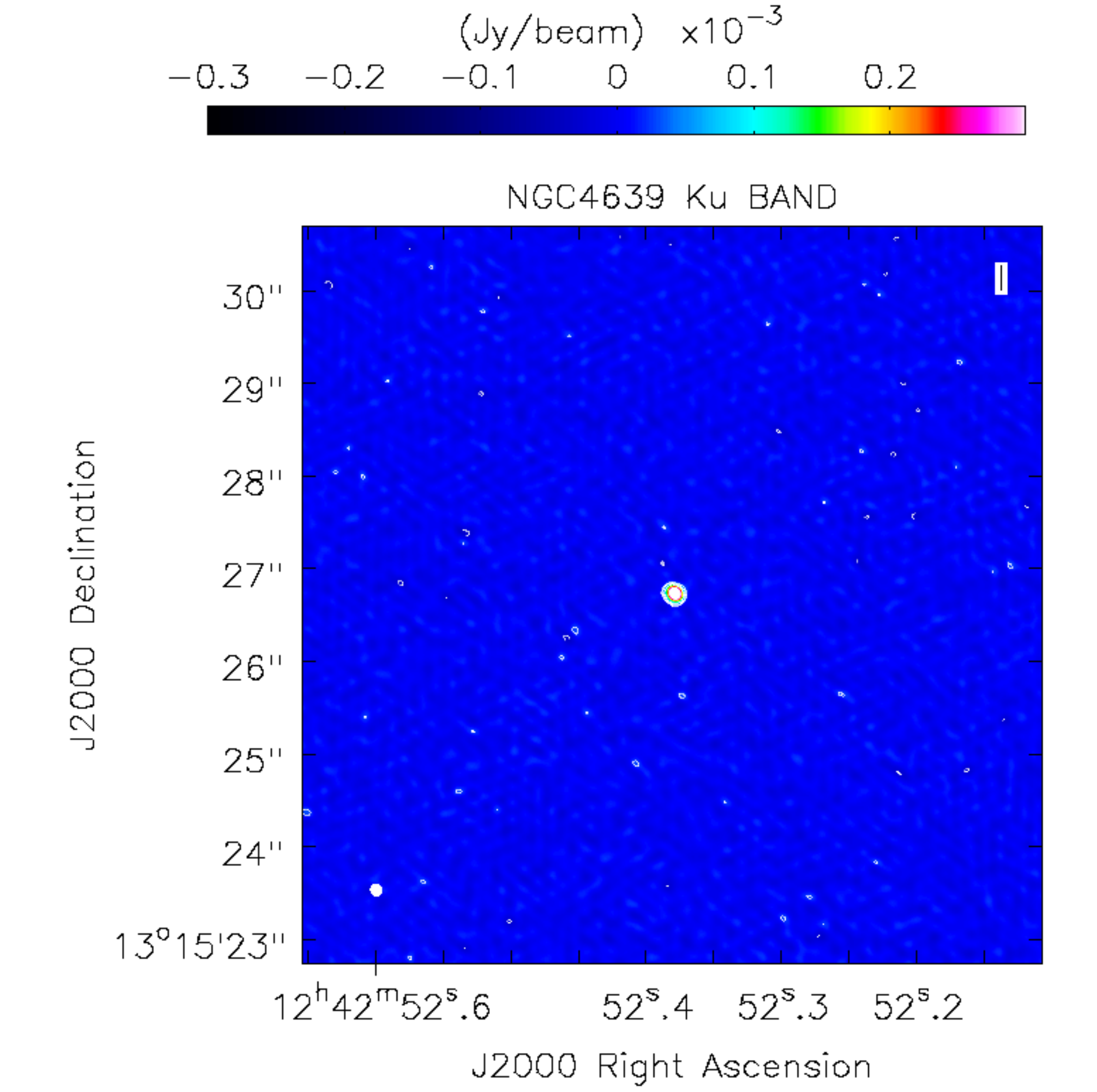}\\ 

\end{tabular}
 \captionof{figure}{Full resolution contour and coloured maps of NGC~4639 in the four frequency bands: L-band (top-left), C-band (top-right), X-band (bottom-left) and Ku-band (bottom-right). The contours are displayed at [-3,3,5,10]$\times\,\sigma_{\rm image}$, where $\sigma_{\rm image}$ is the respective image noise rms given in \Tab~\ref{table:fluxTable}. The restoring beam is depicted as a white-filled ellipse in the left-hand corner of the map.}
\label{fig:contourMaps4639}
\end{table*}

\begin{table*}\scriptsize
\centering
\begin{tabular}{cc}
{\large \textbf{NGC~4698}} & {\large \textbf{NGC~4698}} \\

\includegraphics[scale=0.4]{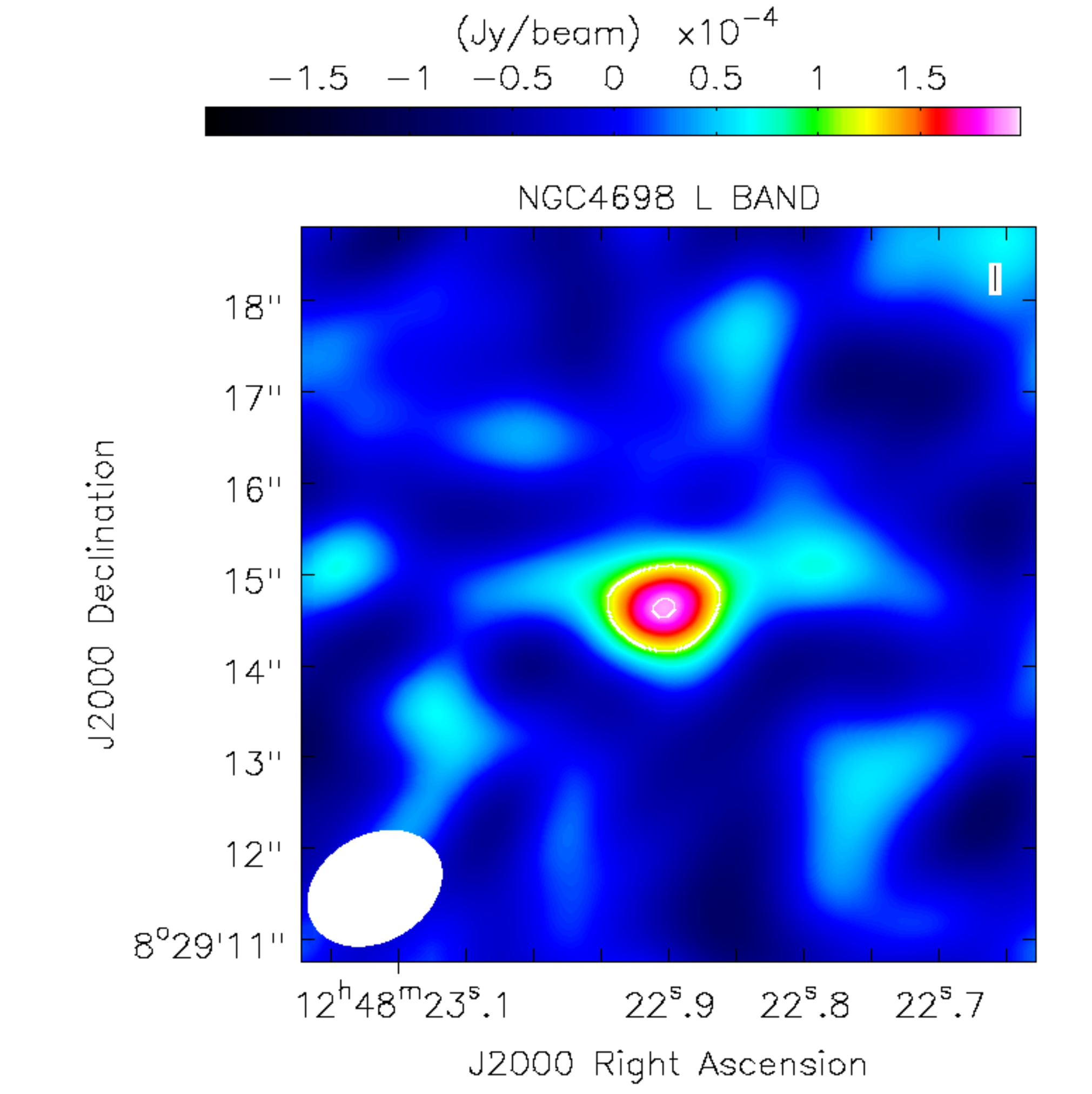} & 
\includegraphics[scale=0.4]{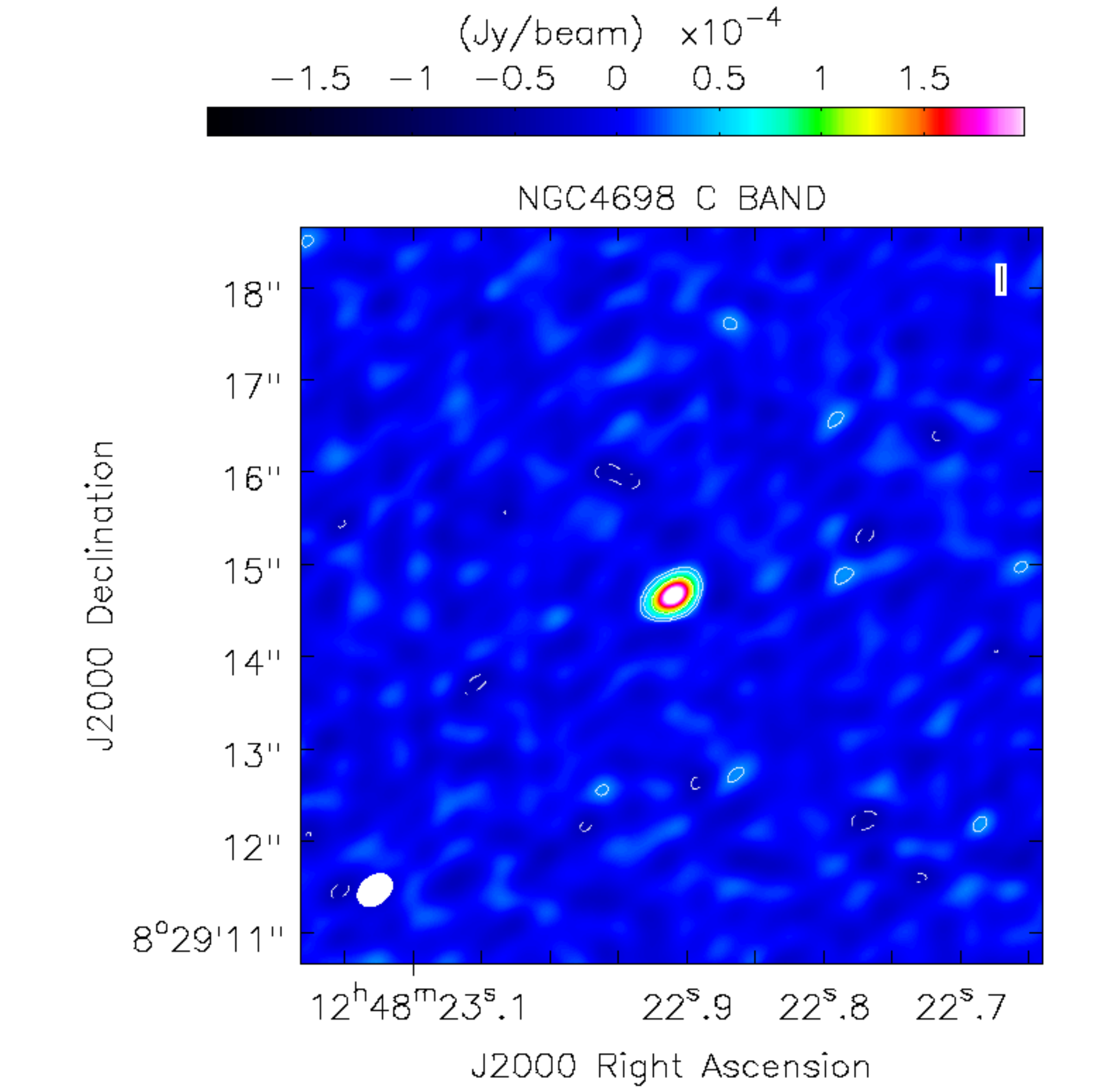} \\

\includegraphics[scale=0.4]{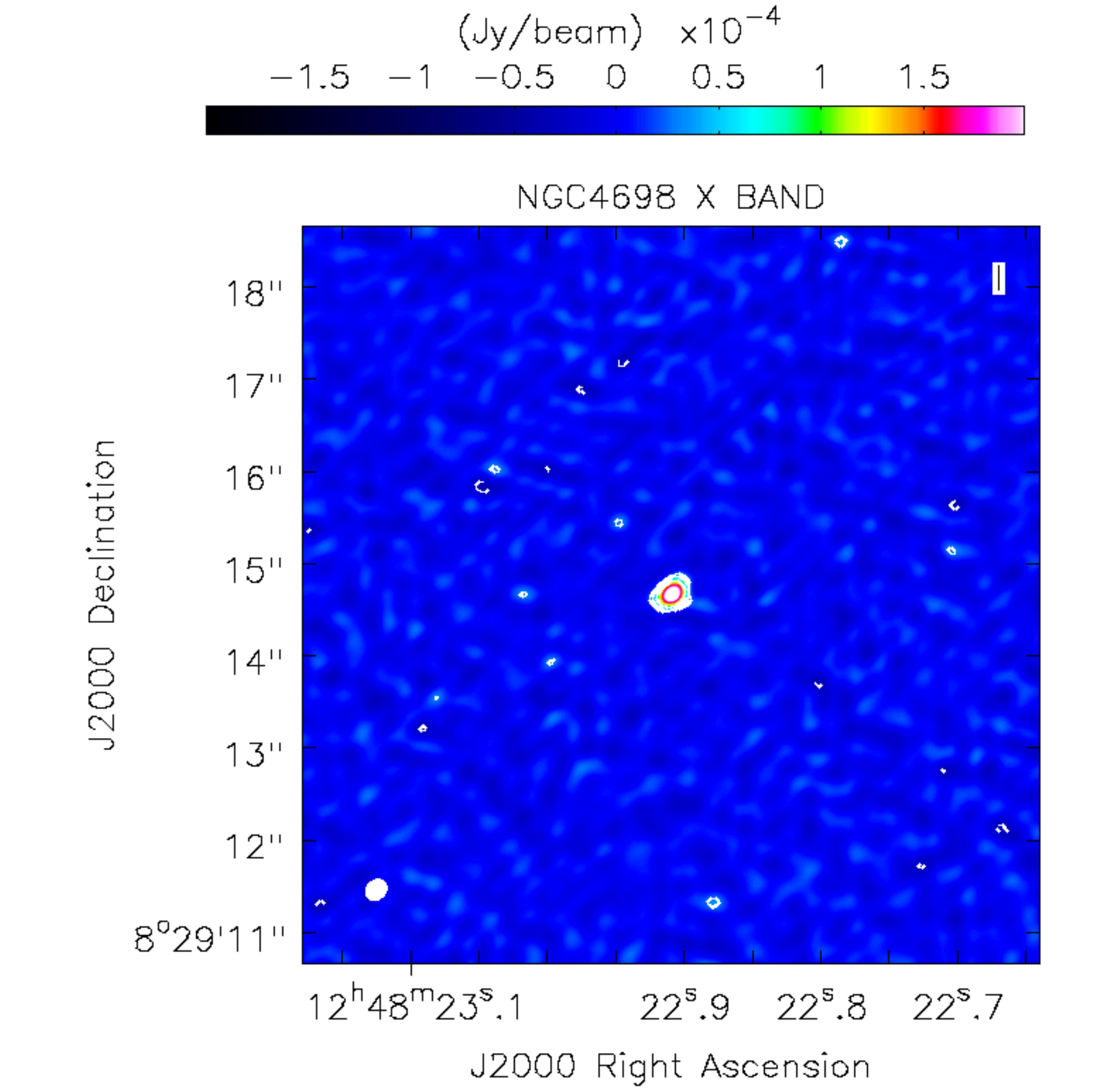} &
\includegraphics[scale=0.4]{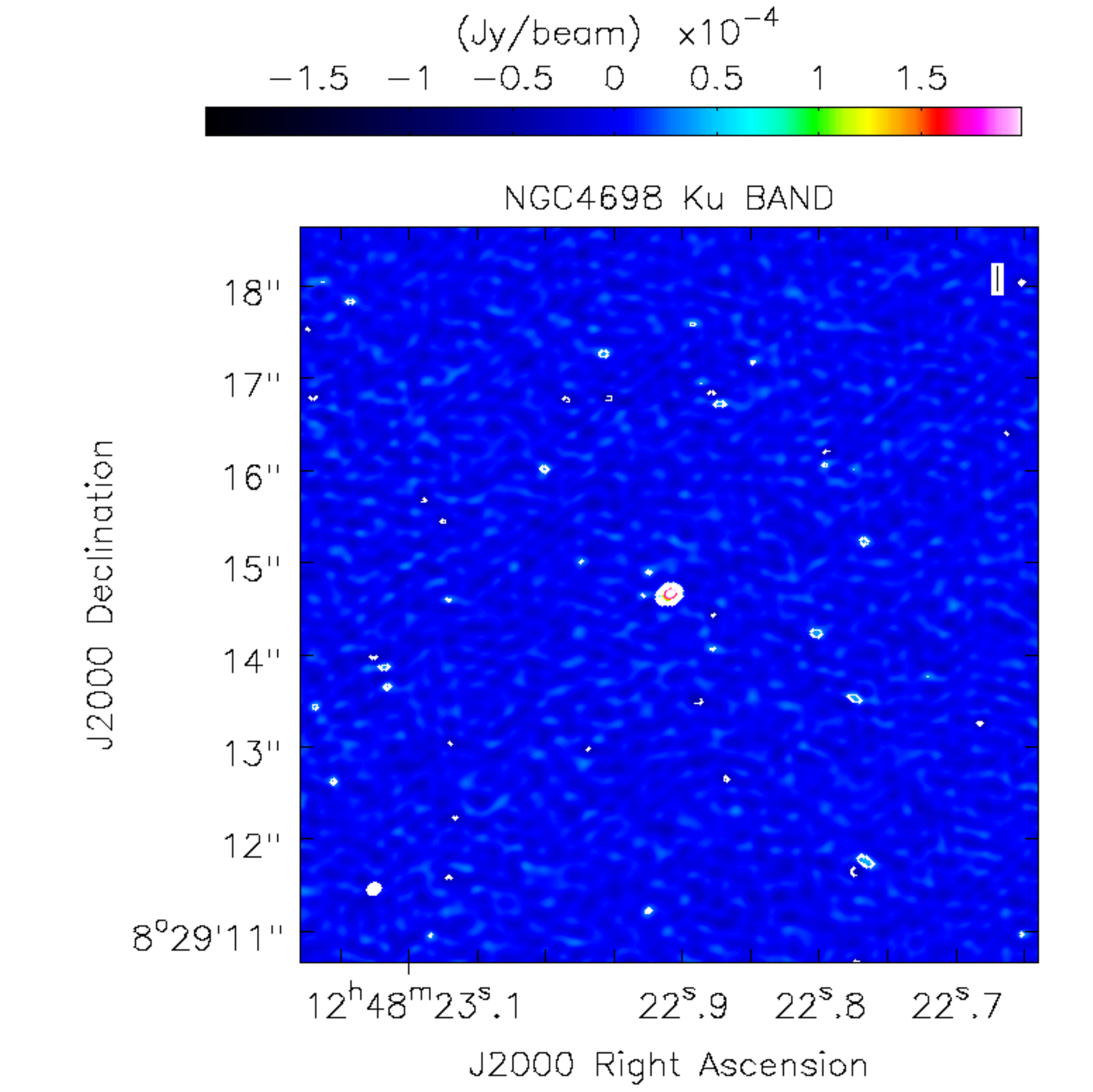}\\ 

\end{tabular}
 \captionof{figure}{Full resolution contour and coloured maps of NGC~4698 in the four frequency bands: L-band (top-left), C-band (top-right), X-band (bottom-left) and Ku-band (bottom-right). The contours are displayed at [-3,3,5,10]$\times\,\sigma_{\rm image}$, where $\sigma_{\rm image}$ is the respective image noise rms given in \Tab~\ref{table:fluxTable}. The restoring beam is depicted as a white-filled ellipse in the left-hand corner of the map.}
\label{fig:contourMaps4698}
\end{table*}

\begin{table*}\scriptsize
\begin{tabular}{cc}
{\large \textbf{NGC~4725}} & {\large \textbf{NGC~4725}} \\

\includegraphics[scale=0.4]{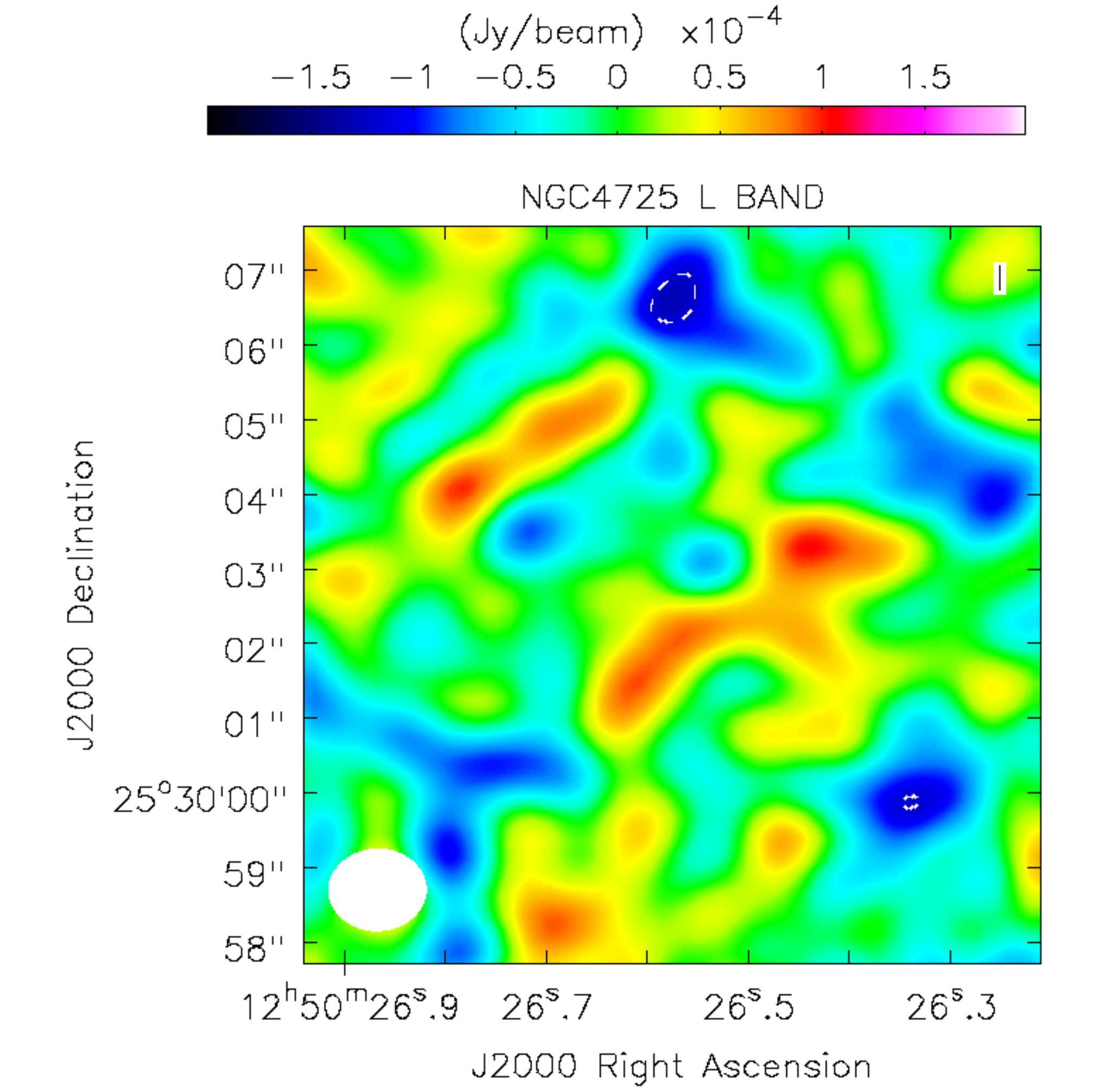} & 
\includegraphics[scale=0.4]{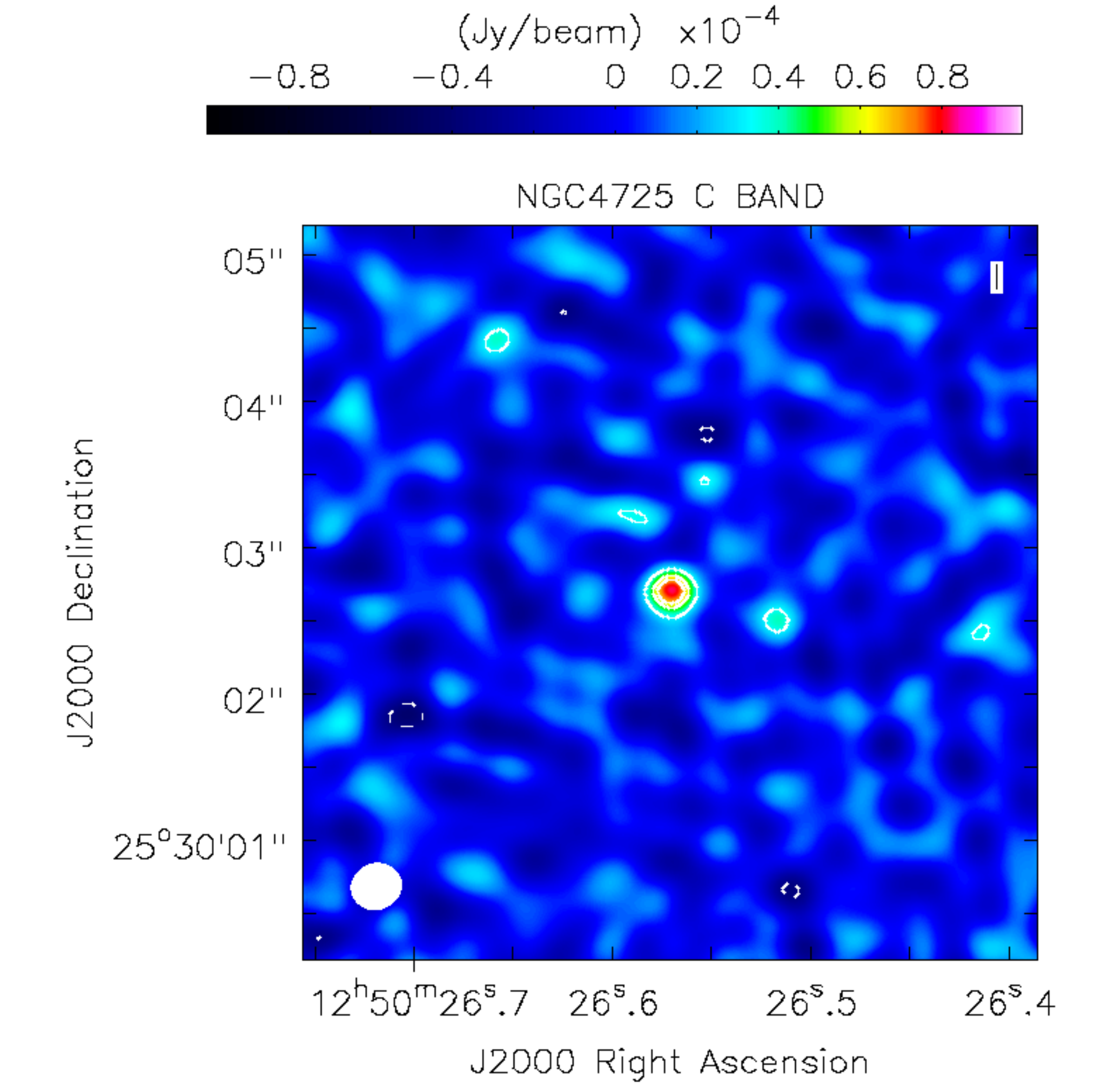} \\

\includegraphics[scale=0.4]{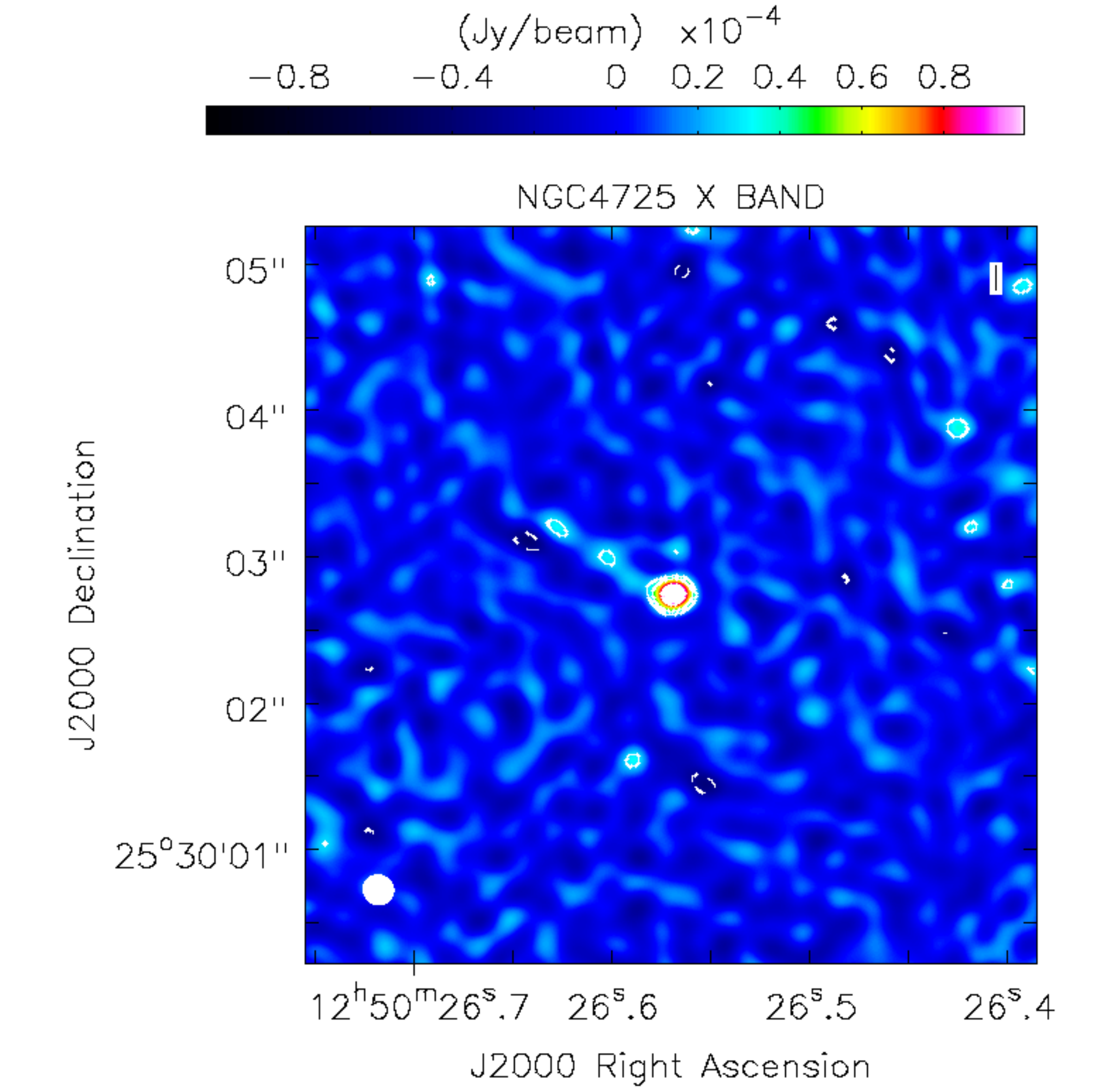} &
\includegraphics[scale=0.4]{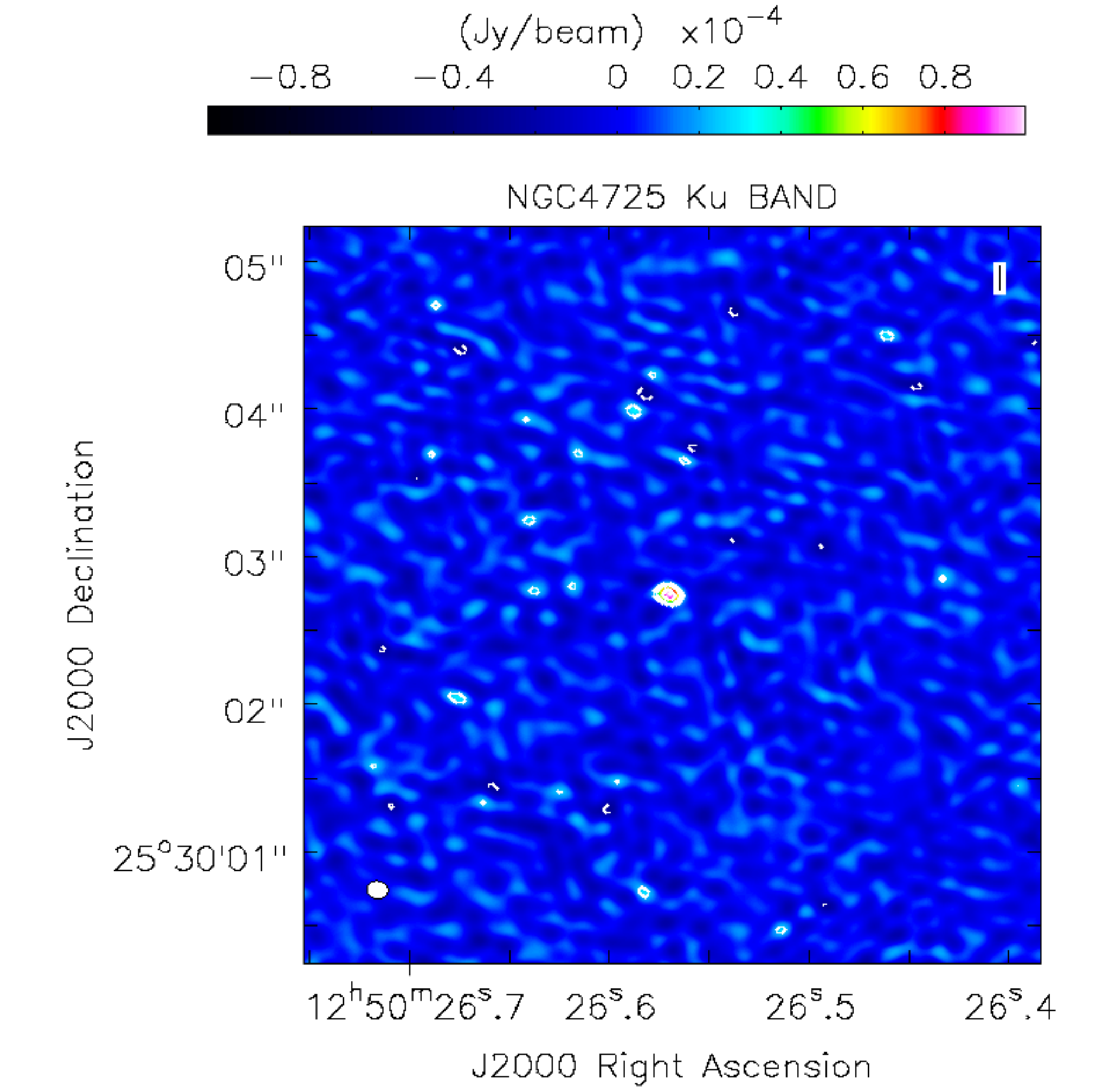}\\ 

\end{tabular}
 \captionof{figure}{Full resolution contour and coloured maps of NGC~4725 in the four frequency bands: L-band (top-left), C-band (top-right), X-band (bottom-left) and Ku-band (bottom-right). The contours are displayed at [-3,3,5,6]$\times\,\sigma_{\rm image}$ for the C band and at [-3,3,5,10]$\times\,\sigma_{\rm image}$ for X and Ku bands, where $\sigma_{\rm image}$ is the respective image noise rms given in \Tab~\ref{table:fluxTable}. The restoring beam is depicted as a white-filled ellipse in the left-hand corner of the map.}
\label{fig:contourMaps4725}
\end{table*}

\bsp	
\label{lastpage}
\end{document}

%% file: journaldefs.tex
\def\aj{AJ}					
\def\araa{ARA\&A}				
\def\apj{ApJ}					
\def\apjl{ApJL}					
\def\apjs{ApJS}					

\def\aap{A\&A}					
\def\aaps{A\&AS}				
\def\mnras{MNRAS}				

\def\ao{Appl.~Opt.}				



%% file: LLAGN.bbl
\begin{thebibliography}{}
\makeatletter
\relax
\def\mn@urlcharsother{\let\do\@makeother \do\$\do\&\do\#\do\^\do\_\do\%\do\~}
\def\mn@doi{\begingroup\mn@urlcharsother \@ifnextchar [ {\mn@doi@}
  {\mn@doi@[]}}
\def\mn@doi@[#1]#2{\def\@tempa{#1}\ifx\@tempa\@empty \href
  {http://dx.doi.org/#2} {doi:#2}\else \href {http://dx.doi.org/#2} {#1}\fi
  \endgroup}
\def\mn@eprint#1#2{\mn@eprint@#1:#2::\@nil}
\def\mn@eprint@arXiv#1{\href {http://arxiv.org/abs/#1} {{\tt arXiv:#1}}}
\def\mn@eprint@dblp#1{\href {http://dblp.uni-trier.de/rec/bibtex/#1.xml}
  {dblp:#1}}
\def\mn@eprint@#1:#2:#3:#4\@nil{\def\@tempa {#1}\def\@tempb {#2}\def\@tempc
  {#3}\ifx \@tempc \@empty \let \@tempc \@tempb \let \@tempb \@tempa \fi \ifx
  \@tempb \@empty \def\@tempb {arXiv}\fi \@ifundefined
  {mn@eprint@\@tempb}{\@tempb:\@tempc}{\expandafter \expandafter \csname
  mn@eprint@\@tempb\endcsname \expandafter{\@tempc}}}

\bibitem[\protect\citeauthoryear{{Anderson} \& {Ulvestad}}{{Anderson} \&
  {Ulvestad}}{2005}]{AndersonUlvestad2005}
{Anderson} J.~M.,  {Ulvestad} J.~S.,  2005, \mn@doi [\apj] {10.1086/430463},
  \href {http://adsabs.harvard.edu/abs/2005ApJ...627..674A} {627, 674}

\bibitem[\protect\citeauthoryear{{Baldi} et~al.,}{{Baldi}
  et~al.}{2018}]{Baldi2018}
{Baldi} R.~D.,  et~al., 2018, \mn@doi [\mnras] {10.1093/mnras/sty342}, \href
  {http://cdsads.u-strasbg.fr/abs/2018MNRAS.476.3478B} {476, 3478}

\bibitem[\protect\citeauthoryear{{Becker}, {White}  \& {Helfand}}{{Becker}
  et~al.}{1995}]{Becker1995}
{Becker} R.~H.,  {White} R.~L.,   {Helfand} D.~J.,  1995, \mn@doi [\apj]
  {10.1086/176166}, \href {http://adsabs.harvard.edu/abs/1995ApJ...450..559B}
  {450, 559}

\bibitem[\protect\citeauthoryear{{Bennett} et~al.,}{{Bennett}
  et~al.}{2003}]{Bennett2003}
{Bennett} C.~L.,  et~al., 2003, \mn@doi [\apjs] {10.1086/377253}, \href
  {http://adsabs.harvard.edu/abs/2003ApJS..148....1B} {148, 1}

\bibitem[\protect\citeauthoryear{{Blandford} \& {K{\"o}nigl}}{{Blandford} \&
  {K{\"o}nigl}}{1979}]{Blandford1979}
{Blandford} R.~D.,  {K{\"o}nigl} A.,  1979, \mn@doi [\apj] {10.1086/157262},
  \href {http://adsabs.harvard.edu/abs/1979ApJ...232...34B} {232, 34}

\bibitem[\protect\citeauthoryear{{Bontempi}, {Giroletti}, {Panessa}, {Orienti}
  \& {Doi}}{{Bontempi} et~al.}{2012}]{Bontempi+2012}
{Bontempi} P.,  {Giroletti} M.,  {Panessa} F.,  {Orienti} M.,   {Doi} A.,
  2012, \mn@doi [\mnras] {10.1111/j.1365-2966.2012.21786.x}, \href
  {http://cdsads.u-strasbg.fr/abs/2012MNRAS.426..588B} {426, 588}

\bibitem[\protect\citeauthoryear{{Briggs}}{{Briggs}}{1995}]{Briggs1995}
{Briggs} D.~S.,  1995, in American Astronomical Society Meeting Abstracts.
  p.~1444

\bibitem[\protect\citeauthoryear{{Buttiglione}, {Capetti}, {Celotti}, {Axon},
  {Chiaberge}, {Macchetto}  \& {Sparks}}{{Buttiglione}
  et~al.}{2010}]{Buttiglione2010}
{Buttiglione} S.,  {Capetti} A.,  {Celotti} A.,  {Axon} D.~J.,  {Chiaberge} M.,
   {Macchetto} F.~D.,   {Sparks} W.~B.,  2010, \mn@doi [\aap]
  {10.1051/0004-6361/200913290}, \href
  {http://adsabs.harvard.edu/abs/2010A%26A...509A...6B} {509, A6}

\bibitem[\protect\citeauthoryear{{Cappi} et~al.,}{{Cappi}
  et~al.}{2006}]{Cappietal06}
{Cappi} M.,  et~al., 2006, \mn@doi [\aap] {10.1051/0004-6361:20053893}, \href
  {http://adsabs.harvard.edu/abs/2006A%26A...446..459C} {446, 459}

\bibitem[\protect\citeauthoryear{{Condon}}{{Condon}}{1992}]{Condon1992}
{Condon} J.~J.,  1992, \mn@doi [\araa] {10.1146/annurev.aa.30.090192.003043},
  \href {http://adsabs.harvard.edu/abs/1992ARA%26A..30..575C} {30, 575}

\bibitem[\protect\citeauthoryear{{Conway}, {Cornwell}  \& {Wilkinson}}{{Conway}
  et~al.}{1990}]{CCW90}
{Conway} J.~E.,  {Cornwell} T.~J.,   {Wilkinson} P.~N.,  1990, \mnras, \href
  {http://adsabs.harvard.edu/abs/1990MNRAS.246..490C} {246, 490}

\bibitem[\protect\citeauthoryear{{Falcke} \& {Biermann}}{{Falcke} \&
  {Biermann}}{1995}]{FalckeBiermann1995}
{Falcke} H.,  {Biermann} P.~L.,  1995, \aap, \href
  {http://adsabs.harvard.edu/abs/1995A%26A...293..665F} {293, 665}

\bibitem[\protect\citeauthoryear{{Falcke} \& {Biermann}}{{Falcke} \&
  {Biermann}}{1999}]{Falcke1999}
{Falcke} H.,  {Biermann} P.~L.,  1999, \aap, \href
  {http://adsabs.harvard.edu/abs/1999A%26A...342...49F} {342, 49}

\bibitem[\protect\citeauthoryear{{Falcke} \& {Markoff}}{{Falcke} \&
  {Markoff}}{2000}]{FalckeMarkoff2000}
{Falcke} H.,  {Markoff} S.,  2000, \aap, \href
  {http://adsabs.harvard.edu/abs/2000A%26A...362..113F} {362, 113}

\bibitem[\protect\citeauthoryear{{Falcke}, {Sherwood}  \& {Patnaik}}{{Falcke}
  et~al.}{1996}]{FalckeSherwoodPatnaik1996}
{Falcke} H.,  {Sherwood} W.,   {Patnaik} A.~R.,  1996, \mn@doi [\apj]
  {10.1086/177956}, \href {http://adsabs.harvard.edu/abs/1996ApJ...471..106F}
  {471, 106}

\bibitem[\protect\citeauthoryear{{Falcke}, {Nagar}, {Wilson}  \&
  {Ulvestad}}{{Falcke} et~al.}{2000}]{Falcke2000}
{Falcke} H.,  {Nagar} N.~M.,  {Wilson} A.~S.,   {Ulvestad} J.~S.,  2000,
  \mn@doi [\apj] {10.1086/309543}, \href
  {http://cdsads.u-strasbg.fr/abs/2000ApJ...542..197F} {542, 197}

\bibitem[\protect\citeauthoryear{{Falcke}, {K{\"o}rding}  \&
  {Markoff}}{{Falcke} et~al.}{2004}]{FalckeKordingMarkoff2004}
{Falcke} H.,  {K{\"o}rding} E.,   {Markoff} S.,  2004, \mn@doi [\aap]
  {10.1051/0004-6361:20031683}, \href
  {http://adsabs.harvard.edu/abs/2004A%26A...414..895F} {414, 895}

\bibitem[\protect\citeauthoryear{{Fender}, {Belloni}  \& {Gallo}}{{Fender}
  et~al.}{2004}]{Fender2004}
{Fender} R.~P.,  {Belloni} T.~M.,   {Gallo} E.,  2004, \mn@doi [\mnras]
  {10.1111/j.1365-2966.2004.08384.x}, \href
  {http://adsabs.harvard.edu/abs/2004MNRAS.355.1105F} {355, 1105}

\bibitem[\protect\citeauthoryear{{Filippenko} \& {Sargent}}{{Filippenko} \&
  {Sargent}}{1985}]{FilippenkoSargent1985}
{Filippenko} A.~V.,  {Sargent} W.~L.~W.,  1985, \mn@doi [\apjs]
  {10.1086/191012}, \href {http://cdsads.u-strasbg.fr/abs/1985ApJS...57..503F}
  {57, 503}

\bibitem[\protect\citeauthoryear{{Gallimore}, {Baum}  \& {O'Dea}}{{Gallimore}
  et~al.}{2004}]{Gallimore2004}
{Gallimore} J.~F.,  {Baum} S.~A.,   {O'Dea} C.~P.,  2004, \mn@doi [\apj]
  {10.1086/423167}, \href {http://adsabs.harvard.edu/abs/2004ApJ...613..794G}
  {613, 794}

\bibitem[\protect\citeauthoryear{{Gallo}, {Fender}  \& {Pooley}}{{Gallo}
  et~al.}{2003}]{Gallo2003}
{Gallo} E.,  {Fender} R.~P.,   {Pooley} G.~G.,  2003, \mn@doi [\mnras]
  {10.1046/j.1365-8711.2003.06791.x}, \href
  {http://cdsads.u-strasbg.fr/abs/2003MNRAS.344...60G} {344, 60}

\bibitem[\protect\citeauthoryear{{Ghisellini}, {Haardt}  \&
  {Matt}}{{Ghisellini} et~al.}{2004}]{Ghisellini2004}
{Ghisellini} G.,  {Haardt} F.,   {Matt} G.,  2004, \mn@doi [\aap]
  {10.1051/0004-6361:20031562}, \href
  {http://adsabs.harvard.edu/abs/2004A%26A...413..535G} {413, 535}

\bibitem[\protect\citeauthoryear{{Giroletti} \& {Panessa}}{{Giroletti} \&
  {Panessa}}{2009}]{GirolettiPanessa2009}
{Giroletti} M.,  {Panessa} F.,  2009, \mn@doi [\apjl]
  {10.1088/0004-637X/706/2/L260}, \href
  {http://cdsads.u-strasbg.fr/abs/2009ApJ...706L.260G} {706, L260}

\bibitem[\protect\citeauthoryear{{Gu} \& {Cao}}{{Gu} \&
  {Cao}}{2009}]{GuCao2009}
{Gu} M.,  {Cao} X.,  2009, \mn@doi [\mnras] {10.1111/j.1365-2966.2009.15277.x},
  \href {http://adsabs.harvard.edu/abs/2009MNRAS.399..349G} {399, 349}

\bibitem[\protect\citeauthoryear{{Hakobyan}, {Adibekyan}, {Aramyan},
  {Petrosian}, {Gomes}, {Mamon}, {Kunth}  \& {Turatto}}{{Hakobyan}
  et~al.}{2012}]{Hakobyan+12}
{Hakobyan} A.~A.,  {Adibekyan} V.~Z.,  {Aramyan} L.~S.,  {Petrosian} A.~R.,
  {Gomes} J.~M.,  {Mamon} G.~A.,  {Kunth} D.,   {Turatto} M.,  2012, \mn@doi
  [\aap] {10.1051/0004-6361/201219541}, \href
  {http://adsabs.harvard.edu/abs/2012A%26A...544A..81H} {544, A81}

\bibitem[\protect\citeauthoryear{{Ho}}{{Ho}}{2008}]{Ho2008}
{Ho} L.~C.,  2008, \mn@doi [\araa] {10.1146/annurev.astro.45.051806.110546},
  \href {http://adsabs.harvard.edu/abs/2008ARA%26A..46..475H} {46, 475}

\bibitem[\protect\citeauthoryear{{Ho} \& {Ulvestad}}{{Ho} \&
  {Ulvestad}}{2001}]{HU01}
{Ho} L.~C.,  {Ulvestad} J.~S.,  2001, \mn@doi [\apjs] {10.1086/319185}, \href
  {http://adsabs.harvard.edu/abs/2001ApJS..133...77H} {133, 77}

\bibitem[\protect\citeauthoryear{{Ho}, {Filippenko}  \& {Sargent}}{{Ho}
  et~al.}{1995}]{HoFilippenkoSargent1995}
{Ho} L.~C.,  {Filippenko} A.~V.,   {Sargent} W.~L.,  1995, \mn@doi [\apjs]
  {10.1086/192170}, \href {http://cdsads.u-strasbg.fr/abs/1995ApJS...98..477H}
  {98, 477}

\bibitem[\protect\citeauthoryear{{Ho}, {Filippenko}  \& {Sargent}}{{Ho}
  et~al.}{1997a}]{HoFilippenkoSargent1997a}
{Ho} L.~C.,  {Filippenko} A.~V.,   {Sargent} W.~L.~W.,  1997a, \mn@doi [\apjs]
  {10.1086/313041}, \href {http://cdsads.u-strasbg.fr/abs/1997ApJS..112..315H}
  {112, 315}

\bibitem[\protect\citeauthoryear{{Ho}, {Filippenko}  \& {Sargent}}{{Ho}
  et~al.}{1997b}]{HoFilippenkoSargent1997}
{Ho} L.~C.,  {Filippenko} A.~V.,   {Sargent} W.~L.~W.,  1997b, \mn@doi [\apj]
  {10.1086/304638}, \href {http://adsabs.harvard.edu/abs/1997ApJ...487..568H}
  {487, 568}

\bibitem[\protect\citeauthoryear{{H{\"o}gbom}}{{H{\"o}gbom}}{1974}]{Hogbom+1974}
{H{\"o}gbom} J.~A.,  1974, \aaps, \href
  {http://cdsads.u-strasbg.fr/abs/1974A%26AS...15..417H} {15, 417}

\bibitem[\protect\citeauthoryear{{Isobe}, {Feigelson}, {Akritas}  \&
  {Babu}}{{Isobe} et~al.}{1990}]{Isobe1990}
{Isobe} T.,  {Feigelson} E.~D.,  {Akritas} M.~G.,   {Babu} G.~J.,  1990,
  \mn@doi [\apj] {10.1086/169390}, \href
  {http://cdsads.u-strasbg.fr/abs/1990ApJ...364..104I} {364, 104}

\bibitem[\protect\citeauthoryear{Jones, Oliphant, Peterson  et~al.}{Jones
  et~al.}{01  }]{scipy}
Jones E.,  Oliphant T.,  Peterson P.,   et~al., 2001--, {SciPy}: Open source
  scientific tools for {Python}, \url {http://www.scipy.org/}

\bibitem[\protect\citeauthoryear{{Kellermann}, {Sramek}, {Schmidt}, {Green}  \&
  {Shaffer}}{{Kellermann} et~al.}{1994}]{Kellerman1994}
{Kellermann} K.~I.,  {Sramek} R.~A.,  {Schmidt} M.,  {Green} R.~F.,   {Shaffer}
  D.~B.,  1994, \mn@doi [\aj] {10.1086/117145}, \href
  {http://adsabs.harvard.edu/abs/1994AJ....108.1163K} {108, 1163}

\bibitem[\protect\citeauthoryear{{Kewley}, {Groves}, {Kauffmann}  \&
  {Heckman}}{{Kewley} et~al.}{2006}]{Kewley2006}
{Kewley} L.~J.,  {Groves} B.,  {Kauffmann} G.,   {Heckman} T.,  2006, \mn@doi
  [\mnras] {10.1111/j.1365-2966.2006.10859.x}, \href
  {http://adsabs.harvard.edu/abs/2006MNRAS.372..961K} {372, 961}

\bibitem[\protect\citeauthoryear{{Kharb}, {O'Dea}, {Baum}, {Colbert}  \&
  {Xu}}{{Kharb} et~al.}{2006}]{Kharb2006}
{Kharb} P.,  {O'Dea} C.~P.,  {Baum} S.~A.,  {Colbert} E.~J.~M.,   {Xu} C.,
  2006, \mn@doi [\apj] {10.1086/507945}, \href
  {http://adsabs.harvard.edu/abs/2006ApJ...652..177K} {652, 177}

\bibitem[\protect\citeauthoryear{{K{\"o}rding}, {Falcke}  \&
  {Corbel}}{{K{\"o}rding} et~al.}{2006}]{Kording2006}
{K{\"o}rding} E.,  {Falcke} H.,   {Corbel} S.,  2006, \mn@doi [\aap]
  {10.1051/0004-6361:20054144}, \href
  {http://adsabs.harvard.edu/abs/2006A%26A...456..439K} {456, 439}

\bibitem[\protect\citeauthoryear{{Laor} \& {Behar}}{{Laor} \&
  {Behar}}{2008}]{LaorBehar2008}
{Laor} A.,  {Behar} E.,  2008, \mn@doi [\mnras]
  {10.1111/j.1365-2966.2008.13806.x}, \href
  {http://adsabs.harvard.edu/abs/2008MNRAS.390..847L} {390, 847}

\bibitem[\protect\citeauthoryear{{Laor}, {Baldi}  \& {Behar}}{{Laor}
  et~al.}{2019}]{LaorBaldiBehar2019}
{Laor} A.,  {Baldi} R.~D.,   {Behar} E.,  2019, \mn@doi [\mnras]
  {10.1093/mnras/sty3098}, \href
  {http://adsabs.harvard.edu/abs/2019MNRAS.482.5513L} {482, 5513}

\bibitem[\protect\citeauthoryear{{Lavalley}, {Isobe}  \&
  {Feigelson}}{{Lavalley} et~al.}{1992}]{Lavalley1992}
{Lavalley} M.,  {Isobe} T.,   {Feigelson} E.,  1992, in {Worrall} D.~M.,
  {Biemesderfer} C.,   {Barnes} J.,  eds,  Astronomical Society of the Pacific
  Conference Series Vol. 25, Astronomical Data Analysis Software and Systems I.
  p.~245

\bibitem[\protect\citeauthoryear{{McMullin}, {Waters}, {Schiebel}, {Young}  \&
  {Golap}}{{McMullin} et~al.}{2007}]{McMullin+07}
{McMullin} J.~P.,  {Waters} B.,  {Schiebel} D.,  {Young} W.,   {Golap} K.,
  2007, in {Shaw} R.~A.,  {Hill} F.,   {Bell} D.~J.,  eds,  Astronomical
  Society of the Pacific Conference Series Vol. 376, Astronomical Data Analysis
  Software and Systems XVI. p.~127

\bibitem[\protect\citeauthoryear{{Merloni}, {Heinz}  \& {di Matteo}}{{Merloni}
  et~al.}{2003}]{Merloni2003}
{Merloni} A.,  {Heinz} S.,   {di Matteo} T.,  2003, \mn@doi [\mnras]
  {10.1046/j.1365-2966.2003.07017.x}, \href
  {http://adsabs.harvard.edu/abs/2003MNRAS.345.1057M} {345, 1057}

\bibitem[\protect\citeauthoryear{{Nagar}, {Falcke}, {Wilson}  \& {Ho}}{{Nagar}
  et~al.}{2000}]{Nagar2000}
{Nagar} N.~M.,  {Falcke} H.,  {Wilson} A.~S.,   {Ho} L.~C.,  2000, \mn@doi
  [\apj] {10.1086/309524}, \href
  {http://adsabs.harvard.edu/abs/2000ApJ...542..186N} {542, 186}

\bibitem[\protect\citeauthoryear{{Nagar}, {Falcke}  \& {Wilson}}{{Nagar}
  et~al.}{2005}]{Nagar2005}
{Nagar} N.~M.,  {Falcke} H.,   {Wilson} A.~S.,  2005, \mn@doi [\aap]
  {10.1051/0004-6361:20042277}, \href
  {http://cdsads.u-strasbg.fr/abs/2005A%26A...435..521N} {435, 521}

\bibitem[\protect\citeauthoryear{{Narayan} \& {Yi}}{{Narayan} \&
  {Yi}}{1994}]{NarayanYi1994}
{Narayan} R.,  {Yi} I.,  1994, \mn@doi [\apjl] {10.1086/187381}, \href
  {http://adsabs.harvard.edu/abs/1994ApJ...428L..13N} {428, L13}

\bibitem[\protect\citeauthoryear{{Orienti} \& {Prieto}}{{Orienti} \&
  {Prieto}}{2010}]{OrientiPrieto2010}
{Orienti} M.,  {Prieto} M.~A.,  2010, \mn@doi [\mnras]
  {10.1111/j.1365-2966.2009.15837.x}, \href
  {http://adsabs.harvard.edu/abs/2010MNRAS.401.2599O} {401, 2599}

\bibitem[\protect\citeauthoryear{{Orienti}, {D'Ammando}, {Giroletti},
  {Giovannini}  \& {Panessa}}{{Orienti} et~al.}{2015}]{Orienti2015}
{Orienti} M.,  {D'Ammando} F.,  {Giroletti} M.,  {Giovannini} G.,   {Panessa}
  F.,  2015, Advancing Astrophysics with the Square Kilometre Array (AASKA14),
  \href {http://adsabs.harvard.edu/abs/2015aska.confE..87O} {p.~87}

\bibitem[\protect\citeauthoryear{{Panessa} \& {Giroletti}}{{Panessa} \&
  {Giroletti}}{2013}]{PG13}
{Panessa} F.,  {Giroletti} M.,  2013, \mn@doi [\mnras] {10.1093/mnras/stt547},
  \href {http://adsabs.harvard.edu/abs/2013MNRAS.432.1138P} {432, 1138}

\bibitem[\protect\citeauthoryear{{Panessa}, {Bassani}, {Cappi}, {Dadina},
  {Barcons}, {Carrera}, {Ho}  \& {Iwasawa}}{{Panessa}
  et~al.}{2006}]{Panessa2006}
{Panessa} F.,  {Bassani} L.,  {Cappi} M.,  {Dadina} M.,  {Barcons} X.,
  {Carrera} F.~J.,  {Ho} L.~C.,   {Iwasawa} K.,  2006, \mn@doi [\aap]
  {10.1051/0004-6361:20064894}, \href
  {http://cdsads.u-strasbg.fr/abs/2006A%26A...455..173P} {455, 173}

\bibitem[\protect\citeauthoryear{{Panessa}, {Barcons}, {Bassani}, {Cappi},
  {Carrera}, {Ho}  \& {Pellegrini}}{{Panessa} et~al.}{2007}]{Panessa2007}
{Panessa} F.,  {Barcons} X.,  {Bassani} L.,  {Cappi} M.,  {Carrera} F.~J.,
  {Ho} L.~C.,   {Pellegrini} S.,  2007, \mn@doi [\aap]
  {10.1051/0004-6361:20066943}, \href
  {http://cdsads.u-strasbg.fr/abs/2007A%26A...467..519P} {467, 519}

\bibitem[\protect\citeauthoryear{{Panessa}, {Baldi}, {Laor}, {Padovani},
  {Behar}  \& {McHardy}}{{Panessa} et~al.}{2019}]{Panessa2019}
{Panessa} F.,  {Baldi} R.~D.,  {Laor} A.,  {Padovani} P.,  {Behar} E.,
  {McHardy} I.,  2019, arXiv e-prints, \href
  {http://adsabs.harvard.edu/abs/2019arXiv190205917P} {}

\bibitem[\protect\citeauthoryear{{Perley} \& {Butler}}{{Perley} \&
  {Butler}}{2013}]{PerleyButler13}
{Perley} R.~A.,  {Butler} B.~J.,  2013, \mn@doi [\apjs]
  {10.1088/0067-0049/204/2/19}, \href
  {http://adsabs.harvard.edu/abs/2013ApJS..204...19P} {204, 19}

\bibitem[\protect\citeauthoryear{{Rau} \& {Cornwell}}{{Rau} \&
  {Cornwell}}{2011}]{RC11}
{Rau} U.,  {Cornwell} T.~J.,  2011, \mn@doi [\aap]
  {10.1051/0004-6361/201117104}, \href
  {http://adsabs.harvard.edu/abs/2011A%26A...532A..71R} {532, A71}

\bibitem[\protect\citeauthoryear{{Reynolds}}{{Reynolds}}{1982}]{Reynolds1982}
{Reynolds} S.~P.,  1982, \mn@doi [\apj] {10.1086/159881}, \href
  {http://adsabs.harvard.edu/abs/1982ApJ...256...13R} {256, 13}

\bibitem[\protect\citeauthoryear{{Saikia}, {K{\"o}rding}, {Coppejans},
  {Falcke}, {Williams}, {Baldi}, {Mchardy}  \& {Beswick}}{{Saikia}
  et~al.}{2018}]{Saikia2018}
{Saikia} P.,  {K{\"o}rding} E.,  {Coppejans} D.~L.,  {Falcke} H.,  {Williams}
  D.,  {Baldi} R.~D.,  {Mchardy} I.,   {Beswick} R.,  2018, \mn@doi [\aap]
  {10.1051/0004-6361/201833233}, \href
  {http://adsabs.harvard.edu/abs/2018A%26A...616A.152S} {616, A152}

\bibitem[\protect\citeauthoryear{{Sandage}, {Tammann}  \& {Yahil}}{{Sandage}
  et~al.}{1979}]{Sandage1979}
{Sandage} A.,  {Tammann} G.~A.,   {Yahil} A.,  1979, \mn@doi [\apj]
  {10.1086/157295}, \href {http://cdsads.u-strasbg.fr/abs/1979ApJ...232..352S}
  {232, 352}

\bibitem[\protect\citeauthoryear{{Terashima} \& {Wilson}}{{Terashima} \&
  {Wilson}}{2003}]{TerashimaWilson2003}
{Terashima} Y.,  {Wilson} A.~S.,  2003, \mn@doi [\apj] {10.1086/345339}, \href
  {http://cdsads.u-strasbg.fr/abs/2003ApJ...583..145T} {583, 145}

\bibitem[\protect\citeauthoryear{{Woo} \& {Urry}}{{Woo} \&
  {Urry}}{2002}]{WooUrry2002}
{Woo} J.-H.,  {Urry} C.~M.,  2002, \mn@doi [\apj] {10.1086/342878}, \href
  {http://adsabs.harvard.edu/abs/2002ApJ...579..530W} {579, 530}

\bibitem[\protect\citeauthoryear{{Wrobel} \& {Ho}}{{Wrobel} \&
  {Ho}}{2006}]{WrobelHo2006}
{Wrobel} J.~M.,  {Ho} L.~C.,  2006, \mn@doi [\apjl] {10.1086/507102}, \href
  {http://adsabs.harvard.edu/abs/2006ApJ...646L..95W} {646, L95}

\bibitem[\protect\citeauthoryear{{Zakamska} \& {Greene}}{{Zakamska} \&
  {Greene}}{2014}]{Zakamska2014}
{Zakamska} N.~L.,  {Greene} J.~E.,  2014, \mn@doi [\mnras]
  {10.1093/mnras/stu842}, \href
  {http://adsabs.harvard.edu/abs/2014MNRAS.442..784Z} {442, 784}

\makeatother
\end{thebibliography}
